\documentclass[a4paper,11pt]{article}
\usepackage{jcappub} 

\makeatletter
\gdef\@fpheader{Preprint}
\makeatother

\usepackage{comment}
\usepackage{cancel}
\usepackage[normalem]{ulem}
\usepackage{xcolor}
\title{Classical Noise in Transient USR Inflation}

\author[a]{Harish D. N.,}
\author[a,b]{Chen-Pin Yeh,}
\author[a]{Da-Shin Lee}
\affiliation[a]{Department of Physics, National Dong Hwa University, Hualien 97401, Taiwan, R.O.C.}
\affiliation[b]{Physics Division, National Center for Theoretical Sciences, Taipei 106319, Taiwan, R.O.C.}

\emailAdd{dnharish0@gmail.com}
\emailAdd{dslee@gms.ndhu.edu.tw}
\emailAdd{chenpinyeh@gms.ndhu.edu.tw}

\abstract{In single-field inflation, linear perturbations originate as quantum fluctuations and undergo a quantum-to-classical transition on super-horizon scales as the decaying mode of the comoving curvature perturbation $\mathcal{R}$ is suppressed. This transition provides a condition for the validity of a classical stochastic description of $\mathcal{R}$. We characterize classicality through the suppression of the quantum commutator relative to phase-space variances and the evolution of the squeezing parameters. In the Starobinsky model, we show that this transition becomes mode-dependent at linear order due to the transient ultra-slow-roll (USR) phase. This is because the decaying mode transiently grows during the USR phase modifying the squeezing dynamics, and shifting the classicalization time depending on when modes exit the horizon relative to the USR phase. The mode corresponding to the dip feature in the power spectrum takes the longest time to become effectively classical.  We then study the implications of this mode-dependent transition for the stochastic description of perturbations. Using the Green's function method, we show that the stochastic spectrum reproduces the power spectrum of the quantum state initialized in the Bunch-Davies vacuum, independently of the mode-dependent transition when the gradient term is retained in the Langevin equation, for both white and colored noise coarse-graining. In contrast, ignoring the gradient term in the Langevin equation introduces an explicit dependence on the coarse-graining parameter $\sigma$. Homogeneous matching procedure can be used to set mode-dependent $\sigma(k)$ that reproduces the power spectrum of the quantum state at the end of inflation. Crucially, however, we show that the matching times precede the classicalization times for modes near the USR transition. Therefore, the transient USR phase in this model leads to a mode-dependent condition for when linear curvature perturbations can be described as classical stochastic noise. }  

\begin{document}
\maketitle
\flushbottom
%%%%%%%%%%%%%%%%%%%%%%%%%%%%
\section{Introduction}
\label{sec:intro}
%%%%%%%%%%%%%%%%%%%%%%%%%%%%
Inflation describes a period of accelerated expansion in the early universe \cite{Starobinsky:1980te,Sato:1981qmu,Guth:1980zm,Linde:1981mu,Albrecht:1982wi} during which quantum fluctuations of the vacuum are stretched to cosmological scales, providing a mechanism for the formation of large-scale structure. The nearly scale-invariant power spectrum generated during slow-roll inflation, as described by linear perturbation theory~\cite{Mukhanov:1981xt, Mukhanov:1982nu, Starobinsky:1982ee, Guth:1982ec, Hawking:1982cz, Bardeen:1983qw} is in excellent agreement with the Cosmic Microwave Background (CMB) observations~\cite{Planck:2015fie}, where the fluctuations remain small on large scales. On smaller scales, which are constrained weakly by the CMB observations, enhanced large-amplitude fluctuations could produce primordial black holes (PBHs). Such large enhancements can make linear calculations unreliable, motivating methods beyond linear order, such as quantum loop corrections and stochastic inflation \cite{Starobinsky:1986fx, Starobinsky:1994bd}.

Stochastic inflation provides an effective description of the evolution of long-wavelength perturbations during inflation by integrating out short-wavelength modes using effective field theory~\cite{Starobinsky:1986fx,Starobinsky:1994bd,Bardeen:1986iq, Rey:1986zk,Goncharov:1987ir,Nambu:1987ef,Nambu:1988je,Kandrup:1988sc,Nakao:1988yi, Nambu:1989uf, Salopek:1990jq,Habib:1992ci,Linde:1993xx,Salopek:1990re,Mollerach:1990zf}. This formalism can be derived systematically from effective field theory using the Schwinger-Keldysh (in-in) path integral~\cite{Morikawa:1989xz,Calzetta:2008iqa,Matarrese:2003ye,PerreaultLevasseur:2013kfq,Wu:2006xp}. The resulting equations describe the evolution of the long-wavelength coarse-grained field after modes have crossed the horizon and undergone a quantum-to-classical transition, with dissipation and noise sourced by short-wavelength quantum modes. This transition is supported by the curvature perturbation and its canonical momentum effectively commuting in the limit of vanishing decaying mode \cite{Polarski:1995jg,  Lesgourgues:1996jc,Kiefer:1998qe,Kiefer:2008ku,dePutter:2019xxv}, which in the Schrodinger picture corresponds to the squeezing of the quantum state \cite{Grishchuk:1990bj, Albrecht:1992kf,  Lee:2004cy}. It is in this sense the quantum-to-classical transition is achieved by squeezing in the linear theory. However, squeezing may not be sufficient to diagnose classicality in nonlinear systems, where the positivity of the Wigner function must be verified~\cite{Ireland:2026txt}. We emphasize that we treat this transition strictly as the emergence of effective classical stochastic behavior for linear perturbations in a closed system, leaving the effects of non-linear dynamics, mode couplings and environmental decoherence to future work.

In the squeezed limit, the coarse-grained quantum field can be replaced by a classical random variable, and its dynamics can be treated as a stochastic Langevin equation with classical stochastic noise~\cite{Starobinsky:1994bd,Polarski:1995jg,  Lesgourgues:1996jc,Kiefer:1998qe,Kiefer:2008ku,dePutter:2019xxv}. In slow-roll (SR) inflation, squeezing occurs shortly after horizon exit, as the decaying mode of the curvature perturbation is rapidly suppressed on super-horizon scales. However, in models with a transient ultra-slow-roll (USR) phase, the decaying mode can transiently grow during the USR phase after horizon exit. The amplitude of this transient growth depends on when a given mode exits the horizon relative to the USR phase. In this work, using the piecewise-linear Starobinsky potential \cite{Starobinsky:1992ts}, we study how the quantum-to-classical transition is modified due to transient USR phase by tracking the evolution of the squeezing parameters. Since squeezing is not invariant under canonical transformations, we specify the canonical variables following Ref.~\cite{Grain:2019vnq}, where large squeezing in these variables indicates effective classicality of linear curvature perturbations. We quantify this transition using the classicality parameter, defined as the ratio of the equal-time commutator of the curvature perturbation and its canonical momentum to the product of their variances in the phase-space~\cite{Assassi:2012et}. We complement this analysis by tracking the growing and decaying modes of the curvature perturbation to show how growing-mode dominance leads to effective classical behavior~\cite{dePutter:2019xxv}. %We also complement this analysis by tracking the growing and decaying mode components of the curvature perturbations and studying how the growing-mode dominance drives the onset of effective classical behavior~\cite{dePutter:2019xxv} in this model.

We then study the implications of this mode-dependent classicalization for the classical noise assumption in stochastic inflation. In USR models, nonlinear stochastic studies in the white-noise case show that the statistics of the comoving curvature perturbation are sensitive to the coarse-graining parameter $\sigma$~\cite{Ahmadi:2022lsm,De:2020hdo,Figueroa:2020jkf,Figueroa:2021zah,Mishra:2023lhe}. In this paper, we restrict to linear stochastic dynamics to isolate the role of coarse-graining procedure (or window function properties and hence $\sigma$) from the non-linear dynamics and mode coupling. We consider both the white noise and the colored noise coarse-graining corresponding to sharp and smooth window function, respectively. We show that retaining spatial gradients in the stochastic Langevin equation exactly reproduces the power spectrum of the curvature perturbation initialized in the Bunch Davies vacuum independently of this classicalization process. However, neglecting spatial gradients in the Langevin equation leads to an explicit $\sigma$ dependence in the stochastic spectrum. A mode-dependent homogeneous matching condition~\cite{Jackson:2023obv} for $\sigma$ can reproduce the power spectrum of the quantum state initialized in the Bunch Davies vacuum. However, we show that the homogeneous matching times occurs before the classicalization times for modes near the USR transition in this model. While this distinction at the level of the two-point correlator is negligible (changing the power spectrum by less than $0.07\%$), mode-dependent classicalization nevertheless provides a criterion for determining when the linear curvature perturbation can be consistently described as a classical stochastic field. 

In this paper, we study how the quantum-to-classical transition is modified due to a transient USR phase in the Starobinsky model and its implications for the classical noise assumption in the stochastic formalism. In Section~(\ref{sec:1}), we derive the power spectrum in the Starobinsky model analytically and then study the quantum-to-classical transition using the squeezing formalism. We give classicality criteria using the classicality parameter, which identifies when perturbations can be described as an effective classical stochastic field. In Section~(\ref{sec:2}), we review the stochastic formalism and discuss when the stochastic equations can be treated as classical Langevin equations. We derive constraints on the coarse-graining parameter using the classicality criteria derived in the previous section. We conclude this section by giving equations for calculating the coarse-grained power spectrum using the Green’s function method. In Section~(\ref{sec:4classical_fluctuations}), we derive the coarse-grained power spectrum for both sharp and smooth window functions and compare it with the power spectrum of quantum fluctuations initialized in the Bunch–Davies state obtained from linear perturbation theory. We also study the stochastic white-noise spectrum when spatial gradients are neglected in the Langevin equation. We end this section by describing the homogeneous matching procedure~and then show that the matching times precede the classicalization times.  Finally, we present our conclusions in Section~(\ref{sec:conclusion}).

\section{The USR model}\label{sec:1}

In this section, we consider the Starobinsky model~\cite{Starobinsky:1992ts} to study how a transient USR phase modifies the quantum-to-classical transition of curvature perturbations. This model allows an analytic treatment of both the background and linear curvature perturbations. We first discuss the background dynamics and compute the power spectrum, and then use the squeezing formalism to study the quantum-to-classical transition. We start from the action for a single, minimally coupled scalar field with potential $V(\phi)$ given by
\begin{equation}\label{eq:single_field_action}
         S = \int d^4x \sqrt{-g} \left[ \frac{M_{\rm Pl}^2}{2} R - \frac{1}{2}   g^{\mu\nu}\partial_\mu\phi,\partial_\nu\phi - V(\phi) \right],
\end{equation}
where $g$ is the metric determinant, $R$ the Ricci scalar, and the reduced Planck mass is set to $M_{\rm Pl}=1$ throughout.

\paragraph{Background dynamics:} We describe the evolution of a spatially homogeneous scalar field on a Friedmann--Lemaître--Robertson--Walker (FLRW) spacetime using the number of e-folds, $N \equiv \ln a$ as the time variable, where $a$ is the scale factor. The background dynamics (Appendix \ref{sec:appendix_a}) are given by the Klein-Gordon equation of motion and the Friedmann constraint equation 
\begin{equation}\label{eq:background_eom}
    \frac{d^2 \bar{\phi}}{dN^2} + (3-\epsilon_1)\frac{d\bar{\phi}}{dN} + \frac{V'(\bar{\phi})}{H^2} = 0,\qquad H^2 = \frac{V(\bar{\phi})}{3 - \epsilon_1}, \qquad
\end{equation} where $H=\dot a/a$ is the Hubble parameter (with dot denoting derivative with respect to cosmic time $t$) and prime on potential denotes the derivative with respect to the field. The inflationary dynamics are characterized by the Hubble flow parameters. The first and second Hubble flow parameters are defined as 
\begin{equation}\label{eq:sr_parmaters}
    \epsilon_1 = -\frac{1}{H}\frac{dH}{dN}=\frac{1}{2}\left(\frac{d\bar{\phi}}{dN}\right)^2\quad\text{and}\quad \epsilon_2= \frac{d\ln{\epsilon_1}}{dN},
\end{equation} where, we use Eq.~(\ref{eq:background_eom}) in the second equality of $\epsilon_1$. The condition for inflation is $\epsilon_1<1$. The slow-roll inflation corresponds to the acceleration term being subdominant in Eq.~(\ref{eq:background_eom}) which requires $\epsilon_1,\epsilon_2\ll1$. The USR inflation corresponds to $dV(\phi)/d\phi$ being subdominant in Eq.~(\ref{eq:background_eom}), which requires\footnote{\label{fn:background}The USR inflation corresponds to $dV/d\phi$ being negligible in Eq.~(\ref{eq:background_eom}), such that the Klein--Gordon equation reduces to $\frac{d^2 \phi}{dN^2} + 3\frac{d\phi}{dN} \simeq 0.$ This equation integrates to $\frac{d\phi}{dN} \propto e^{-3N}$, which in turn implies $\epsilon_1 = \frac{1}{2}\left(\frac{d\phi}{dN}\right)^2 \propto e^{-6N}$ leading to $\epsilon_1 \ll 1$ and $\epsilon_2 \simeq -6$.} $\epsilon_1\ll1$ and $\epsilon_2\simeq-6$. The deviation from the slow-roll requires a feature in the potential, causing sudden transition to USR. In the Starobinsky model, this is achieved by a sudden change in slope of the potential at $\phi_T$. 

The piecewise linear Starobinsky potential with a sudden transition from slow-roll (SR) to USR inflation at $\phi_T$ is described by~\cite{Starobinsky:1992ts}
\begin{equation}\label{usr_model}
V(\phi) =
    \begin{cases}
        V_0 + A_{+} (\phi - \phi_T) & \text{if } \phi \geq \phi_T, \\
        V_0 + A_{-} (\phi - \phi_T) & \text{if } \phi < \phi_T ,
    \end{cases}
\end{equation}
where $A_+ > A_- > 0$. The analytical solution in this model has been studied extensively in the literature
\cite{Starobinsky:1992ts,Leach:2001zf,Martin:2011sn,Martin:2014kja,Ahmadi:2022lsm,Pi:2022zxs,Jackson:2023obv}. The field is initially on the SR attractor with $\phi_i > \phi_T$ and rolls down a region of constant potential slope $A_+$ until it reaches the transition point $\phi_T$. At transition, a sudden change in the slope of the potential from $A_+$ to $A_-$ leads to a rapid change in the field velocity, which temporarily breaks SR and triggers a short USR phase if $A_+\gg A_-$. Eventually, the field relaxes back onto a SR attractor with constant slope $A_-$. The parameters $V_0=2.91719\times10^{-12}$ and $A_+=1\times10^{-14}$ are fixed to match the CMB amplitude~\cite{Planck:2015fie} ($\mathcal{P}_{\mathcal{R}_{cmb}}=2.1\times10^{-9}$ ), while
\begin{equation}\label{eq:model_paramter}
    A_- = 10^{-3} A_+
\end{equation} is chosen to significantly amplify the power
spectrum ($\mathcal{P}_{\mathcal{R}_{PBH}}=5.46\times10^{-3}$) for primordial black hole production~\cite{Pi:2022zxs,Jackson:2024aoo}.

\begin{figure}[htbp]
\begin{center}
\includegraphics[width=0.495\textwidth]{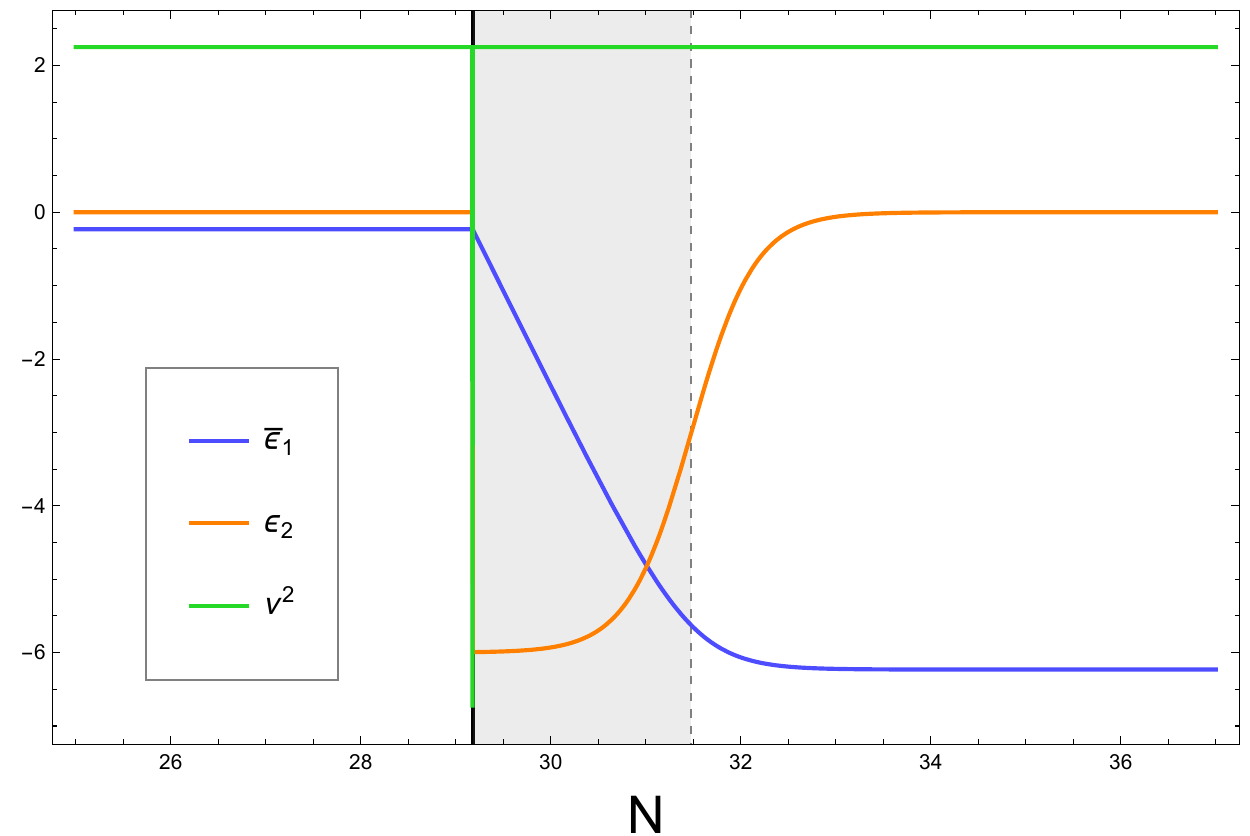}
\includegraphics[width=0.495\textwidth]{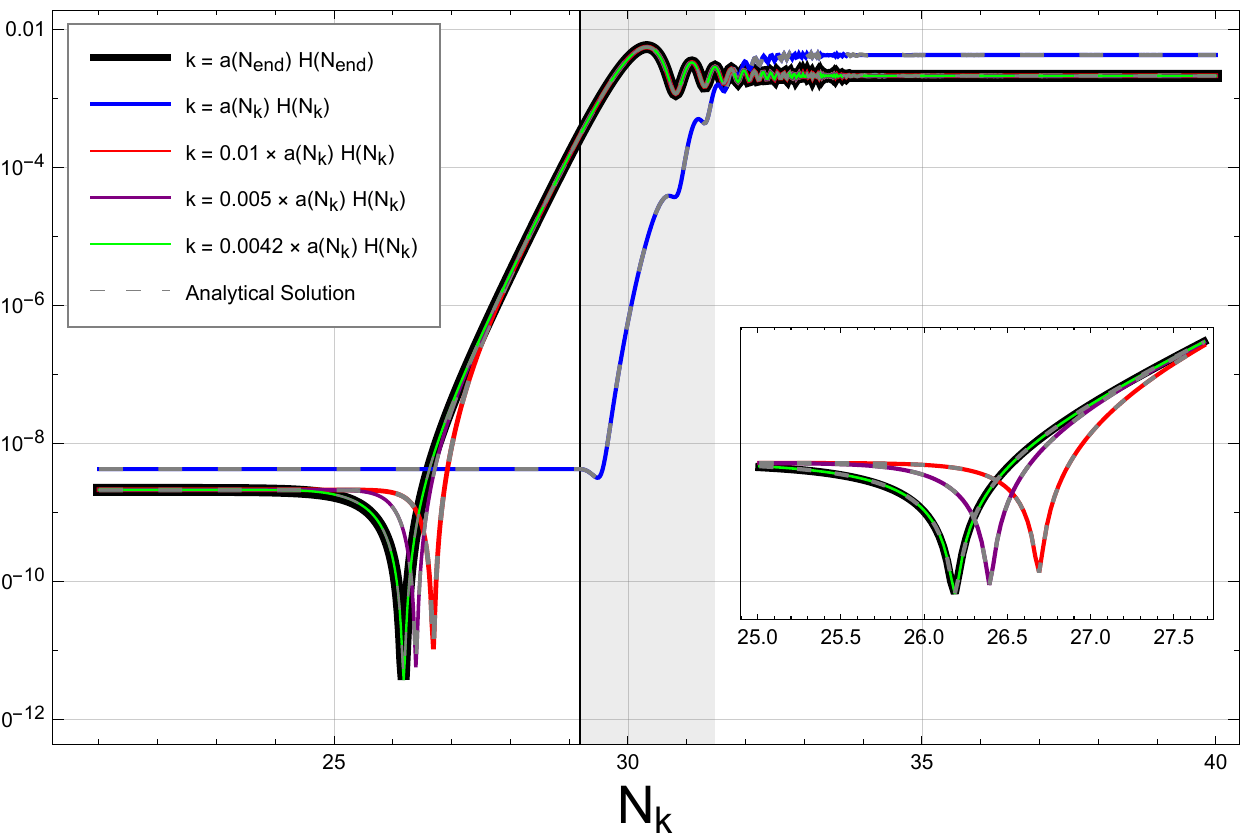}
\caption{Background dynamics and power spectrum of the Starobinsky potential. \textit{Left:}  $\epsilon_1$ (blue) is rescaled for visual clarity $\bar{\epsilon}_1=\log_{10}\epsilon_1+5$ and $\epsilon_2$ is shown in orange. The gray region shows the USR phase and the two vertical lines indicate the transition times from SR$\to$USR ($\epsilon_2=-6$ in black) and USR$\to$SR ($\epsilon_2=-3$ in dashed gray), respectively. The green line shows the Hankel index squared using Eq.~(\ref{eq:z_second_derivative}) (see Eq.~(\ref{eq:index_usr})). \textit{Right:} The power spectrum of curvature perturbations evaluated at different super-horizon times, as a function of the horizon-exit time for each mode. The blue curve represents the numerical solution at horizon exit, the red, magenta and green curves represent the numerical solutions a few e-folds after horizon exit, and the black curve shows the spectrum at the end of inflation. The corresponding analytical spectrum using Eq.~(\ref{eq:usr_qft_spectrum}), is shown as gray dashed lines.}
     \label{fig:sr-usr} 
\end{center}
\end{figure}

We work in the regime where the potential is dominated by the constant term near the transition, $V_0\gg A_\pm |\phi-\phi_T|$, so that $\epsilon_1\ll1$ and the background is quasi-de~Sitter. The Friedmann equation in this regime implies $3H^2\simeq V_0$, and the background field equation reduces to
\begin{equation}
  \frac{d^2 \bar{\phi}}{dN^2}+ 3 \frac{d \bar{\phi}}{dN} =
    \begin{cases}
     -3 \frac{A_+}{V_0}, & \bar{\phi} \geq \phi_T, \\[2mm]
     -3 \frac{A_-}{V_0}, & \bar{\phi} < \phi_T.
     \end{cases}
\end{equation}
 These equations have the solutions
\begin{equation}\label{eq:background_phi}
  \bar{\phi}(N) =
  \begin{cases}
    \bar{\phi}(N_i) - \frac{A_+}{V_0} (N-N_i) - \frac{1}{3}(\frac{d\bar{\phi}(N_i)}{dN} + \frac{A_+}{V_0}) \left(e^{-3(N-N_i)} - 1\right) &
    \text{if } \bar{\phi} \geq \phi_T, \\[2mm]
    \bar{\phi}(N_T) - \frac{A_-}{V_0} (N-N_T) - \frac{1}{3} (\frac{d \bar{\phi}(N_T)}{dN} + \frac{A_-}{V_0}) \left(e^{-3(N-N_T)} - 1\right) &
    \text{if } \bar{\phi} < \phi_T,
  \end{cases}
\end{equation} where $N_i$ corresponds to e-fold at the beginning of inflation and $N_T$ is the e-fold at the transition. The field velocity is given by
\begin{equation}\label{eq:field_velocity}
  \frac{d\bar{\phi}(N)}{dN} =
  \begin{cases}
    -\frac{A_+}{V_0}+ (\frac{d\bar{\phi}(N_i)}{dN} + \frac{A_+}{V_0}) e^{-3(N-N_i)}, & \text{if } \phi \geq \phi_T, \\[2mm]
    -\frac{A_-}{V_0} + (\frac{d\bar{\phi}(N_T)}{dN} + \frac{A_-}{V_0}) e^{-3(N-N_T)}, & \text{if } \phi < \phi_T.
  \end{cases}
\end{equation}
We see that the field and its velocity are continuous at the transition, but the field acceleration is discontinuous due to a jump in $dV/d\phi$ with the pre- and post-transition regions indicated with $+$ and $-$, respectively. This means that while $\epsilon_1$ is continuous, $\epsilon_2$ is discontinuous at the transition. The first Hubble-flow parameter immediately after the transition is~\cite{Martin:2011sn}
\begin{equation}
    \epsilon_1^{-} = \frac{1}{2} \left( \frac{d\bar{\phi}(N)^{-}}{dN} \right)^2
    = \frac{A_-^2}{2V_0^2} \left[ 1 - \left(\frac{A_+ - A_-}{A_-} \right)\, e^{-3 (N - N_T)} \right]^2 .
\end{equation} Here, we use $\frac{d\bar{\phi}(N)}{dN}\Big|_{N_T}\approx\frac{-A_+}{V_0}$ from Eq.~(\ref{eq:field_velocity}), since the field remains in slow-roll for many e-folds before the transition, $N_T\gg N_i$. The duration of the USR phase, $N_{\rm USR}$, can be estimated by finding when $\epsilon_1^{-}$ returns to its slow-roll value~\cite{Jackson:2023obv,Pi:2022zxs}, giving 
\begin{equation}\label{eq:Nusr}
   N_{\rm USR} = N - N_T \simeq \frac{1}{3} \ln\left( \frac{A_+ - A_-}{A_-}\right).
\end{equation} In Fig.~(\ref{fig:sr-usr}), the evolution of the Hubble-flow parameters are shown, where the shaded region corresponds to the USR phase given in Eq.~(\ref{eq:Nusr}), characterized by $\epsilon_2<-3$.
In the next subsection (\ref{sec:quantum_fluctuations}), we study the linear perturbations and compute the power spectrum in this model.

\subsection{Quantum fluctuations in linear perturbation theory}\label{sec:quantum_fluctuations}
In this subsection, following Refs.~\cite{Starobinsky:1992ts,Leach:2001zf,Martin:2011sn}, we derive the mode functions analytically and obtain the corresponding power spectrum in this model. The linear inhomogeneous perturbations are described by the gauge-invariant Mukhanov variable, $f(\eta,\mathbf{x})$. In Fourier space, each mode satisfies the Sasaki-Mukhanov equation~\cite{Sasaki:1986hm,Mukhanov:1988jd} (Appendix~\ref{sec:squeezing_quantization}):
\begin{equation}\label{eq:ms_equation_maintext}
    f_k'' + \left(k^2 - \frac{z''}{z}\right) f_k = 0,
\end{equation} where $z=a\sqrt{2\epsilon_1}$ and the prime denotes the
derivative with respect to conformal time, $\eta\equiv\int dt/a$. In a spatially-flat gauge
\begin{equation}\label{eq:curvature_variable}
f(\eta,\mathbf{x})=a\delta\phi(\eta,\mathbf{x})=z\mathcal{R}(\eta,\mathbf{x}),
\end{equation} where $\delta \phi$ is the field perturbation and $\mathcal{R}$ is the gauge-invariant comoving curvature perturbation \cite{Malik:2008im}. For a general potential, we have
\begin{equation}\label{eq:z_derivative}
    \frac{z'}{z}=\mathcal{H}\Big[1+\frac{\epsilon_2}{2}\Big]=\mathcal{H}\Big[1-\Big(3-\epsilon_1+\frac{a^2}{\mathcal{H}\bar{\phi}'}\frac{dV}{d\phi}\Big)\Big],
\end{equation} where $\mathcal{H}(\eta)=aH$, and we use the background equation in the second equality. The effective mass term is
\begin{align}\label{eq:z_second_derivative}
    \frac{z''}{z} = \left[ 2 - \epsilon_1 + \frac{3}{2} \epsilon_2 - \frac{1}{2} \epsilon_1 \epsilon_2
    + \frac{1}{4} \epsilon_2^2 + \frac{1}{2}\frac{\epsilon_2'}{\mathcal{H}}  \right] \mathcal{H}^2= \left[ 2 - \epsilon_1 + \delta -
    \frac{1}{H^2}\frac{d^2V}{d\phi^2} \right] \mathcal{H}^2,
\end{align} where the background equation is used again to obtain the second equality with $\delta = 6 \epsilon_1 - 2 \epsilon_1^2 + 2 \epsilon_1 \epsilon_2$, denoting the contribution from the metric perturbations. 

For the Starobinsky potential, since $\epsilon_1\ll1$ throughout the dynamics, we have 
\begin{eqnarray}\label{eq:quasi-deSitter}
    \eta\simeq-\frac{1}{\mathcal{H}},
\end{eqnarray} which is the quasi-de~Sitter limit. The function $z$ is continuous but its derivative, Eq.~(\ref{eq:z_derivative}), jumps abruptly due to the change in the potential slope at the transition. This jump generates a delta-function contribution in the effective mass term at the transition. The amplitude of this jump is
\begin{equation}\label{eq:discontinuity}
    \Delta\frac{z'}{z}=-\mathcal{H}\Big[\frac{a^2\Delta dV/d\phi}{\mathcal{H}\bar{\phi}'}\Big]=-3 \mathcal{H} \frac{(A_- - A_{+})}{ -A_{+} +
    (V_0 \frac{d\phi(N_i)}{dN} + A_+) e^{-3( N_T-N_i)} }\simeq \frac{3}{\eta_T} \frac{A_+ - A_-}{A_+},
\end{equation} where, in the second equality we use Eq.~(\ref{eq:field_velocity}). Since the field is initially in the SR attractor and evolves for many e-folds before the transition, $ N_T - N_i \gg 1$, $\frac{d\phi(N_i)}{dN} \simeq -\frac{A_{+}}{V_0}$ is used to obtain the last term.

In the slow-roll regions away from the transition $(\epsilon_{1},\epsilon_2\ll1)$, the effective mass-squared in (\ref{eq:z_second_derivative}) simplifies to
\begin{equation}
    \frac{z''}{z} \simeq\frac{2}{\eta^2}.
\end{equation} The mode functions can then be approximated by the Hankel-function solution with the constant index $\nu\simeq3/2$ given by
\begin{equation}\label{eq:usr_modefunction}
   f_k(\eta) =
\begin{cases}
\frac{1}{\sqrt{2k}} \left(1 - \frac{i}{k\eta}\right) e^{-ik\eta} & \text{if } \eta \leq \eta_T, \\[2mm]
\frac{1}{\sqrt{2k}} \Big[ \alpha_k \left(1 - \frac{i}{k\eta}\right) e^{-ik\eta} + \beta_k \left(1 + \frac{i}{k\eta}\right) e^{+ik\eta} \Big] &
\text{if } \eta > \eta_T,
\end{cases}
\end{equation} where, $\eta_T$ is the transition time and $k_T={-1/\eta_T}$ denotes the corresponding scale at the transition. The mode function before the transition contains only positive-frequency solution to match the Bunch Davies vacuum state in the asymptotic past $-k\eta\to\infty $. Post-transition, the Bogoliubov constants $\alpha_k$ and $\beta_k$ encode the mixing between positive- and negative-frequency modes caused by the sudden change in the potential slope. 
They are determined by the matching conditions at the transition:
\begin{equation}
    f_k(\eta_T^-) = f_k(\eta_T^+), \qquad
    f_k'(\eta_T^-) - f_k'(\eta_T^+) = \Delta \frac{z'}{z} \, f_k(\eta_T).
\end{equation} The resulting Bogoliubov coefficients \cite{Starobinsky:1992ts,Martin:2014kja,Jackson:2023obv},
\begin{equation}\label{eq:alpha_beta}
    \alpha_k = 1 + \frac{3i}{2\eta_Tk} \frac{A_+ - A_-}{A_+} \left( 1 + \frac{1}{\eta_T^2 k^2} \right),
    \quad
    \beta_k = -\frac{3i}{2\eta_Tk} \frac{A_+ - A_-}{A_+}\left(1-\frac{i}{k\eta_T}\right)^2 e^{-2 i k \eta_T},
\end{equation} satisfy the normalization condition $|\alpha_k|^2-|\beta_k|^2=1$. With the mode functions determined, we can now compute the comoving curvature power spectrum. Using Eqs.~(\ref{eq:curvature_variable}) and (\ref{eq:usr_modefunction}), it is given by 
\begin{equation}\label{eq:usr_qft_spectrum}
\mathcal{P}_{\mathcal{R}}^{\rm QFT}(k,N)=\frac{k^3}{2\pi^2}|\mathcal{R}_k|^2=
\frac{H^2(N)}{8\pi^2 \epsilon_1(N)}
\begin{cases} 1+x^2 & N \le N_T, \\[0.2cm] \left|\alpha_k(x+i)e^{ix} + \beta_k(x-i)e^{-ix}\right|^2 & N > N_T ,
\end{cases}
\end{equation} where we define a dimensionless variable 
\begin{equation}\label{eq:x}
    x=-k\eta\simeq\frac{k}{a(N)H(N)}.
\end{equation} In Fig~(\ref{fig:sr-usr}), the power spectrum evaluated at different horizon-crossing times are shown as a function of the horizon-exit time for each mode ($k=a(N_k) H(N_k)$). The analytical spectrum in Eq.~(\ref{eq:usr_qft_spectrum}) (gray dashed lines) is in excellent agreement with the exact numerical solutions shown in colored lines.  The power spectrum evaluated at the horizon exit ($N=N_k$, blue curve) are scale invariant for the modes that cross the horizon before the transition ($k<k_T$), while the modes that cross the horizon after the transition ($k>k_T$) are deviated from the scale invariant amplitude. However, if the spectrum is evaluated after a few e-folds from the horizon exit time (red, magenta and green curves), even modes with $k \lesssim k_T$ experience the deviation due to mixing of positive and negative frequency modes. The power spectrum has a dip when this mixing is destructive and a peak when the mixing is constructive. The spectrum at the end of inflation (black curve) is 
\begin{equation}\label{eq:ps_at_end}
    \mathcal{P}_{\mathcal{R}}^{\rm QFT}(k,N_{\rm end})
\simeq
\frac{H^2(N_{\rm end})}{8\pi^2 \epsilon_1(N_{\rm end})}
\left|\alpha_k
- \beta_k\right|^2,
\end{equation} where $N_{\rm end}$ is the e-fold at the end of inflation. The mode corresponding to the dip in the power spectrum, $k_{\star}$, can be estimated by finding the minima for $k<k_T$ to obtain \cite{Starobinsky:1992ts,Pi:2022zxs}
\begin{equation}\label{eq:dip}
    k_{\star}\approx k_{T}\sqrt{\frac{5}{2}\frac{A_-}{A_+}}\simeq k_{T}\sqrt{\frac{5}{2}}e^{-\frac{3}{2}N_{USR}},
\end{equation} up to $\mathcal{O}\big( ( \frac{k}{k_T} )^4 \big)$. In the second equality, Eq.~(\ref{eq:Nusr}) is used to express the result in terms of the duration of the USR phase with $A_+\gg A_-$. In the next subsection, we study how the transient USR phase affects the quantum-to-classical transition of modes and show that the dip mode takes the longest time to reach classical behavior.

\subsection{Quantum-to-classical transition}\label{sec:squeezing} 
In this subsection, we study how the curvature perturbations become effectively classical using the two-mode squeezing formalism~\cite{Grishchuk:1990bj,Albrecht:1992kf}. Since squeezing is not invariant under canonical transformations, we specify the canonical variables following Ref.~\cite{Grain:2019vnq} in Sec.~(\ref{sec:squeezed_state}), where large squeezing in these variables indicates effective classicality of linear curvature perturbations. Here, we refer to the quantum-to-classical transition as the emergence of effective classical stochastic behavior of curvature perturbations initialized in the Bunch-Davies vacuum on super-horizon scales. Our analysis is restricted to linear theory, with no mode couplings or environmental decoherence. We first review the squeezing formalism and derive the evolution equations for the squeezing parameters $(r_k,\varphi_k,\theta_k)$. In the super-horizon regime, where squeezing becomes large, the commutator between the curvature perturbation and its canonical momentum becomes small compared to their variances in the phase-space. We define the classicality parameter as the ratio of the squared expectation value of the equal-time commutator between the curvature perturbation and its canonical momentum to the product of their variances in the phase-space~\cite{Assassi:2012et}. We then apply this criterion to the Starobinsky model to quantify the quantum-to-classical transition of the curvature perturbations.

We recall that in the super-horizon limit $k\ll aH$, the mode solution of Eq.~(\ref{eq:ms_equation_maintext}) can be decomposed into~\cite{Leach:2001zf}
\begin{equation}\label{eq:growing_and_decaying_decomposition}f^{(\rm h)}_k(\eta)\simeq z(\eta)C_1(k)+z(\eta)C_2(k)\int^{\eta}_{\bar{\eta}}\frac{d\eta'}{z^2(\eta')},
\end{equation} where $\bar{\eta}$ is an arbitrary reference time chosen in the region where the mode is outside the horizon. The first term is called the growing mode solution since $\mathcal{R}^{(\rm h)}_k=f_k^{(\rm h)}/z\to C_1(k)$ becomes constant in the super-horizon limit $\eta\to0^{-}$, while the second term is the decaying-mode solution, which vanishes as $\eta\to0^{-}$. Thus, $\mathcal{R}^{(\rm h)}_k$ becomes constant only after the decaying mode has become negligible. Importantly, although the decaying mode asymptotically vanishes on super-horizon scales, it does not necessarily begin to decay immediately after horizon crossing at $\bar{\eta}$.  In slow roll backgrounds, where $z^{-2}=(a\sqrt{2\epsilon_1})^{-2}\propto\eta^2$, the decaying mode indeed decreases soon after horizon. In ultra-slow-roll backgrounds, however, $z^{-2}=(a\sqrt{2a^{-6}})^{-2}\propto\eta^{-4}$~\textsuperscript{(\ref{fn:background})}, the would-be decaying mode can instead increase soon after horizon exit. In the Starobinsky model, the transient USR phase therefore amplifies this decaying mode before it eventually decays once the USR phase ends. The amount by which it amplifies is mode-dependent because different $k-$modes exit the horizon at different times relative to the transient USR phase. As we will show below, this transient enhancement of the decaying mode during the USR phase modifies the squeezing dynamics and the quantum-to-classical transition of the curvature perturbations.

\subsubsection{Two-mode squeezed state}\label{sec:squeezed_state}

Since the Mukhanov variable is a real field, its Fourier modes satisfy $f_{-\mathbf{k}}=f^{*}_{\mathbf{k}}$ and therefore appear in correlated pairs $(\mathbf{k},-\mathbf{k})$, with their quantum evolution described in terms of the two-mode squeezed states. The conjugate momentum to $f(\mathbf{x},\eta)$ defined from the second-order action (see Appendix~ \ref{sec:squeezing_quantization}) is
\begin{equation}\label{eq:conjugate_momentum}
    P_\mathbf{k}=f_\mathbf{k}'-\frac{z'}{z}f_\mathbf{k}=z\mathcal{R}_{\mathbf{k}}'.
\end{equation} The canonical momentum conjugate to $\mathcal{R}_{\mathbf{k}}$ is $\Pi_{\mathbf{k}}\equiv z^2\mathcal{R}_k'$~\cite{Mukhanov:1990me}, and hence $P_{\mathbf{k}}=\Pi_{\mathbf{k}}/z$. The canonical variables are quantized by promoting them to operators satisfying the canonical commutation relations $[\hat{f}_\mathbf{k}, \hat{P}_{\mathbf{k}'}] = i \, \delta(\mathbf{k}-\mathbf{k}')$ where \cite{Martin:2015qta}
\begin{equation}\label{eq:operator_expansion}
    \hat{f}_{\mathbf{k}}=\frac{1}{\sqrt{2k}}(\hat{a}_{\mathbf{k}}+\hat{a}^{\dagger}_{\mathbf{-k}}),\quad \hat{P}_{\mathbf{k}}=-i\sqrt{\frac{k}{2}}(\hat{a}_{\mathbf{k}}-\hat{a}^{\dagger}_{\mathbf{-k}}).
\end{equation} We note that the operator expansion mixes the creation and annihilation operators associated with the modes $\mathbf{k}$ and $\mathbf{-k}$, which ensure that $\hat f_{-\mathbf{k}}=\hat f^{\dagger}_{\mathbf{k}}$. The Hamiltonian (see Appendix~\ref{sec:squeezing_quantization}) can then be written in the form  %%%%%%%
\begin{equation}\label{eq:quadratic_Hamiltonian}
\hat{H} = \int d^{3}\mathbf{k}\,
            \Bigg[ \frac{k}{2}\Big( \hat{a}_{\mathbf{k}}\hat{a}_{\mathbf{k}}^{\dagger} + \hat{a}_{-\mathbf{k}}\hat{a}_{- \mathbf{k}}^{\dagger} \Big) -\frac{i}{2}\frac{z'}{z} \Big( \hat{a}_{\mathbf{k}}\hat{a}_{-\mathbf{k}} - \hat{a}_{-\mathbf{k}}^{\dagger}\hat{a}_{\mathbf{k}}^{\dagger} \Big)
            \Bigg],
\end{equation} where the first term represents a collection of free oscillators while the second term shows the interaction between the mode $\mathbf{k}$ and $-\mathbf{k}$ due to time dependence of $z'/z$. When $z$ is constant, this interaction disappears. The structure $\hat{a}_{\mathbf{k}}\hat{a}_{-\mathbf{k}} $ implies the particles are created with opposite momenta, thus ensuring momentum conservation. The time evolution generated by this Hamiltonian can be written as a product of a two-mode squeezing transformation and a phase rotation~\cite{Albrecht:1992kf}
\begin{equation}\label{eq:squeezing_decomposition}
        \hat{U}(\eta,\eta_i)=\hat{S}(r_k,\varphi_k)\hat{R}(\theta_k).
\end{equation}  The two-mode squeezing operator is defined by 
\begin{equation}
   \hat{S}(r_k,\varphi_k)=\exp{\left(r_ke^{-2i\varphi_k}\hat{a}_{-  \mathbf{k}}(\eta_i)\hat{a}_{\mathbf{k}}(\eta_i) - r_ke^{2i\varphi_k}\hat{a}^{\dagger}_{-\mathbf{k}}(\eta_i)\hat{a}^{\dagger}_{\mathbf{k}}(\eta_i)\right)},
\end{equation} where $r_k$ is the squeeze amplitude and $\varphi_k$ is the squeezing angle. The two-mode rotation operator is defined by 
\begin{equation}
  \hat{R}(\theta_k)=  \exp{\left(-i\theta_k\hat{a}^{\dagger}_{\mathbf{k}}(\eta_i)\hat{a}_{\mathbf{k}}(\eta_i)-i\theta_k\hat{a}^{\dagger}_{-\mathbf{k}}(\eta_i)\hat{a}_{-\mathbf{k}}(\eta_i)\right)},
\end{equation} and the parameter $\theta_k$ denotes the overall phase of the mode and controls the phase evolution of the state without altering the squeezing amplitude. Acting on the Bunch--Davies vacuum state\footnote{The Bunch--Davies vacuum is selected since in the sub-horizon limit $k\gg z'/z$, the Hamiltonian reduces to that of a free harmonic oscillator~\cite{Albrecht:1992kf}.}
$\vert 0_{\mathbf{k}},0_{-\mathbf{k}}\rangle$, the evolution operator generates the two-mode squeezed state
\begin{equation}
        \hat{S}(r_k,\varphi_k)\hat{R}(\theta_k)\vert 0_{\mathbf{k}},0_{-\mathbf{k}}\rangle = \frac{1}{\cosh r_k} \sum_{n=0}^{\infty} e^{2in\varphi_k}(-1)^n\tanh^n r_k \vert n_{\mathbf{k}},n_{ \mathbf{-k}}\rangle ,
\end{equation}
where $\vert n_{\mathbf{k}},n_{-\mathbf{k}}\rangle$ denotes a two-mode number state. In phase space, the quantum state can be described by the Wigner function, defined as the Wigner--Weyl transform of the density matrix~\cite{Wigner:1932eb}. For Gaussian states such as the two-mode squeezed vacuum, the Wigner function is fully determined by its covariance matrix, and squeezing corresponds to a deformation (becoming an elongated ellipse) of this Gaussian distribution in phase space. We see that the time evolution of the quantum state reduces to finding the evolution of squeeze parameters. To determine the evolution of the squeezing parameters, we begin with the Heisenberg equation of motion
\begin{equation}\label{eq:a_eom}
      i\frac{ d \hat{a}_{\mathbf{k}}}{ d \eta} = k\,\hat{a}_{\mathbf{k}} +i\frac{z'}{z}\hat{a}_{-\mathbf{k}}^{\dagger},
\end{equation} which can be solved using a Bogoliubov transformation
\begin{equation}\label{eq:eom_a_bogoliubov}
    \hat{a}_{\mathbf{k}}(\eta) = \Sigma_k(\eta)\hat{a}_{\mathbf{k}}(\eta_{\rm i}) + \Delta_k(\eta)\hat{a}_{-\mathbf{k}}^{\dagger}(\eta_{\rm i}),
\end{equation} where \(\Sigma_k(\eta)\) and \(\Delta_k(\eta)\) are functions that depend only on the magnitude of the wavevector\footnote{If the initial state is chosen to be the vacuum---which is rotationally invariant and depends only on \(k\)---then these functions remain independent of the direction of \(\mathbf{k}\) at all times as their evolution equation depend only on $|\mathbf{k}|$.}. The canonical commutation relations imply the normalization condition
\begin{equation}\label{eq:normalization}
        \left| \Sigma_k(\eta) \right|^2 - \left| \Delta_k(\eta) \right|^2 = 1.
\end{equation} Furthermore, the combination \((\Sigma_k+\Delta_k^{*})\) satisfies $(\Sigma_k+\Delta_k^{*})'' + \left(k^2-\frac{z''}{z}\right)(\Sigma_k+\Delta_k^{*})=0,$ which is the Sasaki--Mukhanov equation and we therefore identify
\begin{equation}\label{eq:mode_and_squeezing}
    f_k=\frac{1}{\sqrt{2k}}(\Sigma_k+\Delta_k^{*}),\quad P_k=-i\sqrt{\frac{k}{2}}(\Sigma_k-\Delta_k^{*}). 
\end{equation} We now compute the Heisenberg picture operator $\hat{a}_{\mathbf{k}}(\eta)$ in terms of squeezing parameters using Eq.~(\ref{eq:squeezing_decomposition}), which gives
\begin{align}
    \hat{a}_{\mathbf{k}}(\eta)&=\hat{U}^{\dagger}\hat{a}_{\mathbf{k}}           (\eta_i)\hat{U}=\hat{R}^{\dagger}\hat{S}^{\dagger}\hat{a}_{\mathbf{k}}\hat{S}\hat{R}\\
    &=e^{-i\theta_k}\cosh{(r_k)}\hat{a}_{\mathbf{k}}(\eta_i)-e^{i(\theta+2\varphi_k)}\sinh{(r_k)}\hat{a}^{\dagger}_{-\mathbf{k}}(\eta_i)
\end{align} and comparing with equation Eq.~(\ref{eq:eom_a_bogoliubov}), we obtain 
\begin{equation}\label{eq:uandv}
    \Sigma_k(\eta)=e^{-i\theta_k}\cosh{r_k},\quad \Delta_k(\eta)=-e^{i(\theta_k+2\varphi_k)}\sinh{r_k},
\end{equation} which satisfies the normalization given in Eq.~(\ref{eq:normalization}). Finally, using Eq.~(\ref{eq:uandv}) in (\ref{eq:eom_a_bogoliubov}), the evolution equation (\ref{eq:a_eom}) leads to the equation of motion for the squeeze parameters~\cite{Albrecht:1982wi,Albrecht:1992kf,Polarski:1995jg, Martin:2015qta} 
\begin{equation}
    \frac{d r_k}{d\eta} = -\frac{z'}{z} \cos(2\varphi_k),\quad
    \frac{d \varphi_k}{d\eta} = -k + \frac{z'}{z} \coth(2r_k) \sin(2\varphi_k),\quad
    \frac{d \theta_k}{d\eta} = k - \frac{z'}{z} \tanh(r_k) \sin(2\varphi_k).
\end{equation}  The squeeze amplitude $r_k$  characterizes the strength of squeezing and determines the eccentricity of the Wigner function in phase space. The squeezing angle $\varphi_k$  fixes the orientation of the squeezing by specifying the angle between the principal axes of the Wigner function ellipse and the quadrature axes of the canonical variables. The parameter $\theta_k$ denotes the overall phase of the mode and controls the phase evolution of the state without altering the squeezing amplitude.

The equations for $r_k$ and $\varphi_k$ form a closed system and $\theta_k$ can be obtained from their solution. The exact solutions in de~Sitter space with $z'/z=-1/\eta$ are given by~\cite{Polarski:1995jg}
\begin{equation}\label{eq:ds_squeeze_solutions}
r_k(\eta) = \sinh^{-1} \Big(\frac{1}{-2k\eta}\Big),\quad
\varphi_k(\eta) = \frac{\pi}{2} - \frac{\arctan (-2k\eta)}{2},\quad
\theta_k (\eta)= k\eta +\arctan \Big(\frac{1}{2k\eta}\Big).
\end{equation} These solutions also provide an excellent approximation in quasi-de Sitter backgrounds, where $z'/z\simeq-1/\eta$. In the super-horizon limit $x=-k\eta\to0$, the quantum state becomes increasingly squeezed with the squeezing parameters approaching
\begin{align}\label{eq:dS_squeeze_paramters_superhorzion}
    r_k&=-\ln{(x)}+\mathcal{O}(x^2)\to\infty,\nonumber\\
    \varphi_k&=\frac{\pi}{2}-x+\mathcal{O}(x^2)\to\frac{\pi}{2},\nonumber\\
    \theta_k&=-\frac{\pi}{2}+x+\mathcal{O}(x^2)\to-\frac{\pi}{2}.
\end{align} Using Eq.~(\ref{eq:mode_and_squeezing}), the de~Sitter mode functions can be written as
\begin{align}\label{eq:mode_functions_in_terms_of_squeeze_parameters}
   f_k&=\frac{1}{2\sqrt{2k}} \, e^{-i\theta_k} \Big[ e^{r_k} \big(1 - e^{-2i\varphi_k}\big) + e^{-r_k} \big(1 + e^{-2i\varphi_k}\big) \Big],\nonumber\\
   P_k&=\frac{-ik}{2\sqrt{2k}} \, e^{-i\theta_k} \Big[ e^{r_k} \big(1 + e^{-2i\varphi_k}\big) + e^{-r_k} \big(1 - e^{-2i\varphi_k}\big) \Big],
\end{align} and expanding Eq.~(\ref{eq:ds_squeeze_solutions}) in the large-squeezing limit gives at leading-order
\begin{align}
    f_k&=\frac{1}{\sqrt{2k}} \Big[\left(\frac{i}{x}+\frac{i x}{2}-\frac{4 x^2}{3}+O\left(x^3\right)\right)+\left(x^2+O\left(x^3\right)\right) \Big]\sim  \frac{1}{\sqrt{2k}}\frac{i}{x},\\
    P_k&=-i\sqrt{\frac{k}{2}}\Big[\left(1-\frac{x^2}{2}+O\left(x^3\right)\right)+\left( i x+O\left(x^3\right)\right) \Big]\sim -i\sqrt{\frac{k}{2}}.
\end{align} This shows that the dynamics of the curvature perturbation are dominated by the growing-mode components in the large-squeezing limit. To show this explicitly, we decompose the de~Sitter solutions into growing and decaying components~\cite{Polarski:1995jg,Kiefer:2008ku}:
\begin{align}
f_k&=f_k^{D}+if_k^{G} =\frac{\left(\cos{x}-\frac{\sin{x}}{x}\right)}{\sqrt{2k}}+i\frac{\left(\sin{x}-\frac{\cos{x}}{x}\right)}{\sqrt{2k}}\to if_k^{G}=\frac{1}{\sqrt{2k}}\frac{i}{x},\\
P_k&=P_k^{D} + i P_k^{G}=\sqrt{\frac{k}{2}}\sin{x}-i\sqrt{\frac{k}{2}}\cos{x}\to iP_k^{G}=-i\sqrt{\frac{k}{2}}.
\end{align} Thus, in the large squeezing limit, the growing mode of $f_k\sim e^{r_k}$ dominates, while the decaying mode $f_k\sim e^{-2r_k}$ rapidly decays. Similarly, the growing mode of $P_k\sim\text{constant}$ dominates, as its decaying mode $P_k\sim e^{-r_k}$ exponentially decays.

\paragraph{Classical behavior from squeezing:} Now, let us show how large squeezing corresponds to classical behavior of perturbations. To do this, consider the field operators expressed in terms of mode functions
\begin{align}
    \hat{f}_{\mathbf{k}}(\eta)
    &= f_k(\eta)\,\hat{a}_{\mathbf{k}}(\eta_{ i})
    + f_k^*(\eta)\,\hat{a}^{\dagger}_{-\mathbf{k}}(\eta_{ i}), \\
    \hat{P}_{\mathbf{k}}(\eta)
    &= P_k(\eta)\,\hat{a}_{\mathbf{k}}(\eta_{ i})
    + P_k^*(\eta)\,\hat{a}^{\dagger}_{-\mathbf{k}}(\eta_{ i}).
\end{align} Rewriting them in terms of the operators at initial time and taking super horizon limit gives~\cite{dePutter:2019xxv} 
\begin{align}
    \hat{f}_{\mathbf{k}}(\eta) 
                &= \sqrt{2k}\,f_k^{D}\,\hat{f}_{\mathbf{k}}(\eta_{ i}) - \sqrt{\frac{2}{k}}\,f_k^{G}\,\hat{P}_{\mathbf{k}}(\eta_{i})\approx- \sqrt{\frac{2}{k}}\,f_k^{G}\,\hat{P}_{\mathbf{k}} (\eta_{ i})\equiv\hat{f}_\mathbf{k}^{G},\\
    \hat{P}_{\mathbf{k}}(\eta)
                &= \sqrt{2k}\,P_k^{D}\,\hat{f}_{\mathbf{k}}(\eta_{ i}) - \sqrt{\frac{2}{k}}\,P_k^{G}\,\hat{P}_{\mathbf{k}}(\eta_{i})\approx - \sqrt{\frac{2}{k}}\,P_k^{G}\,\hat{P}_{\mathbf{k}}(\eta_{ i})\equiv\hat{P}_\mathbf{k}^{G},
\end{align} where in the last equality we use the large squeezing limit $r_k > 1,~\varphi_k\to\pi/2$, when the decaying components are suppressed. The canonical growing mode operators commute $[\hat{f}^{G}_{\mathbf{k}}(\eta),\hat{P}^{G}_{\mathbf{k}'}(\eta)]~ \approx~ 0$ in this approximate sense and perturbations can be considered effectively classical even though the state remains pure~\cite{Polarski:1995jg,dePutter:2019xxv}. Thus, classical limit corresponds to $f_k^{G}\gg f_k^{D}$ and $P_k^{G}\gg P_k^{D}$. We emphasis that classical here means that we can treat $\hat{f_\mathbf{k}}$ and $\hat{P}_\mathbf{k}$ as stochastic classical variables, drawn from a stochastic ensemble rather than as operators. These variables evolve classically with a probability distribution given by the Wigner function, reproducing the quantum statistics of the system in this limit~\cite{dePutter:2019xxv}. 

\subsubsection{Classicality condition} We consider a mode as effectively classical at late-times if the equal-time commutator between the curvature perturbation and its conjugate momentum becomes small compared to the phase-space variances~\cite{Assassi:2012et}:
\begin{equation}\label{eq:classicality_parameter}
  \mathcal{C}_k(\eta)\equiv \frac{|\langle [\hat{\mathcal{R}}_k, \hat{\Pi}_k] \rangle^2|}{\langle\hat{\mathcal{R}}^2_k\rangle \langle\hat{\Pi}^2_k\rangle}= \frac{|\langle [\hat{f}_k ,\hat{P}_k] \rangle^2|}{\langle\hat{f}^2_k\rangle \langle\hat{P}^2_k\rangle}= \frac{1}{|f_k|^2|P_k|^2}\ll1.
\end{equation} This definition implies that when the classicality parameter $\mathcal{C}_k\ll1$, the dynamics of the mode can be described by classical stochastic variables whose statistical properties are determined by the underlying quantum state. Although this criterion is not invariant under canonical transformations, we use it here specifically to quantify when the curvature perturbation can be effectively described by classical stochastic variables. Since this transition is attained gradually rather than at a sharply defined instant, we introduce a practical definition of the classicalization time. We define $\eta_{\rm cl}$ as the time at which the classicality parameter has decreased to $1\%$ of its initial Bunch--Davies value
\begin{equation}\label{eq:classicality_criteria}
    \mathcal{C}_k(\eta_{\rm cl}) = 0.04.
\end{equation} This corresponds to a two-order-of-magnitude reduction relative to the asymptotic sub-Hubble value $\mathcal{C}_k(\eta\to-\infty)=4$. Since $\mathcal{C}_k(\eta)$ exhibits non-monotonic evolution, the threshold may be crossed multiple times; we therefore define $\eta_{\rm cl}$ as the earliest time such that $\mathcal{C}_k(\eta')<0.04$ for all subsequent times $\eta' \in (\eta_{\rm cl},0^{-})$. This choice corresponds to a regime where $\hat{\mathcal{R}}_k$ and $\hat{\Pi}_k$ (and equivalently $\hat f_k$ and $\hat P_k$) are highly correlated, with the correlation coefficient defined as~\cite{dePutter:2019xxv,Grain:2019vnq}
\begin{align}\label{eq:correlation_coefficient}
    \rho_k^{\rm corr}(\eta)
        &\equiv \frac{\Re\!\left(\langle \hat{\mathcal{R}}_k  \hat{\Pi}_k^{\dagger} \rangle\right)}{\sqrt{\langle \hat{\mathcal{R}}_k \hat{\mathcal{R}}_k^{\dagger} \rangle \, \langle \hat \Pi_k \hat \Pi_k^{\dagger} \rangle}} = \frac{\Re\!\left(\langle \hat f_k \hat P_k^{\dagger} \rangle\right)}{\sqrt{\langle \hat f_k \hat f_k^{\dagger} \rangle \, \langle \hat P_k \hat P_k^{\dagger} \rangle}} = \frac{\Re(f_k P_k^{*})}{|f_k|\,|P_k|}.
\end{align} Using the Wronskian, $ W = f_k P_k^{*} - P_k f_k^{*} = i $, one obtains $\Im(f_k P_k^{*}) = 1/2$, which fixes the modulus $|f_k P_k^{*}|^2 = \Re(f_k P_k^{*})^2 + \Im(f_k P_k^{*})^2$. Expressing the result in terms of $\mathcal{C}_k$ then gives
\begin{equation}
    \rho_k^{\rm corr}(\eta)=\sqrt{1-\frac{\mathcal{C}_k(\eta)}{4}}~\Bigg|_{\eta_{cl}} \simeq0.995 \, .
\end{equation} Therefore, the criterion in Eq.~(\ref{eq:classicality_criteria}) identifies a late-time regime in which $\hat{\mathcal{R}}_k$ and $\hat{\Pi}_k$ (and equivalently $\hat f_k$ and $\hat P_k$) remain highly correlated, and the dynamics of the mode can be effectively described by classical stochastic variables. We apply this classicality criteria to study how USR phase modifies the transition to effectively classical behavior in the Starobinsky model.

\subsubsection{Mode-dependent transition in Starobinsky model}
We now use the classicality criteria defined in Eq.~(\ref{eq:classicality_criteria}) and show how transient USR phase modifies the quantum-to-classical transition in this model. To track the evolution of the squeezing parameters across the USR transition, we invert Eq.~(\ref{eq:uandv}) to obtain \begin{equation}\label{eq:squeeze_paramters} r_k=\sinh^{-1}{\left(|\Delta_k^{*}|\right)},\quad \varphi_k=-\frac{1}{2}\arg{\left(\frac{-\Delta_k^{*}}{\Sigma_k}\right)}. 
\end{equation} The functions $\Delta_k^{*}$ and $\Sigma_k$ can be written in terms of the mode functions using Eq.~(\ref{eq:mode_and_squeezing}) as 
\begin{equation}\label{eq:sigma_and_delta_relation}
    \Sigma_k=\frac{1}{\sqrt{2k}}(kf_k+iP_k),\quad
    \Delta_k^{*}=\frac{1}{\sqrt{2k}}(kf_k-iP_k).
\end{equation}  To complement this description, we also explicitly evolve the growing and decaying mode components by decomposing\footnote{ In the super-horizon limit, the growing mode is constant (while the decaying mode decays) and the spectrum at the end of inflation is proportional to $\propto|\alpha_k-\beta_k|^2$ consistent with Eq.~(\ref{eq:ps_at_end}).} Eq.~(\ref{eq:usr_modefunction}), into  \begin{equation}\label{eq:canonical_transormation_relation}f_k(\eta) = \begin{cases} f_k^{D}+if_k^{G}, & \text{if } \eta \leq \eta_T, \\[2mm] (\alpha_k+\beta_k)f_k^{D}+i(\alpha_k-\beta_k)f_k^{G}, & \text{if } \eta > \eta_T, \end{cases} \end{equation} and similarly for the momentum. We organize the discussion by first considering the case without a USR phase $(\alpha_k=1,\beta_k=0)$ and then comparing it with the USR case.

\begin{figure}[htbp]
    \centering
    \includegraphics[width=0.487\textwidth]{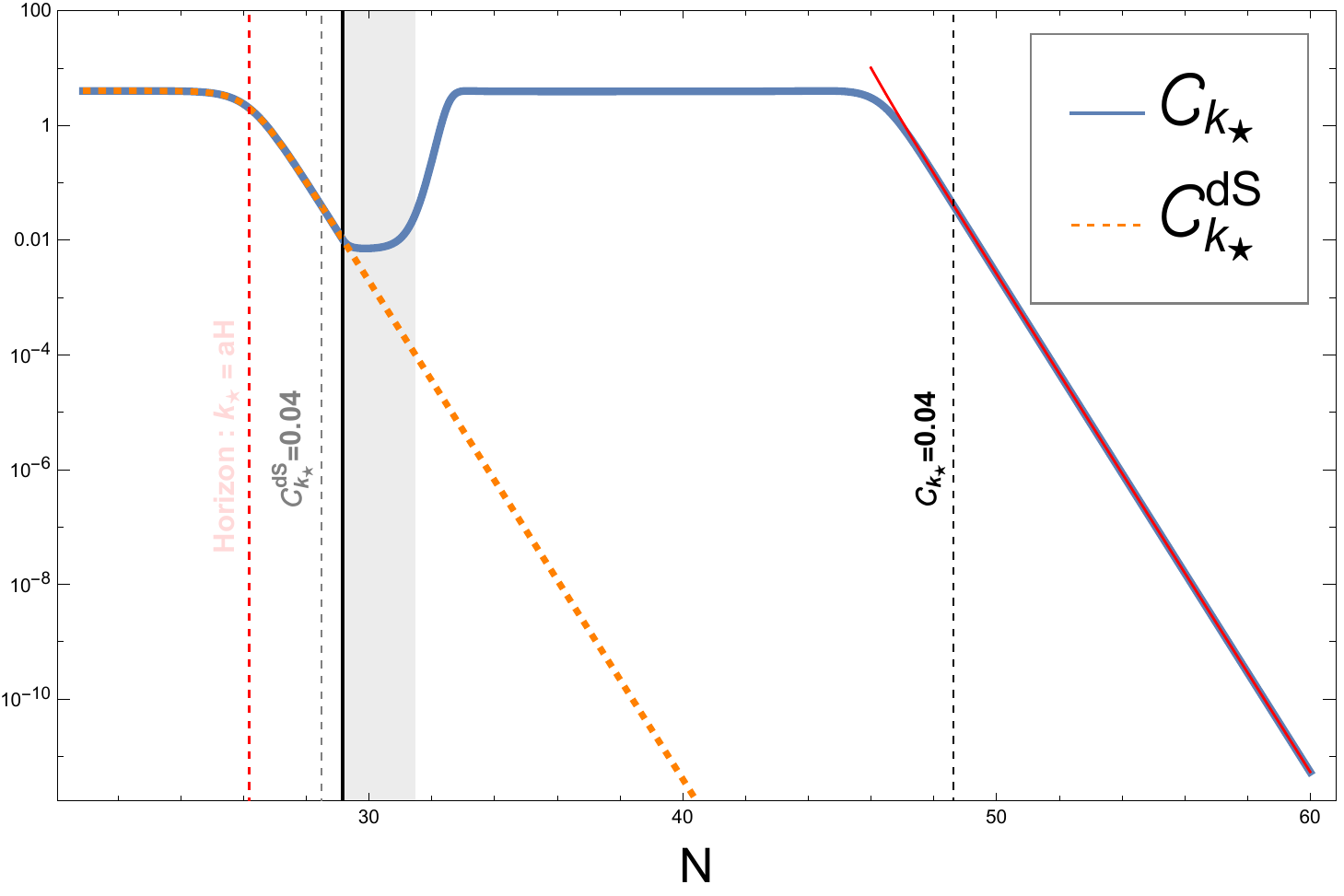}
    \includegraphics[width=0.5\textwidth]{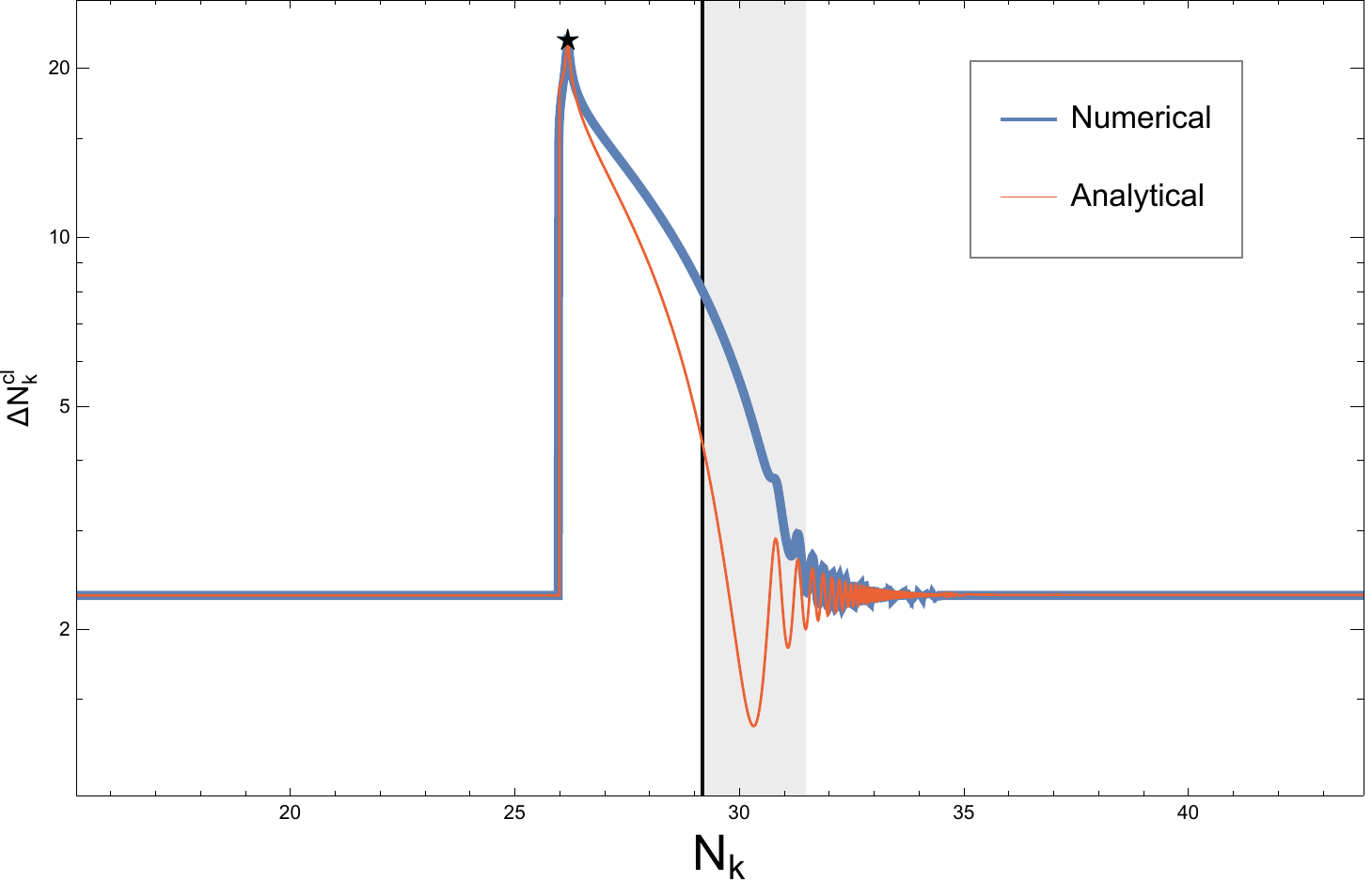}
    \caption{\textit{Left:} Evolution of the classicality parameter for the dip mode. We show Eq.~(\ref{eq:classicality_parameter}) in blue obtained using numerical solution of mode functions. The dashed orange curve shows the corresponding evolution without a USR phase using Eq.~(\ref{eq:classicalization_analytical_pre_26}) and the solid red curve shows the super-horizon approximation from Eq.~(\ref{eq:classicality_after_trantion}). The red vertical dashed line marks horizon exit, while the gray and black dashed lines indicate the classicalization times in the absence and presence of a USR phase, respectively. USR phase is shown in gray shared region with the solid black line indicating the SR-to-USR transition. \textit{Right:} The classicalization time measured from horizon exit, $\Delta N^{cl}_{k}=N^{cl}_k-N_k$ as a function of $N_k$ obtained from classicality condition in Eq.~(\ref{eq:classicality_criteria}). The red curve shows the analytical result using Eq.~(\ref{eq:slowroll_classicalization_time}) for $N_k<26$ and Eq.~(\ref{eq:classical_delay_time}) for $N_k\geq26$ with the black star indicating the dip mode (see Eq.~(\ref{eq:dip})).}
    \label{fig:classicality_parmater}
\end{figure}

\paragraph{In the absence of USR phase:}  The mode function in this case $(\alpha_k=1,\beta_k=0)$ takes the quasi-de Sitter form given in Eq.~(\ref{eq:usr_modefunction})
\begin{equation}\label{eq:mode_ds}
    f_k=\frac{1}{\sqrt{2k}}(1-\frac{i}{k\eta})e^{-ik\eta},\quad P_k=-i\sqrt{\frac{k}{2}}e^{-ik\eta}.
\end{equation} Using the definition in Eq.~(\ref{eq:classicality_parameter}) and condition in Eq.~(\ref{eq:classicality_criteria}), we find
\begin{equation}\label{eq:classicalization_analytical_pre_26}
 \mathcal{C}_k^{dS}=\frac{4}{1+x(N^{cl})^{-2}}=0.04\implies x(N^{cl})=   \frac{1}{\sqrt{99}},
\end{equation} where $x(N^{cl})=-k\eta_{cl}\simeq k/a(N^{cl})H(N^{cl})$. The number of e-folds required for the mode to become classical after the horizon exit can be obtained from the expression of $x(N^{\rm cl})$, which gives
\begin{equation}\label{eq:slowroll_classicalization_time}
    N^{cl}-N_k\simeq-\ln{\left(\frac{1}{\sqrt{99}}\right)}=2.297\,.
\end{equation} This duration is scale-invariant, taking approximately 2.3 e-folds after horizon exit for all modes to become effectively classical in the absence of a USR phase. In the left panel of Fig.~(\ref{fig:classicality_parmater}), we show the evolution of classicality parameter in the absence of USR phase by the orange dashed curve for the dip mode $k_\star$ given in Eq.~(\ref{eq:dip}). The classicality parameter \textit{remains} $\mathcal{C}_{k_{\star}}<0.04$, after $2.3$ e-folds from the horizon exit marked by the gray vertical dashed line.

\begin{figure}[htbp]
\centering
\includegraphics[width=.493\textwidth]{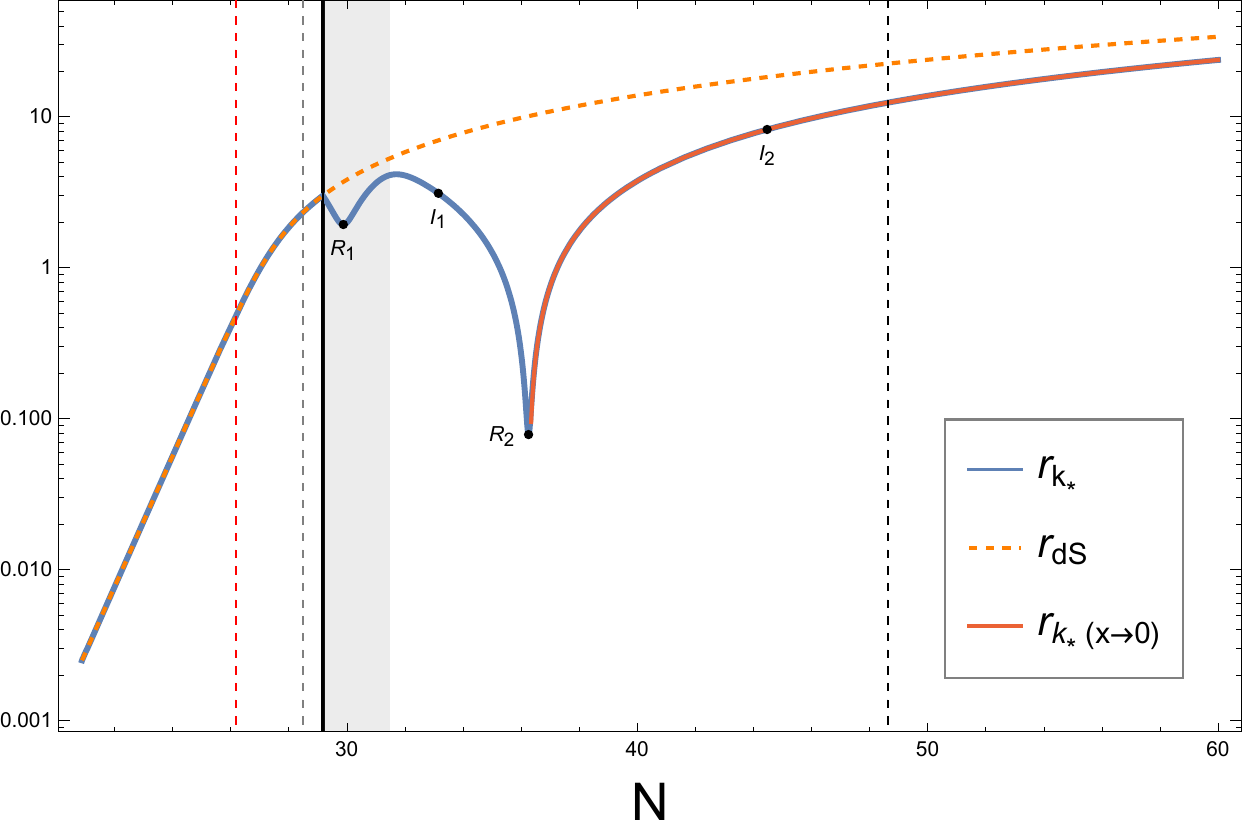}
\hfill
\includegraphics[width=.49\textwidth]{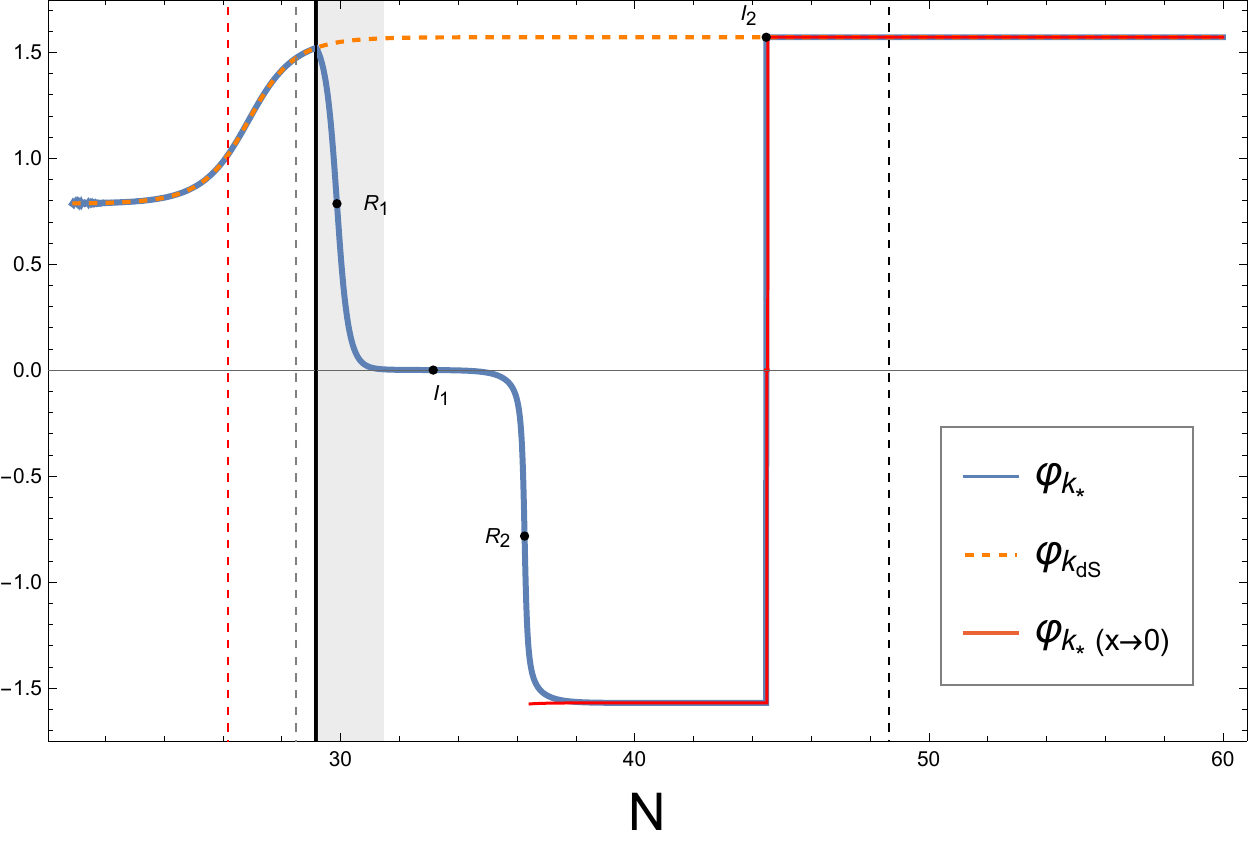}
    \caption{ Evolution of the squeezing parameters for the dip mode $k_{\star}$, showing how the USR phase modifies the squeezing dynamics and delays classicalization. The solid blue curves show the exact numerical solution from Eq.~(\ref{eq:squeeze_paramters}), the dashed orange curves show the slow-roll result from Eq.~(\ref{eq:squeeze_parameter_dS}), and the red curves show the super-horizon approximation from Eq.~(\ref{eq:squeeze_parameters_modified}). The black vertical dashed line marks the onset of the classical regime defined by Eq.~(\ref{eq:classicality_criteria}). For comparison, the gray vertical dashed line shows when the same condition is satisfied in slow-roll. The red dashed vertical line marks horizon exit, while the solid black line marks the SR-to-USR transition. The evolution passes through the phase-space points~\textsuperscript{\ref{fn:phase_space_points}} $R_1 \to I_1 \to R_2 \to I_2$ (corresponding to changes in the dominance of growing and decaying components of $f_{k_\star}$ and $P_{k_\star}$.}
     \label{fig:squeeze_parameters}
\end{figure}

\begin{figure}[htbp]
\centering
\includegraphics[width=.493\textwidth]{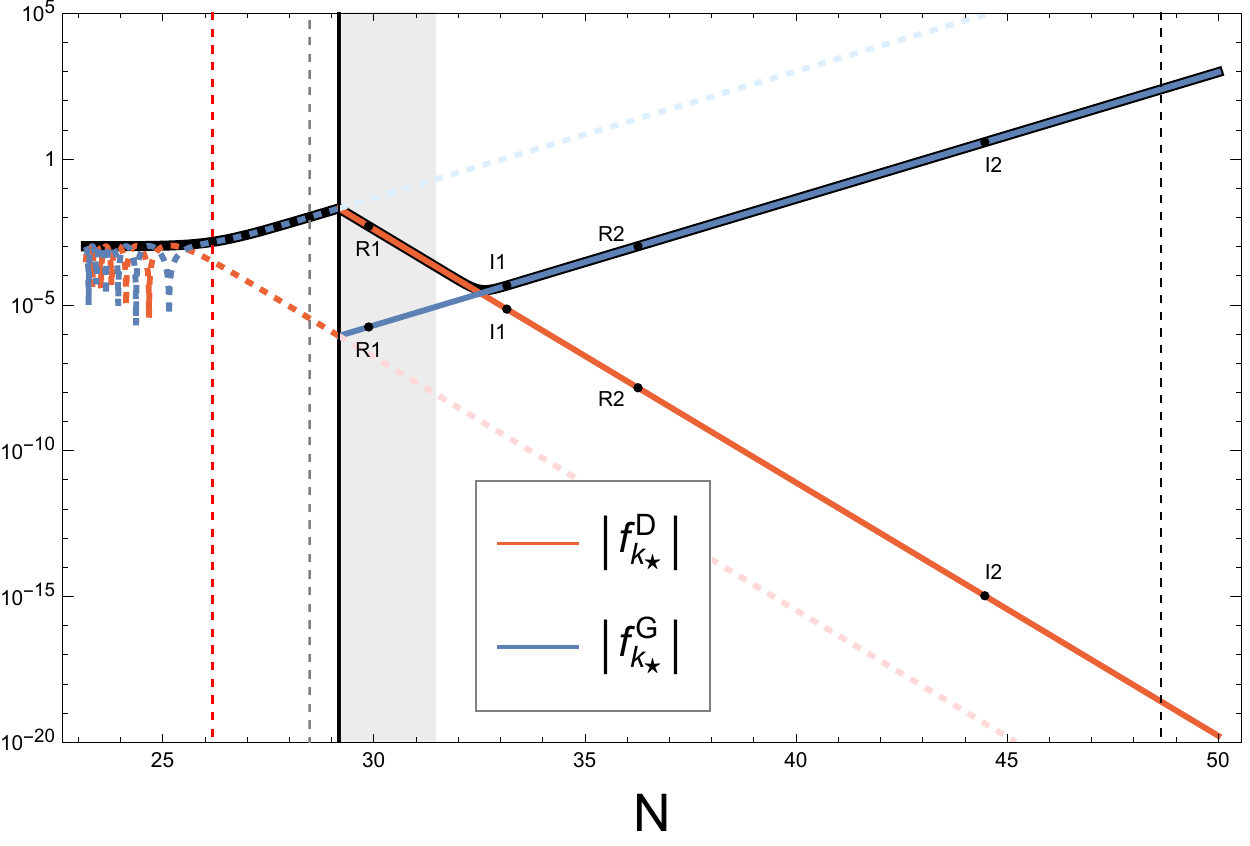}
\hfill
\includegraphics[width=.49\textwidth]{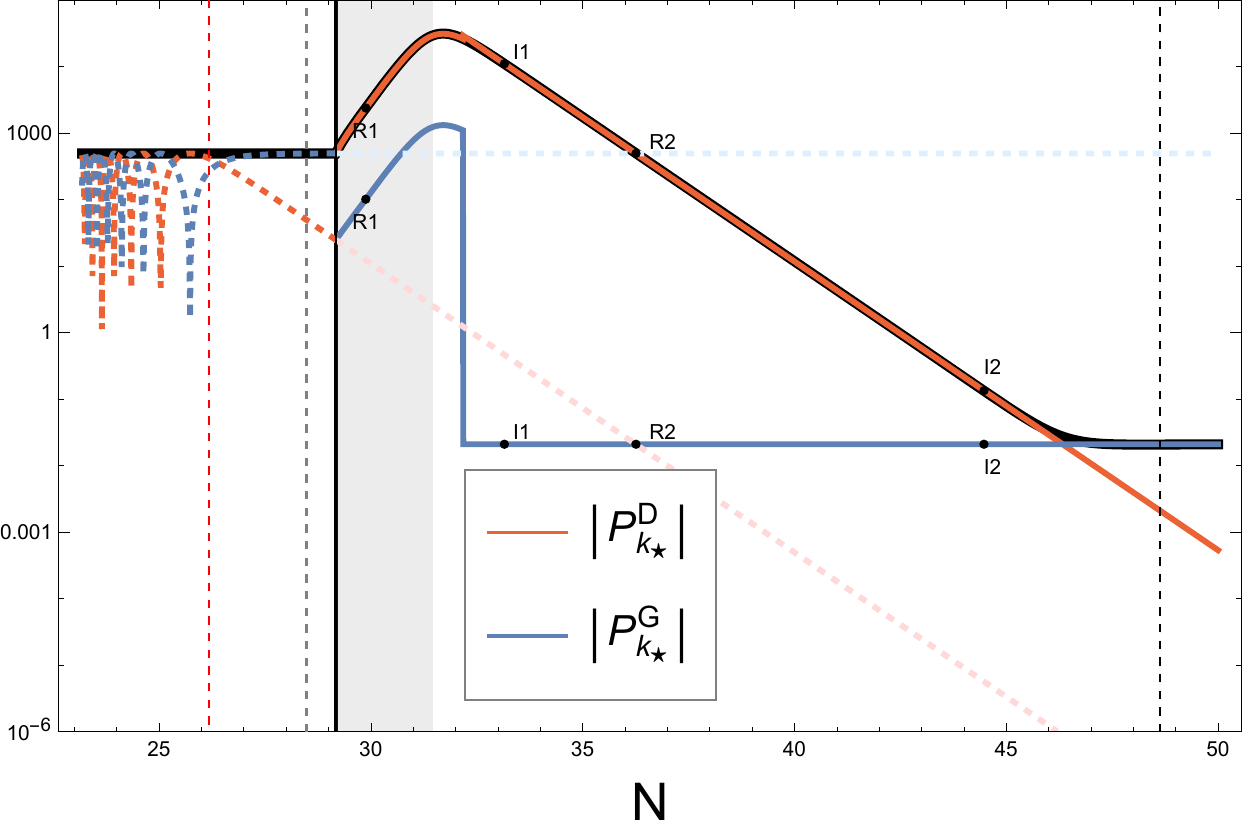}
     \caption{ Evolution of the growing and decaying components of $k_{\star}$ before and after the SR--USR transition using Eq.~(\ref{eq:conjugate_momentum}) and Eq.~(\ref{eq:canonical_transormation_relation}). We use  \(z'/z = \mathcal{H}(1+\epsilon_2/2)\) immediately after the SR--USR transition to ensure continuity of the momentum \(P_{k_\star} = f_{k_\star}' - (z'/z)f_{k_\star}\), and set \(\epsilon_2 = 0\) once the system settles back into the SR regime (after \(\approx 2.3\) e-folds). The sudden SR--USR transition flips the roles of the growing and decaying modes which are shown as dashed blue and red curves before the transition, and as solid blue and red curves after the transition. The thick black curve denotes the full numerical solution for $|f_{k_\star}|$ and $|P_{k_\star}|$. As the state becomes squeezed, $f^{G}_{k_\star} \sim e^{r_{k_\star}}$ grows exponentially while $P^{D}_{k_\star} \sim e^{-r_{k_\star}}$ is exponentially suppressed. The red dashed vertical line marks horizon exit, the thick black vertical line marks the transition time, and the black dashed vertical line indicates the onset of the classical regime defined by Eq.~(\ref{eq:classicality_criteria}). The system evolves through the phase-space points~\textsuperscript{\ref{fn:phase_space_points}} $R_1 \to I_1 \to R_2 \to I_2$.}\label{fig:dip_dynamics} 
     
\end{figure}

We now discuss how the squeezed state evolves in the absence of a USR phase using Eqs.~(\ref{eq:squeeze_paramters})-(\ref{eq:sigma_and_delta_relation}). In this case, the mode functions in Eq.~(\ref{eq:mode_ds}) lead to the corresponding squeeze parameters given by
\begin{equation}\label{eq:squeeze_parameter_dS}
r_k^{\mathrm{dS}} = \sinh^{-1}\Big(\frac{1}{2x}\Big)\simeq-\ln{x}+\mathcal{O}(x),\quad
\varphi_k^{ \mathrm{dS}} = \frac{\pi}{2} - \frac{\arctan 2x}{2}\simeq\frac{\pi}{2}-x+\mathcal{O}(x^3).
\end{equation} The squeezing parameter and the squeeze angle increases monotonically as the modes exit the horizon, with $\varphi_k^{ \mathrm{dS}}$ evolving from $\pi/4$ deep inside the horizon to $\pi/2$ in the super-horizon limit as shown by the orange dashed line in Fig.~(\ref{fig:squeeze_parameters}). They take the values 
\begin{equation}\label{eq:r_threshold}
   r_k^{ \mathrm{dS}}(\eta_{cl})\simeq2.3,\quad \varphi_k^{\mathrm{dS}}(\eta_{cl})\simeq0.93\times\frac{\pi}{2},
\end{equation} for $\mathcal{C}_{k_{\star}}(\eta_{cl})=0.04$, obtained by substituting $x=1/\sqrt{99}$ into Eq.~(\ref{eq:squeeze_parameter_dS}). This corresponds to the regime where the growing mode components dominates the dynamics which can be seen from Fig~(\ref{fig:dip_dynamics}), where the absolute magnitudes of growing (dashed blue curve) and decaying mode components (dashed red curve) are shown. Therefore, modes become effectively classical when the growing mode components of both $\hat f_k$ and $\hat P_k$ dominate, as indicated by the vertical gray dashed line in Fig~(\ref{fig:squeeze_parameters}), Fig~(\ref{fig:dip_dynamics}) and the left panel of Fig~(\ref{fig:classicality_parmater}).

\paragraph{With the transient USR phase:} In this case, we can obtain an approximate analytical estimate of the classicalization time using Eq.~(\ref{eq:classicality_criteria}) together with the post-transition mode function given in Eq.~(\ref{eq:usr_modefunction}). These mode functions undergo Bogoliubov mixing:
\begin{equation}\label{eq:post_trantion_modes}
f_k = \frac{1}{\sqrt{2k}} \Big[ \alpha_k \Big(1 - \frac{i}{k\eta}\Big) e^{-ik\eta} + \beta_k \Big(1 + \frac{i}{k\eta}\Big) e^{ik\eta} \Big],
\quad
P_k = -i\sqrt{\frac{k}{2}} \Big[ \alpha_k e^{-ik\eta} - \beta_k e^{ik\eta} \Big],
\end{equation} with $\alpha_k,\beta_k$ defined in Eq.~(\ref{eq:alpha_beta}) and in the super-horizon limit ($x=-k\eta\to0$), the classicality parameter in Eq.~(\ref{eq:classicality_parameter}) is given by 
\begin{align}\label{eq:classicality_after_trantion}
    \mathcal{C}_k&\approx\frac{4x^2}{|\alpha_k-\beta_k|^4}+\mathcal{O}(x^3)=0.04\implies
    x(N^{cl})\approx\frac{|\alpha_k-\beta_k|^2}{10}.
\end{align} The number of e-folds required for a mode to become effectively classical after horizon exit is approximately given by
\begin{equation}\label{eq:classical_delay_time}
  \Delta  N^{cl}_{k}=N^{cl}_{k}-N_k\approx\ln{\left(\frac{10}{|\alpha_k-\beta_k|^2}\right)}.
\end{equation} The USR phase leads to a mode-dependent classicalization time through the $k$-dependence of $|\alpha_k-\beta_k|^2$. In the absence of a USR phase when $\alpha_k=1,\beta_k=0$, Eq.~(\ref{eq:classical_delay_time}) reduces to Eq.~(\ref{eq:slowroll_classicalization_time}). Therefore, the quantum-to-classical transition becomes mode-dependent due to the transient USR phase in this model. The classicalization time is largest for the mode with smallest value of $|\alpha_k-\beta_k|$, which corresponds to the dip mode. Using Eq.~(\ref{eq:alpha_beta}) and Eq.~(\ref{eq:dip}) for the dip mode, we have 
\begin{equation}\label{eq:classical_delay_dip}
    \lvert \alpha_{k_\star} - \beta_{k_\star} \rvert \approx 
 \left\lvert 
-\frac{5 i}{6}\sqrt{\frac{5}{2}} \left(\frac{A_-}{A_+}\right)^{3/2}
+ \mathcal{O}\!\left[\left(\frac{A_-}{A_+}\right)^{2}\right]
\right\rvert.
\end{equation} Substituting Eq.~(\ref{eq:model_paramter}), we analytically estimate $\Delta N_{k_\star}^{cl}\approx 22.47$ e-folds. To verify this analytical estimate, we solve Eq.~(\ref{eq:ms_equation_maintext}) numerically to obtain the mode functions and determine the classicalization time using the criterion in Eq.~(\ref{eq:classicality_criteria}). This gives
\begin{equation}\label{eq:classicalization_time_dip}
    \Delta N_{k_\star}^{\rm cl}=22.45,
\end{equation} e-folds (see Fig.~(\ref{fig:classicality_parmater})), which shows that the analytical estimate for the dip mode is in good agreement with the numerical result. We show the evolution of $\mathcal{C}_{k_\star}(\eta)$ in the presence of USR phase for the dip mode with blue curve in the left panel of Fig~(\ref{fig:classicality_parmater}). The classicality parameter in this case evolves non-monotonically but \textit{remains} $\mathcal{C}_{k_{\star}}<0.04$ only after 22.45 e-folds beyond which $\hat f_k$ and $\hat P_k$ stays highly correlated as shown by the black vertical dashed line in the left panel of Fig.~(\ref{fig:classicality_parmater}).

We now discuss the evolution of the squeeze parameters using Eqs.~(\ref{eq:squeeze_paramters})-(\ref{eq:sigma_and_delta_relation}). In this case, using Eq.~(\ref{eq:post_trantion_modes}) in super-horizon limit (\(x=-k\eta \to 0\)), the squeezing parameters are given by
\begin{equation}\label{eq:squeeze_parameters_modified}
r_k \simeq -\ln x + \ln|\alpha_k - \beta_k| + \mathcal{O}(x), \quad
\varphi_k \simeq \pm\frac{\pi}{2} - x + 2x^2 \frac{\Im(\alpha_k \beta_k^*)}{|\alpha_k - \beta_k|^2} + \mathcal{O}(x^3).
\end{equation} These analytical solutions reduce to Eq.~(\ref{eq:squeeze_parameter_dS}) in the absence of a USR phase ($\alpha_k=1,~\beta_k=0$). The $\pm$ sign in front of $\pi/2$ in Eq.~(\ref{eq:squeeze_parameters_modified}) comes from the quadrant of $-\Delta^{*}_k / \Sigma_k$ in the complex plane.\footnote{\label{fn:phase_space_points} For example, see Fig.~(\ref{fig:squeeze_parameters}) where the evolution of the dip mode is shown. The real part $-\Delta^{*}_k / \Sigma_k$ is positive between the points $R_1$ and $R_2$ and negative elsewhere, while the imaginary part is positive between $I_1$ and $I_2$ and negative outside this region. In the region between $R_2$ and $I_2$, the real part of $-\Delta^{*}_k / \Sigma_k$ is negative while the imaginary part is positive, which sets $\varphi_k = -\pi/2$. Beyond $I_2$, both the real and imaginary parts are negative, resulting in $\varphi_k = +\pi/2$.}  Clearly, the USR phase modifies the squeezing parameters in a mode-dependent manner through the $k$-dependent Bogoliubov coefficients $\alpha_k$ and $\beta_k$. We see from Eq.~(\ref{eq:squeeze_parameters_modified}), that the squeeze parameters evolve non-monotonically due to the mixing of the positive- and negative-frequency modes which are encoded in the Bogoliubov coefficients (see Fig.~(\ref{fig:squeeze_parameters})). The squeeze angle $\varphi_k$ evolves from $\pi/4$ deep inside the horizon to $\pi/2$ in the super-horizon limit when the growing mode components dominate the dynamics. The values of squeeze parameters when $\mathcal{C}_{k_{\star}}(\eta_{cl})=0.04$ are obtained by using Eq.~(\ref{eq:classicality_after_trantion}) in Eq.~(\ref{eq:squeeze_parameters_modified}) and this gives
\begin{equation}
r_{k_{\star}}\simeq12.4,\quad\varphi_{k_{\star}}\simeq\pi/2.
\end{equation} The squeeze amplitude $r_{k_\star}$ is larger than the value in Eq.~(\ref{eq:r_threshold}), due to the USR phase. From Fig~(\ref{fig:dip_dynamics}), we explicitly see that the growing mode component of curvature perturbation $f_{k_\star}=z\mathcal{R}_{k_\star}\sim e^{r_{k_\star}}$ dominate exponentially while the decaying mode component of the conjugate momentum $P_{k_\star}=\Pi_{k_\star}/z\sim e^{-r_{k_\star}}$ is exponentially suppressed. However, the growing mode components of both $f_{k_\star}$ and $P_{k_\star}$ dominate only after $\approx22.45$ e-folds.

Therefore, for any given mode $k$, the classicality condition in Eq.~(\ref{eq:classicality_criteria}) identifies the point beyond which  $\hat f_k$ and $\hat P_k$ remain highly correlated as defined in Eq.~(\ref{eq:correlation_coefficient}) and the growing-mode components dominate the dynamics. This allows for a stochastic description of inflationary perturbations which will be discussed in the next section. The quantum-to-classical transition becomes strongly mode dependent due to the sudden transition to the USR phase in this model. This is illustrated in the right panel of Fig.~(\ref{fig:classicality_parmater}), where we plot $\Delta N_k^{\rm cl}$, denoting the number of e-folds between horizon exit and the onset of effective
classicality obtained using Eq.~(\ref{eq:classicality_criteria}), as a function of $N_k$. We find significant variation in the classicalization time over the range \(0.04\,k_T\) to \(25\,k_T\), corresponding to horizon-crossing times $N_k\approx N_T-3.2$ to $N_k\approx N_T+3.2$ e-folds. The longest classicalization delay occurs for the dip mode, and this conclusion is robust against the precise choice of the classicality threshold (as long as $\mathcal{C}_k<1$).

In summary, the squeeze parameters evolve monotonically in SR inflation, and modes classicalize uniformly after horizon exit. In models with a transient USR phase however, positive and negative frequency modes mix, producing $k$-dependent Bogoliubov coefficients. This mixing causes squeeze parameters to evolve non-monotonically after horizon exit controlled by $|\alpha_k-\beta_k|$ in the Starobinsky model. Modes with destructive interference (dips in $|\alpha_k-\beta_k|$) experience a temporary delay in the growth of $r_k$ and take longer to become effectively classical. The mode corresponding to the dip in the power spectrum (Eq.~(\ref{eq:dip})) takes the longest time to reach effective classical behavior, as the decaying component of momentum $P_{k_\star}^{D}$ remains dominant for the longest duration (see Fig.~\ref{fig:dip_dynamics}). Therefore, the USR phase leads to a mode-dependent quantum-to-classical transition and this is the main conclusion of this section. We expect this mode-dependent transition to be generic in single-field inflation models producing primordial black holes through an enhancement of the decaying mode of curvature perturbations \cite{Ozsoy:2023ryl}. In the next section, we study the implications of this result to the classical noise assumption in the stochastic description of the curvature perturbation.

\section{Stochastic Formalism}\label{sec:2}
In the rest of this paper, we focus on the implications of mode-dependent classicalization for the stochastic description of linear fluctuations. To isolate the dependence on the coarse-graining scale from non-linear dynamics and mode coupling, we restrict our analysis to linear stochastic dynamics. In this section, we review the stochastic formalism and derive the coarse-grained power spectrum using the Green's function method. 

In the stochastic approach, the field operator in the Heisenberg-picture is split into a long-wavelength $(k< aH)$ coarse-grained component $\hat\phi_{cg}(\mathbf{x},N)$ and a short-wavelength $(k> aH)$ Ultra-Violet (UV) component $\hat\phi_{Q}(\mathbf{x},N)$ as \cite{Starobinsky:1994bd}
\begin{equation}\label{eq:linear_perturbation_split}
        \hat\phi(\mathbf{x},N)=\hat\phi_{cg}(\mathbf{x},N)+\hat\phi_{Q}(\mathbf{x},N).
 \end{equation} The equation of motion for the coarse-grained field can be formally derived by integrating out short-wavelength modes using effective field theory in the Schwinger-Keldysh formalism~\cite{Mukhanov:1981xt, Mukhanov:1982nu, Starobinsky:1982ee, Guth:1982ec, Hawking:1982cz, Bardeen:1983qw}. Here, we derive it from the scalar field equation at leading order (tree-level), neglecting quantum loop corrections~\cite{Syu:2019uwx,Cheng:2021lif,Cheng:2023ikq,Lin:2024gar} due to the self-interaction of the scalar field. To study the evolution of the coarse-grained field, we retain the gradient terms in the scalar field equation derived from the action, Eq.~(\ref{eq:single_field_action}), but neglect them in the constraint equations where metric perturbations are subdominant~\cite{Cheung:2007st,Briaud:2025ayt}. Since the field perturbations in the Starobinsky model are decoupled from the metric perturbations to a very good approximation (see Appendix~\ref{sec:appendix_a}), we use the decoupling limit to describe the field fluctuations in this model. In a spatially flat FLRW background with scale factor $a(t)$, the inhomogeneous scalar field dynamics are governed by the Klein-Gordon equation in terms of e-folds $N=\ln a$:
\begin{equation} \label{eq:KG_full_grad}
    \frac{d^2\hat\phi}{dN^2} + (3 - \epsilon_1)\frac{d\hat\phi}{dN} - \frac{1}{a^2 H^2} \nabla^2\hat\phi + \frac{V'(\hat\phi)}{H^2} = 0,
\end{equation} where $ \epsilon_1$ and $H$ are given by the background Friedmann equation without the gradient term in the decoupling limit~\cite{Cheung:2007st,Briaud:2025ayt}. Substituting Eq.~(\ref{eq:linear_perturbation_split}) in Eq.~(\ref{eq:KG_full_grad}) leads to the long-wavelength coarse-grained field equation~\cite{Grain:2017dqa,Jackson:2024aoo,Briaud:2025ayt}
\begin{equation}\label{eq:stochastic_grad}
    \frac{d^2\hat\phi_{cg}}{dN^2} + (3 - \epsilon_1)\frac{d\hat\phi_{cg}}{dN} - \frac{1}{a^2 H^2} \nabla^2 \hat\phi_{cg} +
    \frac{V'(\hat\phi_{cg})}{H^2} = \mathbf{\hat S}( \mathbf{x},N).
\end{equation} The source term $\mathbf{\hat S}( \mathbf{x},N)$ is approximated to linear order in the short-wavelength field given by
\begin{equation}\label{eq:noise_source}
  \mathbf{\hat S}( \mathbf{x},N)  \equiv-\left[ \frac{d^2\hat\phi_Q}{dN^2} + (3 - \epsilon_1)\, \frac{d\hat\phi_Q}{dN} - \frac{\nabla^2}{a^2 H^2}
  \hat\phi_Q + \frac{V''(\hat\phi_{cg})}{H^2} \hat\phi_Q +\mathcal{O}(\hat\phi_{Q}^2)\right],
\end{equation} whose properties depend on the choice of the coarse-graining window function and the background dynamics. The split in Eq.~(\ref{eq:linear_perturbation_split}) is implemented using a Fourier-space window function:
\begin{equation}\label{eq:phi_Q}
    \hat\phi_Q(\mathbf{x},N) = \int \frac{d^3 \mathbf{k}}{(2\pi)^{3/2}} W\Big(\frac{k}{k_\sigma(N)}\Big) \hat{\phi}_\mathbf{k}(N) e^{i \mathbf{k} \cdot \mathbf{x}},
\end{equation} where $k_{\sigma}(N)$ is the time-dependent cutoff in the Fourier space with $W(k/k_\sigma)\simeq0$ when $k/k_\sigma\ll1$ and $W(k/k_\sigma)\simeq1$ when $k/k_\sigma\gg1$. The field operator $\hat{\phi}_{\mathbf{k}}$ and its conjugate momentum\footnote{The conjugate momentum to $\phi$ in the ADM formalism is given by $ \pi = \frac{\sqrt{\gamma}}{\mathcal{N}}(\frac{\partial \phi}{\partial \eta} - \mathcal{N}^i \partial_i \phi), $ where $\gamma$ is the determinant of the induced spatial metric $\gamma_{ij}$, $\mathcal{N}$ is the lapse function, and $\mathcal{N}^i$ is the shift vector~\cite{Liguori:2004fa}. In the decoupling limit, where metric perturbations are neglected, $\pi=a^2\partial\phi/\partial \eta$ and for the linear perturbations ($\pi=\bar{\pi}+\delta\pi$) this reduces to $ \delta\pi = a^2 \frac{d \delta\phi}{d\eta}=a^3H\frac{d\delta\phi}{dN}$~\cite{Grain:2017dqa} or in Fourier space $\pi_{\mathbf{k}}=a^3H\frac{d\phi_\mathbf{k}}{dN}$.}  $\hat{\pi}_{\mathbf{k}}= a^3H\frac{d\hat{\phi}_{\mathbf{k}}}{dN}$ are expanded with creation and annihilation operators satisfying the usual commutation relation. They define the Bunch-Davies vacuum state, and imposing canonical commutation relation leads to the Wronskian condition for the mode functions $\phi_k\pi_k^{*}-\pi_k\phi_k^{*}=i.$ The mode functions satisfy \begin{equation}\label{eq:mode_equation}
    \frac{d^2 \phi_k}{dN^2} + (3 - \epsilon_1) \frac{d\phi_k}{dN} + \Big[ \frac{k^2}{a^2 H^2} + \frac{V''(\phi_{cg})}{H^2} \Big] \phi_k = 0,
\end{equation} which follows from Eq.~(\ref{eq:KG_full_grad}) with background quantities evaluated at the coarse-grained field $\phi_{cg}$, and the short-wavelength modes exiting the coarse-graining scale act as a stochastic source. The source term can be written in Fourier space using Eq.~(\ref{eq:phi_Q}) in Eq.~(\ref{eq:noise_source}) as 
\begin{equation}
\mathbf{\hat{S}}_{\mathbf{k}}(N) =A(k,N)\hat{\phi}_{\mathbf{k}}(N)+B(k,N)\frac{d\hat{\phi}_{\mathbf{k}}(N)}{dN},
\end{equation} with
\begin{equation}
    A(k,N)\equiv-
\Big( \frac{\partial^2 W(k,N)}{\partial N^2} + (3 - \epsilon_1)\frac{\partial W(k,N)}{\partial N} \Big) ,\quad B(k,N)\equiv-
 2 \frac{\partial W(k,N)}{\partial N}.
 \end{equation} The expectation value of the source term in Fourier space with respect to the initial Bunch–Davies vacuum is
\begin{align}\label{eq:quantum_noise_correlator}
\langle \hat{S}_{\mathbf{k}}(N_1)\hat{S}^{\dagger}_{\mathbf{k'}}       (N_2)\rangle
    &=\delta(\mathbf{k}-\mathbf{k'})\Big[
    A_1 A_2  \phi_{k}(N_1) \phi^*_{k}(N_2)+ B_1B_2\frac{\pi_{k}(N_1)}{a^3(N_1)H(N_1)} \frac{\pi^*_{k}(N_2) }{a^3(N_2)H(N_2)}  \nonumber \\
    &\quad + A_1B_2 \phi_k(N_1) \frac{\pi^{*}_k(N_2)}{a^3(N_2)H(N_2)}
    + B_1A_2\frac{\pi_k(N_1)}{a^3(N_1)H(N_1)}   \phi^{*}_k(N_2)
    \Big],
\end{align} where $A_j = A(k,N_j)$ and $B_j =B(k,N_j)$ for $j=1,2$. The coarse-grained field equation in Eq.~(\ref{eq:stochastic_grad}) can be interpreted as a classical Langevin equation by replacing the quantum fields with stochastic variables, which is justified once the system undergoes a quantum-to-classical transition on super-horizon scales \cite{Polarski:1995jg,  Lesgourgues:1996jc,Kiefer:1998qe,Kiefer:2008ku,dePutter:2019xxv}. The noise term in Eq.~(\ref{eq:stochastic_grad}) is then replaced by a classical stochastic noise $\mathbf{S}(\mathbf{x},N)$. The statistics of this classical noise is determined by the quantum two-point correlation functions of the underlying quantum field 
\begin{equation}
\big<S_{\mathbf{k}}(N_1)S^{*}_{\mathbf{k'}}(N_2)\big>\equiv\big<0|\hat{S}_{\mathbf{k}}(N_1)\hat{S}^{\dagger}_{\mathbf{k'}}(N_2)|0\big>.    
\end{equation} In the next subsection, we constrain the coarse-graining procedure by requiring that the stochastic noise becomes effectively classical when modes undergo the quantum-to-classical transition, as discussed in Section~(\ref{sec:squeezing}). 

\subsection{Coarse-graining procedure}\label{sec:4}

The window function defines the coarse-graining procedure by separating long-wavelength coarse-grained modes from short-wavelength UV modes in Fourier space. The coarse-graining scale is determined by requiring that modes enter the coarse-grained sector only after the quantum-to-classical transition. This transition is associated with strong squeezing of the modes on super-horizon scales. However, squeezing may not be sufficient to diagnose classicality in nonlinear systems, where the positivity of the Wigner function must be verified \cite{Ireland:2026txt}. Since we consider linear dynamics in this work, which preserves the Gaussianity of the initial vacuum state (and hence the Wigner function remains positive), we use the squeezing formalism and Eq.~(\ref{eq:classicality_criteria}) as the classicality criterion.

\subsubsection{Window function}
\paragraph{White-noise coarse-graining:} In the white-noise approach, coarse-graining is implemented with a sharp step-function in momentum space
\begin{equation}\label{eq:white_noise_filter}
    W\Big( \frac{k}{k_\sigma(N)} \Big) = \Theta\Big( \frac{k}{\sigma a(N) H(N)} - 1 \Big),
\end{equation} which transfers modes instantaneously when they cross the coarse-graining scale $k_{\sigma}(N)=\sigma a(N)H(N)$, where $\sigma$ is the coarse-graining parameter. The coarse-graining time $N_\sigma$ is defined by $k=k_\sigma(N_\sigma)$ and $\sigma$ is then related to the number of e-folds by $ \sigma\simeq e^{-(N_\sigma-N_k)},$ where $N_k$ is the horizon exit time with $H(N_\sigma)\simeq H(N_k)$. 

Since different Fourier modes become effectively classical at different times in a transient USR model, the coarse-graining parameter becomes mode-dependent given by
\begin{equation}\label{eq:classical_sigma}
    \sigma^{cl}(k)\simeq e^{-(N^{cl}_{k}-N_k)},
\end{equation} where $N_{k}^{cl}$ is the e-fold at which modes become effectively classical,~Eq.~(\ref{eq:classicality_criteria}).

If instead, we choose $\sigma$ to correspond to the mode $k_\star$, which takes the longest time to become classical in the USR model, then
\begin{align}\label{eq:sigma_constraint_general}
    \sigma&\lesssim e^{-(N^{cl}_{k_{\star}}-N_{k_{\star}})}=\sigma^{cl}(k_{\star}),
\end{align} where $N_{k_{\star}}^{cl}$ is the e-fold at which this mode becomes classical. This choice ensures that all modes coarse-grained at $N_\sigma=N_{k_{\star}}^{cl}$ have completed their quantum-to-classical transition and the stochastic noise can be assumed effectively classical for all the modes.  Using Eq.~(\ref{eq:Nusr}) to relate horizon exit to USR duration while assuming $H\simeq\text{const.}$, the constraint in Eq.~(\ref{eq:sigma_constraint_general}) takes the form
\begin{equation}\label{eq:sigma_constraint_linear_model}
    \sigma \lesssim  e^{-\frac{3}{2}N_{USR}+N_T+\frac{1}{2}\ln{(\frac{5}{2})}-N^{cl}_{k_{\star}}}=\sigma^{cl}(k_{\star}).
\end{equation} 
\paragraph{Colored-noise coarse-graining:} In the colored-noise approach, coarse-graining is implemented using a smooth momentum-space window function, so that modes are transferred continuously over a finite range around $k \sim \sigma aH$. We consider smooth window functions of the form
\cite{Matarrese:2003ye,Casini:1998wr,Winitzki:1999ve,Liguori:2004fa,Wu:2006xp,Mahbub:2022osb}
\begin{equation}\label{eq:colored_noise_filter}
    W\Big(\frac{k}{k_{\sigma}(N)}\Big)
    = 1 - \exp\!\left[-\frac{1}{2}\left(\frac{k}{\sigma a(N)H(N)}\right)^n\right],
\end{equation}
where $n=1,2$ correspond to exponential and Gaussian filters, respectively. Once $\sigma$ is specified, the transition is centered around the coarse-graining time $N_\sigma$, defined by $k=\sigma a(N_\sigma)H(N_\sigma)$. The sharp coarse-graining time of the white-noise prescription is replaced by a finite transition interval
\begin{equation}\label{eq:transtion_interval}
    \Delta N^{(n)} \equiv N - N_\sigma = -\frac{1}{n} \ln \Big[ -2 \ln (1 - \tilde{\delta_c}) \Big] ,
\end{equation} which characterizes the number of e-folds required for a mode to be effectively coarse-grained up to a tolerance $\tilde{\delta_c}$\footnote{For example, $\tilde{\delta_c}=0.01$ corresponds to $1-W=1-\tilde{\delta_c}$, i.e. $99\%$ completion of coarse-graining.}. Importantly, this interval depends on the filter sharpness parameter $n$, and is independent of $\sigma$. For $\sigma$ given by Eq.~(\ref{eq:sigma_constraint_linear_model}), the smooth transition is centered around $N_\sigma=N^{\rm cl}_k$, while $\Delta N^{(n)}$ gives the additional number of e-folds from $N_\sigma=N^{\rm cl}_k$ required to complete the coarse-graining up to the chosen tolerance. We discuss the implications of this transition for the power spectrum in Sec~(\ref{subsec:color}). 
 
\subsection{Linear Langevin equation}

In this subsection, we derive the coarse-grained power spectrum assuming that the modes are effectively classical as defined in Eq.~(\ref{eq:classicality_criteria}). The coarse-grained equation, Eq.~(\ref{eq:stochastic_grad}), can then be treated as a Langevin equation: 
\begin{equation}\label{eq:Langevin_grad}
    \frac{d^2\phi_{cg}}{dN^2} + (3 - \epsilon_1)\frac{d\phi_{cg}}{dN} - \frac{1}{a^2 H^2} \nabla^2 \phi_{cg} +
    \frac{V'(\phi_{cg})}{H^2} = \mathbf{S}( \mathbf{x},N),
\end{equation} where $\phi_{cg}$ is a classical stochastic variable, and $\mathbf{S}(\mathbf{x},N)$ is a classical stochastic noise. The statistical properties of this classical noise are obtained from Eq.~(\ref{eq:quantum_noise_correlator}).  To study the linear evolution of the coarse-grained field and its two-point correlator, we decompose it into a homogeneous background and linear perturbations:
\begin{equation}
    \phi_{cg}(\mathbf{x},N) = \bar{\phi}_{cg}(N)+\delta\phi_{cg}(\mathbf{x},N).
\end{equation} $\bar{\phi}_{cg}(N)$ is the background coarse grained field, and $\delta \phi_{cg}(\mathbf{x},N)$ represents the linear
perturbations. Substituting this decomposition into Eq.~(\ref{eq:Langevin_grad}) gives
\begin{align}
    \frac{d^2 \bar{\phi}_{cg}}{dN^2}+(3-\epsilon_1)\frac{d \bar{\phi}_{cg}}{dN}+\frac{V'(\bar{\phi}_{cg})}{H^2}&=0,\label{eq:1}\\
   \frac{d^2\delta\phi_{cg}}{dN^2} + (3 - \epsilon_1)\frac{d\delta\phi_{cg}}{dN} - \frac{\nabla^2\delta\phi_{cg}}{a^2 H^2}  +
   \frac{V''(\bar{\phi}_{cg})}{H^2}\delta\phi_{cg}&=\mathbf{S}(\mathbf{x},N),\label{eq:2}
\end{align} where the background quantities are evaluated using the coarse-grained field without the stochastic source. The formal solution to Eq.~(\ref{eq:2}) in Fourier-space is
\begin{equation}
\delta\phi_{cg,\mathbf{k}}(N) = \int_{N_i}^{N} dN_1 G_k(N, N_1)  \mathbf{S}_{\mathbf{k}}(N_1),
\end{equation} where $G_k(N, N_1)$ is the retarded Green’s function of the homogeneous equation. The coarse-grained field is initialized at $N_i$ with background value $\bar{\phi}_{\rm cg}(N_i)$ and vanishing initial coarse-grained perturbations $ \delta \phi_{\rm cg}(\mathbf{x},N_i)=0$ and $\frac{d \delta \phi_{\rm cg}(\mathbf{x},N)}{dN}|_{N_i}=0$, while the sub-Hubble modes are placed in the Bunch–Davies vacuum and act as a source of stochastic noise.  The two point correlator is given by
\begin{equation}
\langle \delta\phi_{cg}(\mathbf{x},N)\delta\phi_{cg}^*(\mathbf{x}',N')\rangle
= \int \frac{d^3\mathbf{k}}{(2\pi)^{3/2}}\int \frac{d^3\mathbf{k'}}{(2\pi)^{3/2}}
\, e^{i\mathbf{k}\cdot\mathbf{x}-i\mathbf{k}'\cdot\mathbf{x}'}
\langle \delta\phi_{cg,\mathbf{k}} \, \delta\phi^{*}_{cg,\mathbf{k}'} \rangle.
\end{equation}
Taking the quantum expectation values of canonical variables in the source term, and identifying them as stochastic moments we have
\begin{equation}\label{eq:c_function_delta}
    \langle \delta\phi_{cg,\mathbf{k}} \, \delta\phi^{*}_{cg,\mathbf{k}'} \rangle
= \delta(\mathbf{k}-\mathbf{k}') \, \mathcal{C}_k(N,N'),
\end{equation} where $\mathcal{C}_k(N,N')$ contains all contributions from the canonical variables:
\begin{align}\label{eq:c_function}
\mathcal{C}_k(N,N') &= \int_{N_i}^{N} dN_1 \int_{N_i}^{N'} dN_2 \,
G_k(N,N_1) G_k^*(N',N_2)\langle \mathbf{S}_{\mathbf{k}}(N_1) \mathbf{S}^{*}_{\mathbf{k}}(N_2)\rangle.
\end{align}
We see that the statistics of the coarse-grained field depend on the shape of the window function used in the coarse-graining procedure. Using the
delta function in Eq.~(\ref{eq:c_function_delta}) and performing the angular integrals in spherical coordinates, the correlator reduces to
\begin{equation}
\langle \delta\phi_{cg}(\mathbf{x},N) \delta\phi_{cg}^*(\mathbf{x}',N')\rangle
= \int_0^\infty \frac{dk}{k} \frac{k^3}{2\pi^2} \frac{\sin(kr)}{kr} \, \mathcal{C}_k(N,N'),
\quad r = |\mathbf{x}-\mathbf{x}'|.
\end{equation}

Finally, taking the equal-time and equal-space limit, \(N=N'\) and \(r\to 0\), defines the coarse-grained power spectrum:
\begin{equation}
\langle \delta\phi_{cg}(\mathbf{x},N) \delta\phi_{cg}^*(\mathbf{x},N)\rangle
= \int_0^\infty \frac{dk}{k} \, \mathcal{P}_{\delta\phi_{cg}}(k,N),
\end{equation}
with the power spectrum given by
\begin{equation}\label{eq:stochastic_power_spectrum}
    \mathcal{P}_{\delta\phi_{cg}}(k,N) = \frac{k^3}{2\pi^2} \, \mathcal{C}_k(N,N).
\end{equation}

\section{Coarse-grained Stochastic Power Spectrum}\label{sec:4classical_fluctuations}

In this section, we compute the stochastic power spectrum of curvature perturbations in the Starobinsky model using both the white-noise and the colored-noise window functions. For the white-noise case, we also consider the long-wavelength limit of the coarse-grained equation to study the consequence of dropping the gradient term in Eq.~(\ref{eq:2}). We compare the resulting stochastic spectrum with the standard power spectrum of the quantum state initialized in the Bunch-Davies vacuum state given by Eq.~(\ref{eq:usr_qft_spectrum}).

In this model, $\epsilon_1\ll1$ throughout the evolution and the coarse-grained field equation, Eq.~({\ref{eq:2}}), reduces to 
\begin{equation}\label{eq:stoc_eom}
    \frac{d^2\delta\phi_{cg}}{dN^2} + 3 \frac{d\delta\phi_{cg}}{dN} - \frac{\nabla^2\delta\phi_{cg}}{a^2 H^2} + \frac{V''(\bar{\phi}_{cg})}{H^2}\delta\phi_{cg}
    =\mathbf{S}(\mathbf{x},N).
\end{equation} The formal solution for each Fourier mode is
\begin{equation}
\delta\phi_{cg,\mathbf{k}}(N)
=\int_{x_i}^{x(N)} \frac{dy_1}{-y_1}\,
G_k(x,y_1)\,\mathbf{S}_k(y_1) ,
\end{equation}
where $y_1 = x(N_1)$ is introduced as the integration variable, and $x_i=x(N_i)$ denotes the initial time of integration. The factor $-1/y_1$ comes from the change of integration variable $dN_1 = -dy_1/y_1$ in the quasi–de Sitter background. The Green's function is constructed from the homogeneous solutions of Eq.~(\ref{eq:stoc_eom}) (see Appendix \ref{sec:greens_function} for derivation):
\begin{equation} \label{eq:greensfunction_fullgrad}
    G_k(x,y_1)= \frac{(x-y_1) \cos (x-y_1)-(x y_1+1) \sin (x-y_1)}{y_1^3}\Theta (y_1-x).
\end{equation} The mode functions in the noise source are given by the piecewise matched analytical solution, Eq.~(\ref{eq:usr_modefunction}), with the background transition encoded in the Bogoliubov coefficients, Eq.~(\ref{eq:alpha_beta}). We then compute the stochastic power spectrum of field fluctuations using Eq.~(\ref{eq:stochastic_power_spectrum}) and use the relation in Eq.~(\ref{eq:curvature_variable}) to obtain the stochastic power spectrum of curvature perturbations. We show only the final piecewise results of the coarse-grained power spectrum here, with detailed derivations given in Appendix~\ref{sec:coarse_grained_power_spectrum}.

\subsection{White noise spectrum }\label{subsec:white}
In this subsection, we derive the stochastic power spectrum for the white-noise case using the sharp window function defined in Eq.~(\ref{eq:white_noise_filter}). Differentiating the window function with respect to e-fold time (ignoring terms of order $
\mathcal{O}(\epsilon_1),\mathcal{O}{(\epsilon_1\epsilon_2})$), we obtain
\begin{align} \label{eq:white_noise}
    A(k,N) &= \frac{-2k}{\sigma a(N) H(N)}  \delta\big(\frac{k}{\sigma a(N) H(N)}-1\big) + \frac{k^2}{\sigma^2 a^2 H^2}
    \frac{d}{dN}\delta\big(\frac{k}{\sigma a(N) H(N)}-1\big),\nonumber\\
    B(k,N)&=-\frac{2 k}{\sigma a(N)H(N) } \delta \big(\frac{k}{\sigma a(N)H(N)}-1\big).
\end{align} Substituting them into Eq.~(\ref{eq:stochastic_power_spectrum}), the delta functions in the filter collapse the time integrals (see Appendix~\ref{appendix:pre-transition_correlator}), giving \begin{equation}\label{eq:white_noise_general}
\mathcal{P}_{\delta\phi_{cg}}(k,N) = \frac{k^3}{2\pi^2} \Big| G(x,\sigma) \left( 3\,\phi_k(\sigma) + \frac{\pi_k(\sigma)}{a^3H} \right) + \sigma\,\phi_k(\sigma) \frac{\partial G(x,y_1)}{\partial y_1}\Big|_{y_1=\sigma} \Big|^2,
\end{equation} where $\pi_k$ is the conjugate momentum (see Eq.~(\ref{eq:conjugate_field_momentum})). This shows that the coarse-grained power spectrum of the field fluctuation depends explicitly on the noise amplitudes of both the field and its canonical momentum, and on the Green's function encoding the linear dynamics of the coarse-grained field. The corresponding curvature power spectrum is obtained using the relation given in Eq.~(\ref{eq:curvature_variable}).

Substituting the noise amplitude from Eq.~(\ref{eq:usr_modefunction}) and Green's function from  Eq.~(\ref{eq:greensfunction_fullgrad}) into Eq.~(\ref{eq:white_noise_general}), and then using the relation in Eq.~(\ref{eq:curvature_variable}), we obtain the coarse-grained curvature power spectrum (see Appendix~\ref{sec:coarse_grained_power_spectrum}):
\begin{equation}\label{eq:white_noise_curvature_spectrum}
  \mathcal{P}_{\mathcal{R}_{cg}}(k,N) = \dfrac{H(N)^2}{8\pi^2\epsilon_1(N)}\Theta(N-N_\sigma)
  \begin{cases}
  (1+x(N)^2)
  & \text{if } N \leq N_T, \\[0.2cm]
  \left|\alpha_k(x(N)+i)e^{ix}+\beta_k(x(N)-i)e^{-ix}\right|^2
  & \text{if } N > N_T .
  \end{cases}
\end{equation} The coarse-grained power spectrum reproduces the power spectrum of the quantum state given in Eq.~(\ref{eq:usr_qft_spectrum}) and is insensitive to the exact value of the coarse-graining parameter $\sigma$. This implies that to compute the stochastic two-point correlator of $\mathcal{R}$, the coarse-grained equation in this model can be considered as a Langevin equation with classical stochastic noise independently of how squeezed the state is. This is consistent with the fact that, for linear dynamics and Gaussian initial states, the stochastic two-point correlator reproduces the quantum two-point correlator of any Hermitian operators~\cite{Martin:2015qta}.

\subsection{Colored noise spectrum}\label{subsec:color}
In this subsection, we consider the colored-noise case corresponding to a smooth window function of the form given in Eq.~(\ref{eq:colored_noise_filter}). Taking its time derivative and using $y_1=x(N_1)$, we obtain
\begin{equation}\label{eq:color_filter_exponential}
 \bar{A}_1(k,y_1) =\frac{y_1 e^{-\frac{y_1}{2 \sigma }} (4 \sigma +y_1)}{4 \sigma ^2},
 \qquad
 \bar{B}_1(k,y_1)=\frac{y_1e^{-\frac{y_1}{2 \sigma }}}{\sigma },
\end{equation}
\begin{equation}\label{eq:color_filter_gaussian}
 \bar{A}_2(k,y_1) =\frac{y_1^2  (\sigma ^2+y_1^2)}{\sigma ^4}
 e^{-\frac{y_1^2}{2 \sigma ^2}},
 \qquad
 \bar{B}_2(k,y_1)=\frac{2 y_1^2 e^{-\frac{y_1^2}{2 \sigma ^2}}}{\sigma ^2},
\end{equation}
for $n=1$ and $n=2$, respectively. Using the Green’s function from Eq.~(\ref{eq:greensfunction_fullgrad}) together with the noise amplitudes from Eq.~(\ref{eq:usr_modefunction}), the coarse-grained power spectrum\footnote{To get this analytical form, we use $x_i\gg\sigma$ in the integration, see Appendix~(\ref{sec:coarse_grained_power_spectrum}) for details.} for colored noise is given by  
\begin{equation}\label{eq:colored_noise_spectrum}
  \mathcal{P}^{color}_{\mathcal{R}_{cg}}(k,N) \simeq e^{-\left(\frac{k}{\sigma aH}\right)^n}\dfrac{H(N)^2}{8\pi^2\epsilon_1}\Theta(N-N_\sigma)
  \begin{cases}
  (1+x^2)
  & \text{if } N \leq N_T, \\[0.2cm]
  \left|\alpha_k(x+i)e^{ix}+\beta_k(x-i)e^{-ix}\right|^2
  & \text{if } N > N_T .
  \end{cases}
\end{equation} The coarse-grained spectrum for the colored noise differs from the white-noise case given in Eq.~(\ref{eq:white_noise_curvature_spectrum}) by a multiplicative factor $e^{-\left(\frac{k}{\sigma aH}\right)^n}$. Although the spectrum depends on $\sigma$ through this multiplicative factor, it approaches the power spectrum of the quantum state, Eq.~(\ref{eq:usr_qft_spectrum}), in the super-horizon limit, since 
$e^{-\left(\frac{x}{\sigma}\right)^n}\to1$ as $x\to0$.

The smooth window function leads to a gradual coarse-graining over a finite interval determined by its shape, in contrast to the instantaneous coarse-graining of the white-noise case. Therefore, the spectrum is suppressed by a factor of $e^{-n}$ at the coarse-graining scale $k_\sigma$. The suppression factor interpolates from $e^{-n}$ at scale $k_\sigma$ to $1$ in the super-horizon limit over the interval $N-N_\sigma$, controlled by the sharpness parameter $n$ and independent of $\sigma$. For $\delta_c=0.01$, Eq.~(\ref{eq:transtion_interval}), gives the transition intervals of $\Delta N^{(1)}\simeq 3.9$ and $\Delta N^{(2)} \simeq 1.95$ e-folds for $n=1$ and $n=2$, respectively. This means that as long as coarse-graining is performed sufficiently early, $N_\sigma<N_{\rm end}-\Delta N^{(n)}$, the multiplicative factor $e^{-\left(\frac{k}{\sigma aH}\right)^n}\to1$ by the end of inflation and Eq.~(\ref{eq:colored_noise_spectrum}) reduces to Eq.~(\ref{eq:white_noise_curvature_spectrum}).

\subsection{Long-wavelength approximation without the gradient}\label{subsec:LW_cg}
\begin{figure}[htbp]
\begin{center}
    \includegraphics[width=0.495\textwidth]{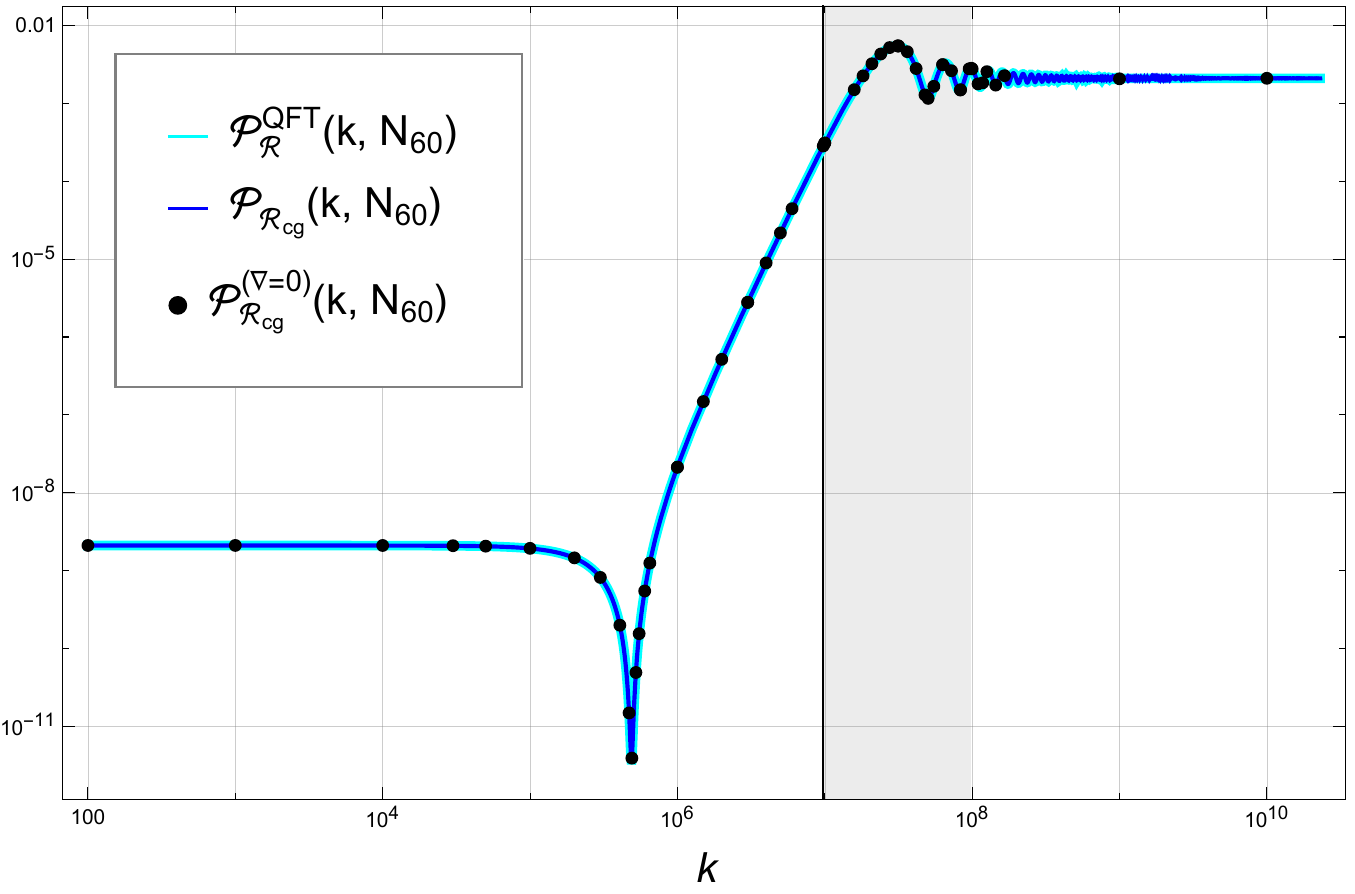}
    \includegraphics[width=0.495\textwidth]{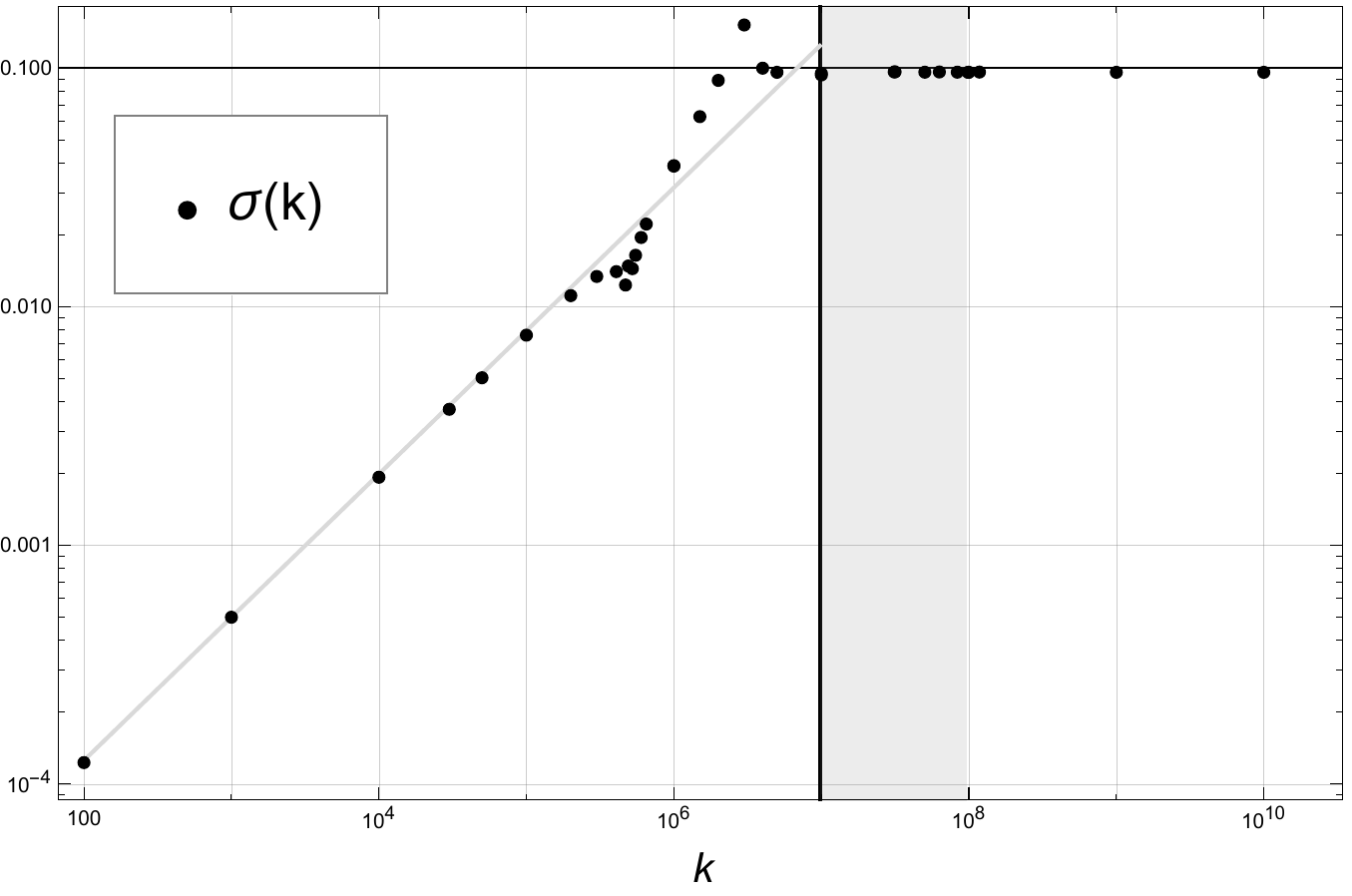}
    \caption{The curvature power spectrum at the end of inflation.
            Left: The cyan curve shows the numerical Sasaki--Mukhanov result, while the blue curve and black points show the stochastic spectrum with and without  the gradient term,  given by Eqs.~(\ref{eq:white_noise_curvature_spectrum}) and (\ref{eq:white_long_wave_approx}), respectively.
            Right: Coarse-graining parameter as a function of wavenumber. We show $\sigma(k)$ for which the stochastic spectrum in Eq.~(\ref{eq:white_long_wave_approx}) agrees with the quantum spectrum at $N_{\rm end}=60$ to within $1\%$. The gray line for $k<k_T$ shows the scaling $\sigma(k)\propto(k/k_T)^{3/5}$, with the proportionality constant $0.285595$ chosen for illustration. 
            }
        \label{fig:white_noise}
\end{center}
\end{figure}

In this subsection, we consider the white-noise coarse-grained dynamics in the long-wavelength limit  by neglecting the gradient term in Eq.~(\ref{eq:stoc_eom}). The Green's function under this approximation takes the form (see Appendix~\ref{sec:greens_function})
\begin{equation}
    G_{k}^{(\nabla=0)}(x(N),y_1(N_1))=\frac{y_1^3-x^3}{3y_1^3}\Theta(y_1-x).
\end{equation} Substituting this Green's function and the noise amplitudes from Eq.~(\ref{eq:usr_modefunction}) into Eq.~(\ref{eq:white_noise_general}), and then using the relation in Eq.~(\ref{eq:curvature_variable}),  we obtain the coarse-grained comoving curvature power spectrum (Appendix \ref{sec:coarse_grained_power_spectrum}):
\begin{equation}\label{eq:white_long_wave_approx}
    \mathcal{P}^{(\nabla=0)}_{\mathcal{R}_{cg}}(k,N) = \frac{H(N)^2}{8\pi^2\epsilon_1}\Theta(N-N_\sigma)
    \begin{cases}
        \left|  F(x,\sigma) \right|^2 , & N \le N_T, \\[0.2cm]
        \left| \alpha_k F(x,\sigma) + \beta_k F^*(x,\sigma) \right|^2 , & N > N_T ,
    \end{cases}
\end{equation} where $F(x,\sigma)$ is a complex function given by
\begin{equation}\label{eq:f}
     F(x,\sigma) = e^{i\sigma} \Big(i+\sigma - i\frac{\sigma^2}{3} + i\frac{x^3}{3\sigma}\Big).
\end{equation} The stochastic spectrum in this case depends explicitly on the coarse-graining parameter $\sigma$. It approaches the power spectrum of the quantum state  given in Eq.~(\ref{eq:usr_qft_spectrum}) in the long-wavelength limit ($x\to0$), provided that $\sigma\ll1$ as shown in the left panel of Fig.~(\ref{fig:white_noise}). Therefore, the cost of this approximation is that the coarse-grained dynamics cannot be described by a Langevin equation with classical stochastic noise for arbitrary values of $\sigma$. Instead, the stochastic noise in the Langevin equation should be introduced only sufficiently long after horizon crossing.

For each mode, we determine the value of $\sigma$ for which the stochastic spectrum agrees with the power spectrum of the quantum state given in Eq.~(\ref{eq:usr_qft_spectrum}) at the end of inflation $N_{\rm end}=60$ to within $1\%$. The coarse-graining parameter is scale dependent and for the modes with $k \geq k_T$, the stochastic spectrum approaches the power spectrum of the quantum state for $\sigma\sim0.1$, while for modes with $k< k_T$, it requires $\sigma\ll1$ as illustrated by the black dots in the right panel of Fig.~(\ref{fig:white_noise}). To understand this scaling behavior, we write the stochastic spectrum in Eq.~(\ref{eq:white_long_wave_approx}) in terms of the real and imaginary parts of $F(x,\sigma)$ as 
\begin{equation}\label{eq:LW_stochastic_spectrum_in_terms_of_growing_and_decaying_contributions}
    \mathcal{P}^{(\nabla=0)}_{\mathcal{R}_{cg}}(k,N)=\frac{H(N)^2}{8\pi^2\epsilon_1}\Theta(N-N_\sigma) \Big|\Re[F(x,\sigma)](\alpha_k + \beta_k)+ i\times \Im[F(x,\sigma)](\alpha_k - \beta_k)\Big|^2.
\end{equation} In the super-horizon limit ($x\to0$), the stochastic spectrum retains a nonzero contribution from the $(\alpha_k+\beta_k)$ term, since $\Re[F(0,\sigma)]\neq0$. This is in contrast to the case when the gradient term is retained in Eq.~(\ref{eq:white_noise_curvature_spectrum}), where the coefficient of $(\alpha_k+\beta_k)$ vanishes as $\lim_{x\to0}\Re[(x+i)e^{ix}]=0.$

Expanding Eq.~(\ref{eq:f}) in the super-horizon limit for $\sigma\ll1$ gives
    \begin{equation}
        \Re[F(0,\sigma)] = -\frac{\sigma ^5}{45} + \mathcal{O}(\sigma^6) ,\quad \Im[F(0,\sigma)] = 1+ \frac{\sigma ^2}{6} + \mathcal{O}(\sigma^3).
    \end{equation}

\begin{itemize}
    \item \textbf{$\boldsymbol{k\ge k_T}$ case:} For modes with $k\ge k_T$, expanding the Bogoliubov coefficients in Eq.~(\ref{eq:alpha_beta}) gives
        \begin{equation} 
            \Big| \frac{(\alpha_k+\beta_k)}{(\alpha_k-\beta_k)}\Big|\approx  \Big| \frac{2 i (\frac{k}{k_T}) }{3+3 e^{2 i (\frac{k}{k_T}) }+2 i (\frac{k}{k_T}) }\Big|\approx~\mathcal{O}(1),
        \end{equation} whose magnitude oscillates around 1, and approaches 1 for $k\gg k_T$. Therefore, $|\alpha_k+\beta_k|\sim|\alpha_k-\beta_k|$ and the relative contributions of the two terms in Eq.~(\ref{eq:LW_stochastic_spectrum_in_terms_of_growing_and_decaying_contributions}) are determined by the coefficients  $\Im[F]$ and $\Re[F]$. Since $\Im[F]\gg\Re[F]$, with $\Im[F]=1+\frac{\sigma^2}{6}+\mathcal{O}(\sigma^3)$, the stochastic spectrum is dominated by $(\alpha_k-\beta_k)$ term and approaches the power spectrum of the quantum state for $\sigma\sim0.1$.
    \item \textbf{$\boldsymbol{k<k_T}$ case:}  Whereas for modes with $k<k_T$, the behavior is qualitatively different since the ratio at leading order\footnote{Expanding the ratio for $k<k_T$ gives at leading order 
        \begin{equation*} 
            \Big|\frac{(\alpha_k+\beta_k)}{(\alpha_k- \beta_k)}\Big|\simeq \Big|3i \left(1-\frac{A_+}{A_-}\right)\Big(\frac{k}{k_T}\Big)^{-3}+(\frac{A_+}{A_-})^2-\frac{3}{5} i \left(2 \Big(\frac{A_+}{A_-}\Big) ^2+\Big(\frac{A_+}{A_-}\Big) -3\right)\Big(\frac{k}{k_T}\Big)^{-1}+\mathcal{O}\Big(\frac{k}{k_T}\Big)\Big|
        \end{equation*} For the $k_\star \lesssim k \lesssim k_T$, the $k^{-1}$ term dominates and for $k \lesssim k_\star$, $k^{-3}$ term dominates. The crossover scale is determined by equating these two terms giving $k=\sqrt{\frac{5A_-}{2A_+}}k_T\approx k_\star$.}
        \begin{equation}
        \Big|\frac{(\alpha_k+\beta_k)}{(\alpha_k-\beta_k)}\Big|\approx   
            \begin{cases}\label{eq:lw_scaling}
                \left|  -\frac{3}{5} i \left(2 \big(\frac{A_+}{A_-}\big)^2+\big(\frac{A_+}{A_-}\big) -3\right) \Big(\frac{k}{k_T}\Big)^{-1} \right| , & k_\star \lesssim k \lesssim k_T, \\[0.2cm] ~\left|3i \left(1-\frac{A_+}{A_-}\right)\Big(\frac{k}{k_T}\Big)^{-3}\right|, & k \lesssim k_\star ,
            \end{cases}
        \end{equation} scales as $k^{-1}$ for $k_\star \lesssim k \lesssim k_T$ and as $k^{-3}$ for $k\lesssim k_{\star}$ where $k_\star$ is the dip mode defined in Eq.~(\ref{eq:dip}). 
        \begin{itemize}
            \item[(a)] $k_\star\lesssim k\lesssim k_T:$ Although  $|\alpha_k+\beta_k|\gg|\alpha_k-\beta_k|$ for these modes, the relative contribution of the $(\alpha_k+\beta_k)$ term scales only as $k^{-1}$, while $\Re[F]$ is suppressed by $\mathcal{O}(\sigma^5)$ and therefore, the two contributions in Eq.~(\ref{eq:LW_stochastic_spectrum_in_terms_of_growing_and_decaying_contributions}) can both remain important.

            \item[(b)] $k\ll k_T~(k \lesssim k_\star):$ In contrast, the large amplitude of the ratio scales as $k^{-3}$ and compensates for the $\mathcal{O}(\sigma^5)$ suppression of $\Re[F]$, so this term dominates unless $\sigma$ is very small. The relative contribution of $(\alpha_k+\beta_k)$ term in Eq.~(\ref{eq:LW_stochastic_spectrum_in_terms_of_growing_and_decaying_contributions}) is \begin{equation}\label{eq:lw_scaling_behaviour}
                \Big|\frac{\Re[F(0,\sigma)](\alpha_k+\beta_k)}{\Im[F(0,\sigma)](\alpha_k-\beta_k)}\Big|\propto \frac{\sigma^5}{k^3}. 
            \end{equation} Therefore, increasingly smaller values of coarse-graining parameter with the scaling $\sigma(k)\propto k^{3/5}$ is required for this contribution to remain negligible, so that the stochastic spectrum is dominated by the $(\alpha_k-\beta_k)$ term and hence reproduce the power spectrum of the quantum state at $N_{\rm end}$.

            We note that this strong scale dependence of $\sigma(k)$ comes from the residual $(\alpha_k+\beta_k)$ contribution to the stochastic spectrum since the transition-modified mode functions ($\alpha_k\neq1,\ \beta_k\neq0$) are used as the noise source in the Langevin equation. Since these modes leave the horizon well before the transition, $\mathcal{R}_k$ is dominated by its \textit{growing} mode when the transition occurs and hence does not significantly modify $\mathcal{R}_k$ at the end of inflation.\footnote{For $k\ll k_T$, one has $k/(a_{\rm T}H_{\rm T})\ll1$ at the transition, so the gradient term in the Sasaki--Mukhanov equation is negligible. The long-wavelength solution is therefore of the form $\mathcal{R}_k^{\rm h}=C_1+C_2\int dN/(aHz^2)$, where $C_1$ is the growing (constant) mode and the second term is the decaying mode. Since these modes have been outside the horizon for a sufficiently long time before the transition, the decaying mode rapidly decays and its contribution to the $\mathcal{R}_k$ is very small. Thus, the transition does not significantly modify the dominant growing mode, and hence does not significantly change the value of $\mathcal{R}_k$ at the end of inflation.} If the pre-transition mode function with $(\alpha_k\approx1,\beta_k\approx0)$ is used as the noise source in Eq.~(\ref{eq:white_noise_general}) instead, the stochastic spectrum is of the form \begin{equation}\label{eq:LW_stochastic_spectrum_growing_only}
              \mathcal{P}^{(\nabla=0)}_{\mathcal{R}_{cg}}(k,N)\simeq\frac{H(N)^2}{8\pi^2\epsilon_1}\Theta(N-N_\sigma) \Big|\Re[F(x,\sigma)]+ i\times \Im[F(x,\sigma)]\Big|^2.
            \end{equation} Since $\Re[F]\ll\Im[F]=1+\frac{\sigma^2}{6}+\mathcal{O}(\sigma^3)$, the stochastic spectrum  approaches the power spectrum of the quantum state for $\sigma\sim0.1$ for these modes. This is consistent with the so-called analytical homogeneous-matching procedure described in Ref.~\cite{Jackson:2023obv}, where the pre-transition mode functions are used as the noise source for $k\ll k_T$. In the next subsection, we use the numerical homogeneous-matching procedure to identify the range of modes for which this approximation is valid. 
            
        \end{itemize}
        
\end{itemize}

In general, to determine whether the scale dependence of $\sigma$ reflects the time at which the gradient term can be neglected in the Langevin equation, we compare it with the homogeneous matching time obtained from the long-wavelength solution of the Sasaki--Mukhanov equation in the next subsection.

\subsubsection{Homogeneous matching procedure}\label{sec:homogeneous_match} The homogeneous-matching procedure consists of matching the full Sasaki--Mukhanov solution for $\mathcal{R}_k$ and its first derivative to the corresponding long-wavelength homogeneous solution at a suitable matching time $N_{\rm h}$~\cite{Leach:2001zf,Jackson:2023obv}. In the long-wavelength limit, the solution to Eq.~(\ref{eq:ms_equation_maintext}) reduces to the homogeneous solution given by Eq.~(\ref{eq:growing_and_decaying_decomposition}). This homogeneous solution can be reconstructed by matching it to the exact numerical solution, $\mathcal{R}_k$, at $N_{\rm h}$~\cite{Jackson:2023obv}:
\begin{equation}\label{eq:homogeneous_matching}
    \mathcal R_k^{(\rm h)}(N) = \mathcal R_k(N_{\rm h}) + z^2(N_{\rm h})a(N_{\rm h})H(N_{\rm h}) \left.\frac{d\mathcal R_k}{dN}\right|_{N_{\rm h}} \int_{N_{\rm h}}^{N} \frac{dN'}{a(N')H(N')z^2(N')}.
\end{equation} The homogeneous matching time $N_{\rm h}(k)$ is determined by identifying the earliest time at which the homogeneous solution, $\mathcal{R}_k^{(\rm h)}$, reproduces the power spectrum obtained from the full Sasaki--Mukhanov solution at $N_{\rm end}$ to within $1\%$~\cite{Jackson:2024aoo}: 
\begin{equation}\label{eq:homogeneous_matching_condition}
        \left| \frac{
         \frac{k^3}{2\pi^2}|\mathcal{R}^{(\rm h)}_{k}(N_{\rm end},N_{\rm h}(k))|^2 - \frac{k^3}{2\pi^2}|\mathcal{R}_{k}(N_{\rm end})|^2 }{ \frac{k^3}{2\pi^2}|\mathcal{R}_{k}(N_{\rm end})|^2} \right| \leq 0.01.
\end{equation} The corresponding matching time defines $\sigma_{\rm matching}(k)\equiv k/a(N_{\rm {h}})H(N_{\rm {h}})$, so that homogeneous equation ($k=0$) can be used to describe $\mathcal{R}_k$ from $N_{\rm {h}}(k)$ to the end of inflation.

\begin{figure}[htbp]
\begin{center}
    \includegraphics[width=0.494\textwidth]{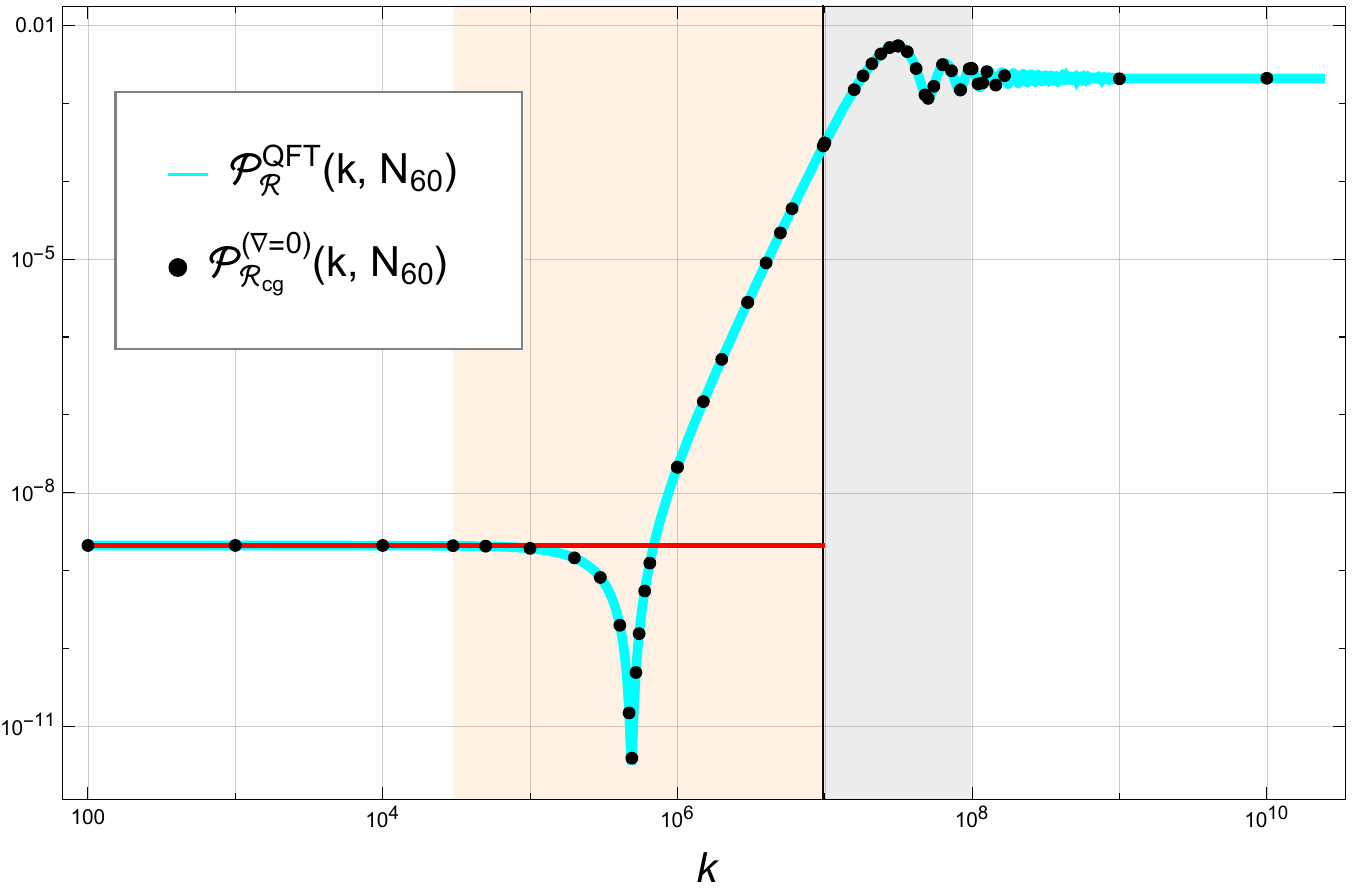}
    \includegraphics[width=0.497\textwidth]{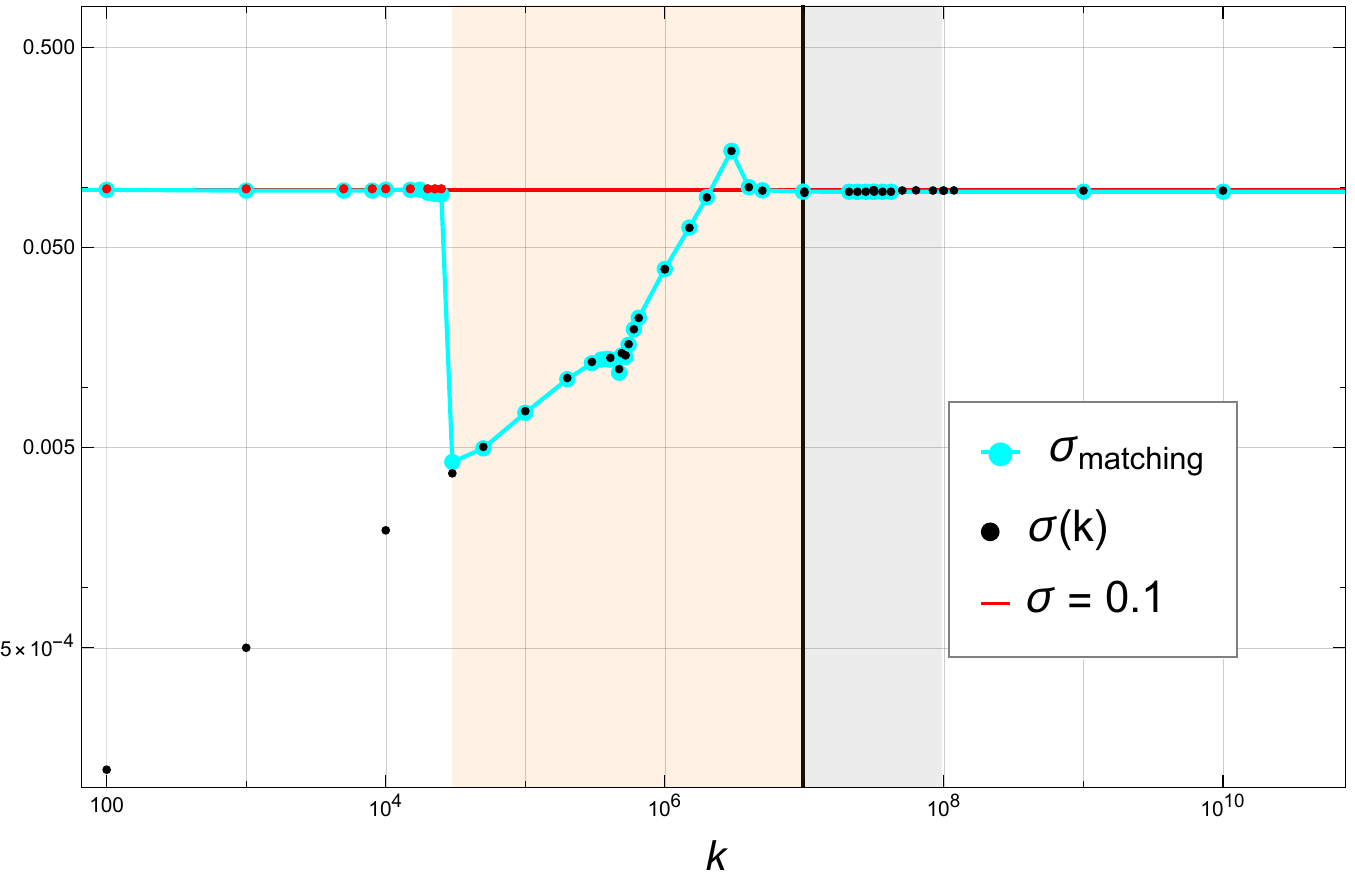}
    \caption{
            Left: The curvature power spectrum at the end of inflation. The cyan curve shows the numerical Sasaki--Mukhanov result, while the black points show the stochastic spectrum without the gradient term. The red line for $k\leq k_T$ shows the stochastic spectrum  given in Eq.~(\ref{eq:LW_stochastic_spectrum_growing_only}) with $\sigma=0.1$~,obtained by using pre-transition mode functions as source. The orange shaded region marks the range of modes where the stochastic spectrum,  Eq.~(\ref{eq:LW_stochastic_spectrum_growing_only}) with $\sigma=0.1$~, obtained from pre-transition mode, deviates more than $1\%$ from the exact numerical Sasaki--Mukhanov power spectrum.
            Right: Coarse-graining parameter as a function of wavenumber. The cyan curve shows \(\sigma_{\rm match}(k)\) obtained from the homogeneous matching procedure, while the black points show the value of  $\sigma(k)$ for which the stochastic spectrum in Eq.~(\ref{eq:white_long_wave_approx}) agrees with the quantum spectrum at $N_{\rm end}=60$ to within $1\%$.
            The red line (and dots) shows $\sigma=0.1$ for comparison.}
        \label{fig:homogeneous_matching}
\end{center}
\end{figure}

In Fig.~(\ref{fig:homogeneous_matching}), the cyan curve in the left panel shows the numerically computed power spectrum at $N_{\rm end}=60$ by solving Eq.~(\ref{eq:ms_equation_maintext}), while the cyan line in the right panel shows the corresponding $\sigma_{\rm matching}(k)$ obtained from the matching times by numerically solving Eq.~(\ref{eq:homogeneous_matching}). The black points in the left panel show the stochastic spectrum from Eq.~(\ref{eq:white_long_wave_approx}), while those in the right panel show the corresponding $\sigma(k)$ for which the stochastic spectrum agrees with the power spectrum of the quantum state at $N_{\rm end}=60$ to within $1\%$. Finally, the stochastic spectrum obtained using the analytical homogeneous-matching procedure in Eq.~(\ref{eq:LW_stochastic_spectrum_growing_only}) is shown in red in the left panel for modes $k<k_T$, while the corresponding $\sigma\simeq0.1$ is shown by the red line (and dots) in the right panel.

For modes $k\ll k_T$,  the stochastic spectrum using the pre-transition mode function, Eq.~(\ref{eq:LW_stochastic_spectrum_growing_only}), reproduces the power spectrum of the quantum state at the end of inflation with $\sigma\simeq0.1$. However, the orange shaded region marks the range of modes where Eq.~(\ref{eq:LW_stochastic_spectrum_growing_only}) with $\sigma=0.1$,~deviates more than $1\%$ from the power spectrum of the quantum state (see left panel of Fig.~(\ref{fig:homogeneous_matching})). This is because the pre-transition mode functions fails to account for transition effects\footnote{As described below Eq.~(\ref{eq:x}), the positive and negative frequency modes mix destructively leading to the characteristic dip in the power spectrum for modes near the transition ($k\lesssim k_T$). The pre-transition mode functions ($\alpha_k\approx1,\beta_k\approx0$) fails to account for this mixing and therefore homogeneous matching procedure fails for these modes~\cite{Jackson:2023obv}.} for these modes. Therefore, using the full mode function given in Eq.~(\ref{eq:usr_modefunction}) as noise source in Eq.~(\ref{eq:white_noise_general}), leads to the long-wavelength stochastic spectrum given in Eq.~(\ref{eq:white_long_wave_approx}) and we find $\sigma(k)\simeq\sigma_{\rm matching}(k)$ for modes in the orange region as well as for $k\ge k_T$ (right panel of Fig.~(\ref{fig:homogeneous_matching})). Therefore, homogeneous matching procedure with scale dependent $\sigma_{\rm matching}(k)$ can be used to reproduce the power spectrum of the quantum state when gradient terms are ignored in the Langevin equation.

These results are consistent with the homogeneous matching procedure described in \cite{Jackson:2023obv}. In particular, \cite{Jackson:2023obv,Jackson:2024aoo} showed that this procedure can break down for finite modes near transitions in the Starobinsky model unless coarse-graining is performed beyond the transition, and \cite{Domenech:2023dxx} noted that the $\delta N$ formalism is reliably applicable only sufficiently after sudden transition. Furthermore, it was shown in \cite{Jackson:2023obv,Briaud:2025ayt} that including gradient corrections or their interactions reduces the sensitivity of the stochastic power spectrum to $\sigma$. While these studies focus on gradient expansion method and numerical solution to nonlinear evolution of the coarse-grained field using Langevin equations with linear classical noise, we restrict our analysis to linear dynamics using Green's function method. Finally, we conclude this paper by addressing whether the linear curvature perturbations $\mathcal{R}_k$ are effectively classical at homogeneous matching time.

\subsubsection{Perturbations at homogeneous matching times}
On super-horizon scales, quantum fluctuations of $\mathcal{R}$ are usually treated as classical noise in the stochastic inflation even in USR models. In this subsection, we study whether this assumption holds at the homogeneous matching time by comparing $\sigma_{\rm matching}(k)$ with the classicality bound $\sigma^{\rm cl}(k)$ in this model. We recall that $\sigma^{\rm cl}(k)$ in Eq.~(\ref{eq:classical_sigma}) gives the upper bound on the coarse-graining parameter for each mode to be considered effectively classical in the large-squeezing limit as defined in Eq.~(\ref{eq:classicality_criteria}). Interestingly, although the homogeneous matching procedure in the stochastic description can reproduce the power spectrum of the quantum state initialized in the BD vacuum, we find that $\sigma_{\rm matching}(k)$ does not coincide with $\sigma^{\rm cl}(k)$.

\begin{figure}[htbp]
    \centering
    \includegraphics[width=0.75\linewidth]{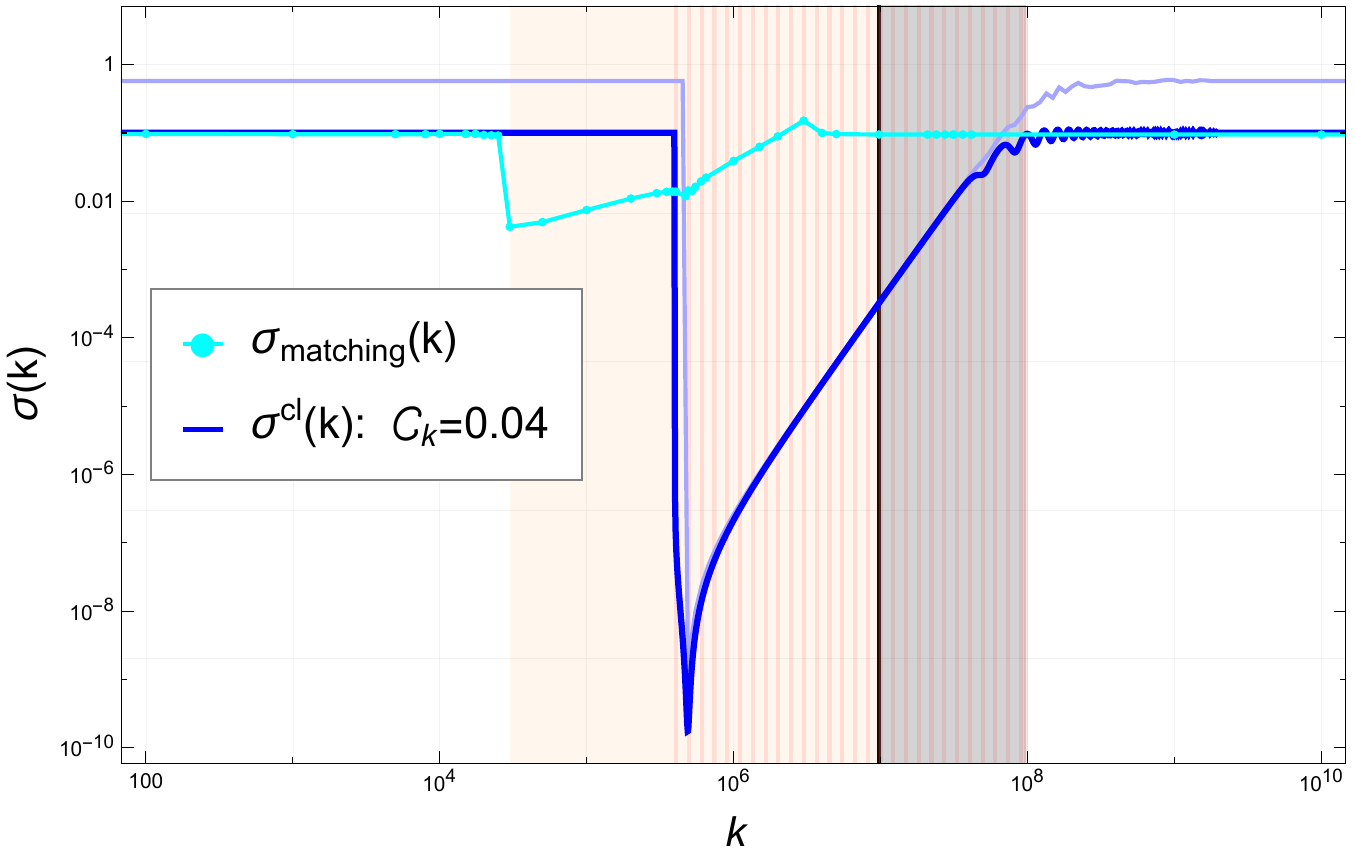}
    \caption{Comparison of the mode-dependent coarse-graining parameters corresponding to the homogeneous matching procedure and the classicality condition. The cyan curve represents $\sigma_{\rm matching}(k)$ obtained from Eq.~(\ref{eq:homogeneous_matching}). The light and dark blue curves shows $\sigma^{\rm cl}(k)$ at the thresholds $\mathcal{C}_k=1$ and $\mathcal{C}_k=0.04$, respectively. For modes near the USR transition indicated by the region with red vertical lines, the hierarchy $\sigma_{\rm matching}(k) \gg \sigma^{\rm cl}(k)$ for both $\mathcal{C}_k=1~\&~0.04$ indicates that homogeneous matching time occurs well before the classicalization time. The gray shaded region shows the USR phase and the orange region is as described in Fig.~(\ref{fig:homogeneous_matching}).}
    \label{fig:classical_noise}
\end{figure}

For the modes near the transition, the homogeneous matching time precedes the classicalization time ($N_{\rm h}<N_k^{cl}$), with $\sigma_{\rm matching}(k)>\sigma^{\rm cl}(k)$ as shown by the region marked with red vertical lines in Fig.~(\ref{fig:classical_noise}). Therefore, for modes near the transition, the curvature perturbation remains in the quantum regime at the homogeneous-matching time, implying that the commutator of the canonical variables $\mathcal{R}_k$ and $\Pi_k$ is comparable to their phase-space variances. Thus, a classical stochastic description of these perturbations is not yet justified at the homogeneous-matching time in this model.

To see this distinction explicitly, consider the dip mode, for which the matching time occurs $N_{\rm h}(k_{\star})-N_{k_\star}\simeq4.7$ e-fold from horizon exit (this value is obtained numerically from Eq.~(\ref{eq:homogeneous_matching})), corresponding to $\sigma_{\rm matching}(k_{\star})\sim\mathcal{O}(10^{-2})$. Whereas the classicalization requires $\Delta N^{\rm cl}_{k_\star}=22.45$ e-folds from horizon exit (see Eq.~(\ref{eq:classicalization_time_dip})), which gives a much smaller value of $\sigma^{\rm cl}(k_{\star})\sim\mathcal{O}(10^{-10})$. Although $\sigma_{\rm matching}(k_\star)$ and $\sigma^{\rm cl}(k_\star)$ give different values of the coarse-graining parameter, both choices\footnote{The stochastic power spectrum for the dip mode in Eq.~(\ref{eq:white_long_wave_approx}) obtained using $\sigma_{\rm matching}(k_\star)\simeq10^{-2}$ reproduces the power spectrum of the quantum state with a relative error of $\simeq0.03323\%$. Using $\sigma^{\rm cl}(k_\star)\simeq10^{-10}$ instead gives a similarly accurate result, with a relative error of $\simeq0.03226\%$.} reproduce the quantum two-point correlator (power spectrum) with a relative error of less than $0.07\%$.

While the distinction between $\sigma_{\rm matching}(k)$ and $\sigma^{\rm cl}(k)$ does not significantly affect the two-point correlator, $\sigma^{\rm cl}(k)$ provides a mode-dependent condition for when quantum fluctuations can be consistently described as classical stochastic noise. Even in linear theory with a Gaussian initial state,  $\sigma^{\rm cl}(k)$ becomes relevant for higher-order correlators involving momentum, for which the stochastic description requires the perturbations to be in the large-squeezing regime to reproduce the corresponding quantum correlators~\cite{Martin:2014kja,dePutter:2019xxv}. 

However, the $ n$-point functions of $\mathcal{R}$ are of primary interest, as these higher-order statistics are used to construct the full probability distribution function (PDF) of $\mathcal{R}$. The tail of this distribution determines the predicted abundance of PBHs, which are formed from rare, large fluctuations~\cite{Pattison:2017mbe, Figueroa:2020jkf, Ezquiaga:2019ftu}. These $n$-point functions are completely determined by two-point functions in linear theory with a Gaussian initial state. Beyond linear theory however, non-linear dynamics of perturbations could couple $\mathcal{R}$ to its conjugate momentum. For instance, in the separate-universe approach, long-wavelength perturbations are treated as locally homogeneous background fluctuations with their nonlinear evolution described by Langevin equations, sourced by linear stochastic noise at the homogeneous matching time~\cite{Pattison:2019hef,Jackson:2023obv,Jackson:2024aoo,Briaud:2025ayt}. This raises the question of how the higher-order statistics of $\mathcal R$ may be affected by treating the linear stochastic noise as classical before the perturbations have become effectively classical as defined in Eq.~(\ref{eq:classicality_parameter}). It is therefore interesting to study how the higher-order statistics of $\mathcal R$ are affected when the stochastic noise is treated as classical at the homogeneous matching times compared with the classicality bound $\sigma^{\rm cl}(k)$.

\section{Conclusion}\label{sec:conclusion}

We study a transient ultra-slow-roll (USR) Starobinsky model and show that the quantum-to-classical transition of curvature perturbations is mode-dependent. The quantum-to-classical transition refers to when the linear curvature perturbation can be effectively described as a classical stochastic variable in a closed system. We characterize this transition through the suppression of the equal-time commutator of curvature perturbation and its conjugate momentum relative to their variances on super-horizon scales and the evolution of the two-mode squeeze state. While all Fourier modes eventually become effectively classical on super-horizon scales, the time required for this transition depends on when they exit the horizon relative to the USR phase. This is because the decaying mode, which would decay on super-horizon scales in a slow-roll background, instead grows transiently during the USR phase. This growth modifies the evolution of the conjugate momentum and hence the squeezing and effective classicality of the linear curvature perturbation. In particular, the mode corresponding to the dip feature in the power spectrum takes the longest time to become effectively classical. We expect this mode-dependent transition to be a generic feature of single-field inflation models in which primordial black holes (PBHs) are produced through an enhancement of the decaying mode of the linear curvature perturbations~\cite{Ozsoy:2023ryl}.

We then study the implications of this mode-dependent effective classicalization for the stochastic description of linear perturbations. The mode-dependent classicalization time leads to a mode-dependent constraint on the coarse-graining parameter $\sigma^{cl}(k)$, with the dip mode requiring the smallest value, $\sigma^{cl}(k_\star),$ defined in Eq.~(\ref{eq:sigma_constraint_linear_model}). This constraint ensures that the modes included in the coarse-grained description are sufficiently squeezed and effectively classical.

We analytically derive the stochastic power spectrum of curvature perturbations from the Langevin equation using the Green's function method for both sharp and smooth window functions, corresponding to white and colored noise, respectively. In the white-noise case, the stochastic power spectrum is independent of $\sigma$ and reproduces the power spectrum of the quantum state initialized in the Bunch Davies vacuum. For colored noise, the spectrum differs from the white-noise case by a multiplicative filter factor $e^{-\left(\frac{k}{\sigma aH}\right)^n}$, and approaches the power spectrum of the quantum state in the super-horizon limit $k/(\sigma aH)\to0$. Thus, the stochastic description reproduces the the power spectrum of the quantum state independently of the mode-dependent squeezing required for classicalization.

In contrast, when the gradient terms are neglected in the Langevin equation, the stochastic white noise spectrum depends explicitly on $\sigma$. In this approximation, the homogeneous matching procedure can determine a mode-dependent $\sigma(k)$ that reproduces the power spectrum of the quantum state initialized in the Bunch-Davies vacuum. Crucially, however, for modes near the USR transition, we show the homogeneous matching times precede the classicalization times. This shows that linear perturbations remain quantum at the homogeneous matching times for these modes. Therefore, $\sigma^{\rm cl}(k)$ gives a mode-dependent condition under which linear curvature perturbations can be treated as classical stochastic noise. While this distinction has little effect on the power spectrum, $\sigma^{\rm cl}(k)$ can become important for higher-order correlators involving momentum, even in linear theory. This is because large squeezing is required for the stochastic description to reproduce the corresponding higher-order quantum correlators~\cite{Martin:2014kja,dePutter:2019xxv}.

In linear theory, the $n$-point functions of $\mathcal{R}$ are completely determined by the two-point function. However, since non-linear dynamics may couple $\mathcal{R}$ to its conjugate momentum, whether the stochastic description of $n$-point functions of $\mathcal{R}$ requires large squeezing beyond linear theory requires further study. These higher-order statistics are used to construct the full probability distribution function (PDF) of $\mathcal{R}$. The far tail of this distribution governs the predicted abundance of PBHs, which form from rare, large fluctuations~\cite{Carr:2020gox,Pattison:2017mbe, Figueroa:2020jkf, Ezquiaga:2019ftu}. This also raises an interesting question of how to distinguish genuine non-linear effects in the PDF of coarse-grained $\mathcal{R}$ from possible artifacts introduced by prematurely treating the quantum noise in the coarse-grained equation as classical.

Our analysis is limited to linear dynamics of the perturbations in a closed system, where the Wigner function remains positive for a Gaussian initial state. Beyond linear theory, the initial Gaussian state is no longer preserved, and the Wigner function can develop interference fringes and lose its positive-definite property~\cite{Ireland:2026txt}. Consequently, the squeezing formalism and $\mathcal{C}_k$ in Eq.~(\ref{eq:classicality_parameter}) inferred from liner dynamics becomes insufficient to determine when modes become effectively classical. In general, curvature perturbations beyond linear theory cannot be treated as an isolated system because interactions with other fields, mode couplings, gravitational self-interactions, and entanglement across spatial regions require an open quantum system description subject to decoherence~\cite{Joos:1984uk}. A detailed study of the quantum-to-classical transition and implications for stochastic inflation including the effects of non-linear dynamics and open-system interactions, is beyond the scope of this paper and is left for future work. 

\acknowledgments
We thank Prof. Kin-Wang Ng for carefully reading the manuscript and for his valuable suggestions and insightful discussions. We also thank Prof. De-Chang Dai for generously allowing us to use his workstation for our numerical computations. The work of D.S. L. was supported in part by the National Science and Technology Council, R.O.C. (NSTC 114-2112-M-259-007-MY3). The work of C.P. Y. was supported in part by the National Science and Technology Council, R.O.C. (NSTC 115-2112-M-259-009).
\appendix
\section{Linear Perturbations}\label{sec:squeezing_quantization}
In linear perturbation theory, the field and the metric is split into a homogeneous background and a small perturbation:
\begin{equation}
    \phi(\mathbf{x},\eta) = \bar{\phi}(\eta) + \delta\phi(\mathbf{x},\eta) ,\quad g_{\mu\nu}(\mathbf{x},\eta)=\bar{g}_{\mu\nu}(\eta)+\delta g_{\mu\nu}(\mathbf{x},\eta).
\end{equation} Substituting this decomposition into the action Eq.~(\ref{eq:single_field_action}) and varying with respect to the field and the metric gives the Klein--Gordon and Einstein equations, respectively. The background dynamics are given by
\begin{equation}\label{eq:conformal_background_dynamics}
 3 \mathcal{H}^2 = \frac{\bar{\phi}'^2}{2} +a^2V(\bar{\phi}),\qquad
  \bar{\phi}'' + 2\mathcal{H}\bar{\phi}' + a^2\frac{d V(\bar{\phi})}{d\phi} = 0,
\end{equation} where \(\mathcal{H}\equiv a'/a\) is the conformal Hubble parameter and primes denote derivatives with respect to conformal time \(\eta\). The corresponding background equations in terms of the number of e-folds $N$ are given in Eq.~(\ref{eq:background_efolds_equation}).

The scalar field perturbations can be described in terms of the gauge-invariant Mukhanov variable $f(\eta, \mathbf{x})$. In spatially-flat gauge \cite{Mukhanov:1988jd,Sasaki:1986hm}
\begin{equation}\label{eq:curvature_variable_appendix}
    f(\eta,\mathbf{x})=a\delta\phi(\eta,\mathbf{x})=z\mathcal{R}(\eta,\mathbf{x}),
\end{equation} where $\delta \phi$ is the field perturbation and $\mathcal{R}$ is the gauge-invariant comoving curvature perturbation \cite{Malik:2008im}. The function $z$ is defined as $z\equiv a \sqrt{2\epsilon_1}$, where $\epsilon_1$ is the first slow-roll Hubble parameter $\epsilon_1=1-\mathcal{H}'/\mathcal{H}^2.$ Expanding the action, Eq.~(\ref{eq:single_field_action}) to second order gives \cite{Mukhanov:1990me}
\begin{equation}
    ^{(2)}\delta S = \frac{1}{2} \int d^4x \left[ (f')^2 - (\nabla f)^2 + \frac{z''}{z} f^2 \right].
\end{equation} In Fourier space,
\begin{equation}
    f(\eta, \mathbf{x}) = \frac{1}{(2\pi)^{3/2}} \int d^3\mathbf{k}\, f_\mathbf{k}(\eta) e^{i \mathbf{k} \cdot \mathbf{x}}.
\end{equation}  Since $f(\eta, \textbf{x})$ is real, we have $f_{-\textbf{k}}=f_{\textbf{k}}^*$, where a star denotes the complex conjugation. This relation shows that the Fourier modes are not independent. We split the integral in the action into two parts and change $\textbf{k}$ into $-\textbf{k}$ in the second part, in order to deal with independent variables only. This gives~\cite{Albrecht:1992kf,Martin:2015qta}
\begin{eqnarray}
    ^{(2)}\delta S &=& \frac{1}{2}
        \int_{\mathbb{R}^{3+}} d\eta \,  d^3\textbf{k}
        \Biggl[f_{\textbf{k}}'f_{\textbf{k}}^{*}{}' + f_{\textbf{k}}^*{}'f_{\textbf{k}}' -2\frac{z'}{z}\bigl(f_{\textbf{k}}'f_{\textbf{k}}^* \nonumber 
        +f_{\textbf{k}}f_{\textbf{k}}^*{}'\bigr) +\left(\frac{z'^2}{z^2}-k^2\right)\left(f_{\textbf{k}}f_{\textbf{k}}^*
        +f_{\textbf{k}}^*f_{\textbf{k}}\right) \Biggr]\, ,
\end{eqnarray} where the integral over $\textbf{k}$ is now performed in half the Fourier space, $\textbf{k} \in \mathbb{R}^{3+}$. The Euler-Lagrange equations give the Sasaki--Mukhanov equation
\begin{equation}\label{eq:ms_appendix}
    f_\mathbf{k}'' + \left(k^2 - \frac{z''}{z}\right) f_\mathbf{k} = 0.
\end{equation} Since the mode functions depend only on the magnitude $k \equiv |\mathbf{k}|$, the vector notation is dropped in Eq.~(\ref{eq:ms_equation_maintext}) in the main text. For a general potential $V(\phi)$, using the background equation, we have 
\begin{eqnarray}
    \frac{z'}{z}=\mathcal{H}\big[1+\frac{\epsilon_2}{2}\big]=\mathcal{H}\Big[1-\Big(3-\epsilon_1+\frac{a^2}{\mathcal{H}\bar{\phi}'}\frac{dV(\bar\phi)}{d\phi}\Big)\Big],
\end{eqnarray} and
\begin{equation}\label{eq:effective_mass}
    \frac{z''}{z} = \left[ 2 - \epsilon_1 + \frac{3}{2} \epsilon_2 - \frac{1}{2} \epsilon_1 \epsilon_2
    + \frac{1}{4} \epsilon_2^2 + \frac{1}{2} \frac{d\epsilon_2}{dN} \right] \mathcal{H}^2
    = \left[ 2 - \epsilon_1 + \delta - \frac{1}{H^2}\frac{d^2V(\bar{\phi})}{d\phi^2} \right] \mathcal{H}^2,
\end{equation} with $\delta = 6 \epsilon_1 - 2 \epsilon_1^2 + 2 \epsilon_1 \epsilon_2$ representing the contribution from metric perturbations. The corresponding Hankel index is
\begin{equation}\label{eq:index_usr}
    \nu^2 = \frac{1}{4} + \frac{z''}{z} \, \eta^2, \qquad \eta \equiv  \int_{N}^{N_e} \frac{dN'}{a(N') H(N')},
\end{equation} where $N_e$ denotes the end of inflation with $N\equiv\ln{a}$. In general, $\nu$ is $\textit{instantaneous}$, since $z''/z$ varies with time for an arbitrary potential, and a constant $\nu$ with an exact Hankel-function solution is only valid if $z''/z$ remains approximately constant over the relevant time interval. This formulation is fully general and can be applied to any potential $V(\phi)$.

The conjugate momentum is
\begin{equation}
    P_\mathbf{k} = \frac{\delta \mathcal{L}}{\delta f_\mathbf{k}^{*'}} = f_\mathbf{k}' - \frac{z'}{z} f_\mathbf{k} = z \, \mathcal{R}_\mathbf{k}'.
\end{equation} The Hamiltonian then follows as
\begin{equation}
    H = \int_{\mathbb{R}^{3+}} d^3\mathbf{k} \left[ P_\mathbf{k} P_\mathbf{k}^* + \frac{z'}{z} (P_\mathbf{k}^* f_\mathbf{k} + P_\mathbf{k} f_\mathbf{k}^*) + k^2 f_\mathbf{k} f_\mathbf{k}^* \right].
\end{equation}  Promoting the classical Fourier components $(f_\mathbf{k}, P_\mathbf{k})$ to quantum operators $(\hat{f}_\mathbf{k}, \hat{P}_\mathbf{k})$ satisfying the canonical commutation relation
\begin{equation}
    [\hat{f}_\mathbf{k}, \hat{P}_{\mathbf{k}'}] = i \, \delta(\mathbf{k}-\mathbf{k}'),
\end{equation} and we expand the mode operators in terms of creation and annihilation operators:
\begin{equation}
    \hat{f}_\mathbf{k} = \frac{1}{\sqrt{2k}} \left(\hat{a}_\mathbf{k} + \hat{a}_{-\mathbf{k}}^\dagger\right), \qquad \hat{P}_\mathbf{k} = -i \sqrt{\frac{k}{2}} \left(\hat{a}_\mathbf{k} - \hat{a}_{-\mathbf{k}}^\dagger\right).
\end{equation} The creation and annihilation operators obey the standard commutation relations
\begin{equation}
    [\hat{a}_\mathbf{k}, \hat{a}_{\mathbf{k}'}^\dagger] = \delta(\mathbf{k}-\mathbf{k}'), \qquad [\hat{a}_\mathbf{k}, \hat{a}_{\mathbf{k}'}] = [\hat{a}_\mathbf{k}^\dagger, \hat{a}_{\mathbf{k}'}^\dagger] = 0.
\end{equation} The Hamiltonian then takes the form~\cite{Albrecht:1992kf}
\begin{equation}
    \hat{H} = \int d^3\mathbf{k} \Bigg[ \frac{k}{2} (\hat{a}_\mathbf{k} \hat{a}_\mathbf{k}^\dagger + \hat{a}_{-\mathbf{k}} \hat{a}_{-\mathbf{k}}^\dagger) - \frac{i}{2} \frac{z'}{z} (\hat{a}_\mathbf{k} \hat{a}_{-\mathbf{k}} - \hat{a}_{-\mathbf{k}}^\dagger \hat{a}_\mathbf{k}^\dagger) \Bigg].
\end{equation}

\section{Decoupling Limit}\label{sec:appendix_a}
In this appendix, we rewrite the equation of motion in terms of the e-fold variable $N=\ln{a}$, and show that in the quasi-de Sitter limit the field perturbations effectively decouple from the metric perturbations.  The background dynamics, Eq.~(\ref{eq:conformal_background_dynamics}), in e-folds take the form
\begin{equation}\label{eq:background_efolds_equation}
  H^2 = \frac{V}{3  - \epsilon_1} , \qquad
  \frac{d^2 \bar{\phi}}{dN^2} + (3-\epsilon_1)\frac{d\bar{\phi}}{dN} + \frac{1}{H^2}\frac{dV}{d\phi} = 0,
\end{equation} where the prime on $V'(\bar{\phi})$ is the derivative with respect to background field and $\epsilon_1=-\frac{1}{H}\frac{dH}{dN}$.

In spatially-flat gauge, using Eq.~(\ref{eq:curvature_variable_appendix}) and working in Fourier space with $f_k=a\phi_k$, the mode equation Eq.~(\ref{eq:ms_appendix}) written in terms of $N$, for the field perturbation takes the form
\begin{equation}
    \frac{d^2\phi_k}{dN^2} + (3 - \epsilon_1)\frac{d\phi_k}{dN} + \Big[ \frac{k^2}{a^2 H^2}  +
    \frac{1}{H^2}\frac{d^2V}{d\phi^2}-\frac{1}{a^3H}\frac{d}{dN}\left(a^3H\big(\frac{d\bar{\phi}}{dN}\big)^2\right)\Big]\phi_k(N) = 0.
\end{equation} 

 The metric perturbations contribute to the effective mass through
\begin{align*}
    \delta\equiv \frac{1}{H a^3}\frac{d}{dN}\!\left(a^3 H \big(\frac{d \bar\phi}{dN}\big)^2\right)
    &=6 \epsilon_1 - 2 \epsilon_1^2 + 2 \epsilon_1 \epsilon_2.
\end{align*} In a quasi--de~Sitter background ($\epsilon_1 \ll 1$), we have $\eta \simeq -1/\mathcal{H}$. This implies $\delta \ll 1$ even when $|\epsilon_2| \sim \mathcal{O}(1)$, since it depends only on the combination $\epsilon_1 \epsilon_2$, which is subdominant. 
For the Starobinsky potential, the background is quasi-de~Sitter with $\epsilon_1\ll1$ throughout the dynamics. However, we have $d^2V/d\phi^2 = 0$ everywhere except at transition, which are treated separately by matching the solutions across them. Consequently, metric perturbations are subdominant, and field perturbations can be considered to effectively decouple from metric perturbations. In this limit, the canonical conjugate momentum associated with the field perturbation is given by~\cite{Langlois:1994ec,Grain:2017dqa}
\begin{equation}\label{eq:conjugate_field_momentum}
     \pi_k\equiv a^3H\frac{d\phi_k}{dN}.
\end{equation}

\section{Green's Function}\label{sec:greens_function}
In this appendix, we derive the Green's function corresponding to the coarse-grained equation of motion.  The coarse-grained equation, Eq.~(\ref{eq:stoc_eom}), can be rewritten in terms of conformal time, which makes it convenient to find solutions.  In conformal time $\eta$, we define $v_k = a \, \delta \phi_{\rm cg, k}$ and the equation of motion for $v_k$ then becomes 
\begin{equation}
    \left[\frac{d^2}{d\eta^2} + k^2 - \frac{a''}{a} + a^2 \frac{d^2V}{d\phi^2} \right] v_k = 0.
\end{equation}
Using the slow-roll parameter $\epsilon_1$ and the conformal Hubble rate $\mathcal{H} = a H$, with $a\simeq-1/\eta H$ ($\epsilon_1\ll1$ throughout the dynamics of this model) this can be rewritten as
\begin{equation}\label{eq:conformal_cgequation}
    \left[\frac{d^2}{d\eta^2} + k^2 - \big(2 - \epsilon_1 - \frac{1}{H^2}\frac{d^2V}{d\phi^2}\big) \mathcal{H}^2  \right] v_k = 0.
\end{equation}
The potential is linear except at the transition, where
$d^2V/d\phi^2$ induces a delta-function contribution due to jump in $dV/d\phi$. The solutions for $v_k$ must be matched across this sharp transition in the potential at $\eta = \eta_T$, the matching condition reads
\begin{align}\label{eq:appendix_matching}
v_k(\eta_T^-) = v_k(\eta_T^+),\qquad    v_k'(\eta_T^-) - v_k'(\eta_T^+) &= - \frac{\Delta d^2V/d\phi^2}{H^2} \mathcal{H}^2 \, v_k(\eta_T)\nonumber\\
    &=- \frac{\Delta dV/d\phi}{|\bar\phi_{cg}'|H^2}\mathcal{H}^2v_k(\eta_T)=3\frac{(A_+-A_-)}{V_0}(-\frac{V_0}{\mathcal{H}A_+})\mathcal{H}^2v_k(\eta_T)\nonumber\\
    &=-3\mathcal{H}(\frac{A_+-A_-}{A_+})v_k(\eta_T),
\end{align} where the jump magnitude is same as obtained in Eq.~(\ref{eq:discontinuity}) since the metric perturbations are subdominant. Away from the transition, $d^2V/d\phi^2=0$ and the two linearly independent homogeneous solutions to Eq.~(\ref{eq:conformal_cgequation}) (with $\epsilon_1=0$) can be written as
\begin{equation}
u_2(\eta)=u_1^{*}(\eta),\quad u_1(\eta)=
\begin{cases}
    \displaystyle \frac{1}{\sqrt{2k}}
    \left(1-\frac{i}{k\eta}\right)e^{-ik\eta},
    & \eta\le\eta_T, \\[4pt]
    \displaystyle \frac{1}{\sqrt{2k}} \left[ \alpha_k\left(1-\frac{i}{k\eta}\right)e^{-ik\eta} +\beta_k\left(1+\frac{i}{k\eta}\right)e^{ik\eta}
\right], & \eta>\eta_T .
\end{cases}
\end{equation} The Bogoliubov coefficients satisfy $|\alpha_k|^2-|\beta_k|^2=1$ and the corresponding field fluctuations $u_{1,2}(N)=u_{1,2}(\eta)/a$ obey the conserved Wronskian
\begin{equation}\label{eq:wronskian}
W(u_1(N'),u_2(N'))=
\begin{cases}
\displaystyle \frac{i}{a^3H},
& N\le N_T, \\[4pt]
\displaystyle \frac{i}{a^3H}(|\alpha_k|^2-|\beta_k|^2), & N>N_T .
\end{cases}
\end{equation} We then construct from the field fluctuations in e-fold $u_{1,2}(N)$, a set of two linearly independent solutions
\begin{align}
    y_1(N)&=u_1(N_i)u_2(N)-u_2(N_i)u_1(N),\\
    y_2(N)&=\frac{du_1}{dN}\Big|_{N_i} u_2(N)-\frac{du_2}{dN}\Big|_{N_i}u_1(N),
\end{align}  and construct the Green's function
\begin{align}\label{eq:appendix_greens_fucntion_full}
    G_k(N,N')&=\frac{y_1(N')y_2(N)-y_1(N)y_2(N')}{W(y_1(N'),y_2(N'))}\nonumber\\&= \frac{(x-y) \cos (x-y)-(x y+1) \sin (x-y)}{y^3}\Theta (N-N').
\end{align} which gives Eq.(\ref{eq:greensfunction_fullgrad}) with step function written in terms of $ x=k/a(N)H,~~y=k/a(N')H$ there. 

\subsection{Long-wavelength limit }
In this subsection, we further ignore the gradient term in the coarse-grained equation Eq.~(\ref{eq:conformal_cgequation})
\begin{equation}\label{eq:coarse-grained:ms}
    \left[\frac{d^2}{d\eta^2} - \big(2  -\epsilon_1- \frac{1}{H^2}\frac{d^2V}{d\phi^2}\big) \mathcal{H}^2  \right] v_k = 0,
\end{equation} and the jump magnitude remains unchanged but the homogeneous solutions with $\epsilon_1=0$ is now 
\begin{equation}
    v_2(\eta)=v_1^{*},\qquad v_1(\eta)=
\begin{cases}
    \displaystyle -\frac{\eta ^2}{3}\sqrt{\frac{k^3}{2}}-\frac{i}{ \eta }\frac{1}{\sqrt{2k^3}}, & \eta\le\eta_T, \\[4pt]
\displaystyle \tilde\alpha_k(-\frac{\eta ^2}{3}\sqrt{\frac{k^3}{2}}-\frac{i}{ \eta }\frac{1}{\sqrt{2k^3}})+\tilde \beta_k(-\frac{\eta ^2}{3}\sqrt{\frac{k^3}{2}}-\frac{-i}{ \eta }\frac{1}{\sqrt{2k^3}}), & \eta>\eta_T .
\end{cases}
\end{equation} Here, we choose the constants\footnote{Although any normalization works for the construction of the Green's function as long as we use two-linearly independent solutions of homogeneous
equation.} of solutions so that the constant mode of the field fluctuation $\delta\phi_k=v_k/a$ corresponds to the late-time power spectrum amplitude and matches the values at the horizon up to an overall phase and the Wronskian is of the same form as Eq~(\ref{eq:wronskian}).  These solutions are then matched across the transition using the same conditions Eq~(\ref{eq:appendix_matching}), as the discontinuity is in the potential. The Bogoliubov normalization remains $ |\tilde \alpha_k|^2 - |\tilde \beta_k|^2 = 1.$ Then we construct from the field fluctuations in e-fold $v_{1,2}(N)=v_{1,2}(\eta)/a$, a set of two linearly independent solutions
\begin{align}
\bar y_1(N)&=v_1(N_i)v_2(N)-v_2(N_i)v_1(N),\\
   \bar  y_2(N)&=\frac{dv_1}{dN}\Big|_{N_i} v_2(N)-\frac{dv_2}{dN}\Big|_{N_i}v_1(N),
\end{align} and construct the Greens function
\begin{align} \label{eq:dSGreen_noGrad}
    G_k(N,N')&=\frac{\bar{y}_1(N')\bar{y}_2(N)-\bar{y}_1(N)\bar{y}_2(N')}{W(\bar{y}_1(N'),\bar{y}_2(N'))}\nonumber\\&=\frac{y^3-x^3}{3y^3}\Theta (N-N'),
\end{align} for the case without the gradient terms in the Langevin equation.

\section{Coarse-grained Power Spectrum}\label{sec:coarse_grained_power_spectrum}

In this appendix, we calculate the linear coarse-grained power spectrum before and after the transition. For convenience, we introduce
\begin{equation}
    x=-k\eta = \frac{k}{a(N)H},
\end{equation} so that time integrals can be expressed in terms of $x$. We also define
\begin{equation}\label{eq:int_variables}
    y_1 \equiv x(N_1), \quad y_2 \equiv x(N_2), \quad x_i \equiv x(N_i),
\end{equation} where $N_i$ is the beginning of inflation, and the integrals are evaluated in the regime $0 \le x(N) < \sigma < x_i$ with $\sigma$ being the coarse-graining scale.
\subsection{Pre-transition case}
\label{appendix:pre-transition_correlator}

\subsubsection{White noise case}
We now compute the coarse-grained correlator (see Eq~(~\ref{eq:c_function})) in the pre-transition regime. Rewriting it in terms of the variables in Eq.~(\ref{eq:int_variables}), the $\phi\phi$ contribution to the correlator is then given by
\begin{equation}\label{eq:dS_phiphi_term}
\bar{\mathcal{C}}_{k}^{\phi\phi}(x(N),x(N)) = \int_{x_i}^{x(N)} \frac{dy_1}{-y_1} \int_{x_i}^{x(N)} \frac{dy_2}{-y_2} \, G_k(x, y_1) A(k, y_1) \phi_k(y_1) \, G_k(x, y_2) A(k, y_2) \phi_k^*(y_2),
\end{equation}
where $G_k$ is the Green's function.  Using the white-noise window function (Eq.~\ref{eq:white_noise}), the delta functions collapse the integrals\footnote{We have used the identities $\delta(g(z))=\sum_j |g'(z_j)|^{-1}\delta(z-z_j)$, with $g(z_j)=0$, and $\int dz\,w(z)\delta'(z-z_j)=-w'(z_j)$ for generic functions $g$ and $w$.} to give
\begin{equation}
    \bar{\mathcal{C}}_{k}^{\phi\phi}(x,x) = \Big| \Big( 3 \phi_k(\sigma)+\sigma  \frac{d\phi_k(y)}{dy} \big|_{y=\sigma}  \Big) G_k(x,\sigma) + \sigma \phi_k(\sigma) \frac{\partial G_k(x,y)}{\partial y} \Big|_{y=\sigma} \Big|^2.
\end{equation}  The mode functions are yet to be specified. Similarly, we get the $\pi\pi$ and mixed contributions 
\begin{align}
    \bar{\mathcal{C}}_{k}^{\pi\pi}(x,x) &= \Big| -2 \sigma \frac{d\phi_k(y)}{dy} \big|_{y=\sigma} G_k(x,\sigma) \Big|^2, \\
    \bar{\mathcal{C}}_{k}^{\phi\pi}(x,x) &= \Big(-2 \sigma \frac{d\phi_k(y)}{dy} \big|_{y=\sigma} G_k(x,\sigma)\Big) \Big[\Big(\sigma \frac{d\phi_k(y)}{dy} \big|_{y=\sigma} + 3 \phi_k(\sigma) \Big) G_k(x,\sigma) \\&+ \sigma \phi_k(\sigma) \frac{\partial G_k(x,y)}{\partial y} \Big|_{y=\sigma}\Big],\nonumber\\ \bar{\mathcal{C}}_{k}^{\pi\phi}(x,x)&=(\bar{\mathcal{C}}_{k}^{\phi\pi}(x,x))^*
\end{align} Summing them, the correlator reduces to a single squared modulus, giving the coarse-grained power spectrum\footnote{In Eq.~(\ref{eq:white_noise_general}), this is expressed in terms of the canonical momentum $\pi_k = a^3 H \frac{d\phi_k}{dN} = a^3 H (-y \frac{d\phi_k}{dy})$.} in white noise case:
\begin{equation}\label{eq:appendix_white_noise_general}
    \mathcal{P}_{\delta\phi_{cg}} = \frac{k^3}{2\pi^2} \bar{\mathcal{C}}_k(N,N) = \frac{k^3}{2\pi^2} \Big| G_k(x,\sigma) \Big(3 \phi_k(\sigma) - \sigma \frac{d\phi_k(y)}{dy} \big|_{y=\sigma} \Big) + \sigma \phi_k(\sigma) \frac{\partial G_k(x,y)}{\partial y} \Big|_{y=\sigma} \Big|^2.
\end{equation} We see the power spectrum depends on noise amplitudes of both the canonical variables as well as the coarse-grained dynamics encoded in the greens function. We substitute the noise amplitudes using the pre-transition mode functions from Eq.~(\ref{eq:usr_modefunction}) 
\begin{equation}\label{eq:modefunction_pretransition}
    \phi_k(y) =\frac{f_k}{a} =\frac{H}{\sqrt{2 k^3}} (i + y) e^{i y}, \qquad \frac{\pi_k(y)}{a^3 H} = -y\frac{d\phi_k}{dy} = \frac{H}{\sqrt{2 k^3}} (- i y^2) e^{i y},
\end{equation} and the Green's function from Eq.~(\ref{eq:appendix_greens_fucntion_full}) to obtain
\begin{align}
    \mathcal{P}_{\delta\phi_{cg}}(k,N) &= \frac{k^3}{2\pi^2} \Big| \frac{H e^{i x} (x+i)}{\sqrt{2} k^{3/2}} \Theta(N - N_\sigma) \Big|^2
\nonumber\\&= \frac{H^2}{4\pi^2} (1 + x^2) \Theta(N - N_\sigma).
\end{align}  We notice that $\sigma$ terms cancels and the spectrum is independent of $\sigma$ and reproduces the power spectrum of the quantum state initialized in the Bunch-Davies vacuum in the pre-transition case in Eq.~(\ref{eq:usr_qft_spectrum}). In the super-horizon limit $x \to 0$, this reduces to the scale-invariant spectrum.

However, in the long-wavelength limit, we neglect the gradient terms in the coarse-grained equation, corresponding to the Green’s function in Eq.~(\ref{eq:dSGreen_noGrad}). Using the mode functions in Eq.~(\ref{eq:modefunction_pretransition}) again, we get 
\begin{align}
    \mathcal{P}_{\delta\phi_{cg}}(k,N) &= \frac{k^3}{2\pi^2}
    \Big| \frac{H e^{i \sigma} \big( i + \sigma- \frac{i \sigma^2}{3}  + \frac{i x^3}{3 \sigma}  \big)}{\sqrt{2} k^{3/2}} \Big|^2 \Theta(N - N_\sigma)\\
    &= \frac{H^2}{4\pi^2} \Big(1 + \frac{\sigma^2}{3} + \frac{\sigma^4}{9} + \frac{x^6}{9 \sigma^2} - \frac{2 \sigma x^3}{9} + \frac{2 x^3}{3 \sigma} \Big) \Theta(N - N_\sigma),
\end{align} where the stochastic spectrum in this approximation depends explicitly on $\sigma$. This $\sigma-$dependence is a consequence of dropping the gradient term, and the coarse-grained spectrum approaches the power spectrum of the quantum state for $\sigma \ll 1$.

\subsubsection{Colored noise case}
To compute the correlator for the colored-noise case, we use the exponential and Gaussian window functions (Eqs.~(\ref{eq:color_filter_exponential}) and (\ref{eq:color_filter_gaussian})) with the Green's function in Eq.~(\ref{eq:greensfunction_fullgrad}). Using the noise amplitudes from the pre-transition mode functions from Eq.~(\ref{eq:modefunction_pretransition}), we obtain the power spectrum: 
\paragraph{Exponential filter $(n=1)$:}
\begin{align}
    \frac{k^3}{2\pi^2}\mathcal{C}_{k}(N,N) =& \frac{H^2}{4\pi^2} \Bigg[ (1+x^2)\, e^{-\frac{x+x_i}{\sigma}}  \Big(e^{\frac{x}{2\sigma}} - e^{\frac{x_i}{2\sigma}}\Big)^2  \nonumber\\ 
    &+ \frac{(x_i^2+1)\, e^{-x_i/\sigma} \big[(1+x_i x) \sin(x-x_i) - (x-x_i) \cos(x-x_i)\big]^2}{4 x_i^4 \sigma^2}  \nonumber\\ 
    &+ \frac{(e^{x_i/(2\sigma)} - e^{x/(2\sigma)})\, e^{-(2x_i+x)/(2\sigma)}}{2 x_i^2 \sigma}  \Big[ (x_i + (x_i-1)x + 1)(x_i(x-1)+x+1)\sin 2(x-x_i)  \nonumber\\ 
    &\quad - 2(x-x_i)(x_i x+1)\cos 2(x-x_i) \Big] \Bigg].
\end{align}

\paragraph{Gaussian filter $(n=2)$:}
\begin{align}
    \frac{k^3}{2\pi^2}\mathcal{C}_{k}(N,N) &= \frac{H^2}{4\pi^2} \Bigg[ (1+x^2) \, e^{-(x^2+x_i^2)/\sigma^2}  \Big(e^{x^2/(2\sigma^2)} - e^{x_i^2/(2\sigma^2)} \Big)^2  \nonumber\\ 
    &\quad + \frac{e^{-(2x_i^2+x^2)/(2\sigma^2)} \, \big[ (x-x_i)\cos(x-x_i) - (x_i x +1) \sin(x-x_i) \big] }{x_i^2 \sigma^6} \nonumber\\ 
    &\qquad \times \Big( f_1(x,x_i,\sigma) \cos(x-x_i) + f_2(x,x_i,\sigma) \sin(x-x_i) \Big) \Bigg], \\[2mm] f_1(x,x_i,\sigma) &= e^{x^2/(2\sigma^2)} \Big[ 2\sigma^4 x_i (x x_i + 1) + (x-x_i)(x_i^2+1) \Big]  - 2 \sigma^4 x_i (x x_i +1) e^{x_i^2/(2\sigma^2)}, \nonumber\\ 
    f_2(x,x_i,\sigma) &= 2 \sigma^4 x_i (x_i - x) e^{x_i^2/(2\sigma^2)}  - e^{x^2/(2\sigma^2)} \Big[ x_i \big( 2\sigma^4(x_i - x) + x x_i^2 + x_i + x \big) + 1 \Big]. \nonumber
\end{align}
The coarse-grained spectrum in this case depends on both the coarse-graining parameter $\sigma$ and the initial time $x_i$ at which coarse-graining begins, indicating sensitivity to earlier times through non-local contributions. Therefore, the colored-noise case differs from the white-noise case by retaining sensitivity to sub-horizon mode amplitudes. However, for $x_i \gg \sigma$, the memory of initial conditions is exponentially suppressed as the window function decays for modes far from the coarse-graining scale $k = \sigma a H$.

In this limit $x_i\to\infty$, the above formulas simplifies to
\begin{equation}\label{eq:pre_color_spectrum}
    \mathcal{P}^{(n=1)}_{\delta\phi_{cg}}(k,N) \simeq e^{-\frac{x}{\sigma}} \frac{H^2}{4\pi^2} (1+x^2), \qquad
    \mathcal{P}^{(n=2)}_{\delta\phi_{cg}}(k,N) \simeq e^{-\frac{x^2}{\sigma^2}} \frac{H^2}{4\pi^2} (1+x^2),
\end{equation} for exponential and Gaussian window functions, respectively. In the super-horizon limit $x \to 0$, both reduce to the scale-invariant spectrum $\mathcal{P}_{\delta\phi_{cg}}^{(n=1,2)} = H^2/(4\pi^2)$.

\subsection{Post-transition case}\label{appendix:post-transition_correlator}

For the post-transition case, the mode functions appearing in the noise source are linear combinations of the pre-transition modes,
\begin{equation}
    \phi_k^{\rm post}(y)=\alpha_k\phi_k^+(y)+\beta_k\phi_k^-(y),
    \qquad
    \frac{\pi_{k}^{\rm post}(y)}{a^3H}=\alpha_k\phi_N^+(y)+\beta_k\phi_N^-(y),
\end{equation} where the Bogoliubov coefficients $\alpha_k$ and $\beta_k$ are defined in Eq.~(\ref{eq:alpha_beta}) and 
\begin{equation}
    \phi^\pm\equiv\frac{H}{\sqrt{2k^3}}(y\pm i)e^{\pm iy},\quad
    \quad\phi_{N}^\pm\equiv\frac{H}{\sqrt{2k^3}}(\mp iy^2)e^{\pm iy}.
\end{equation} This allows us to write the correlator compactly as
\begin{equation}\label{eq:post_transition_correlator}
    \mathcal{C}_k(N,N) = (|\alpha_k|^2 + |\beta_k|^2)\,\mathcal{I}_{+-} + \alpha_k \beta_k^*\, \mathcal{I}_{++} + \beta_k \alpha_k^*\, \mathcal{I}_{++}^*,
\end{equation} where $\mathcal{I}_{-+}^{*}=\mathcal{I}_{+-}=\bar{\mathcal{C}}_{k}(N,N)$ is the pre-transition correlator and hence it remains to evaluate $\mathcal{I}_{++}$ to obtain full correlator for post-transition case. To obtain the above relation, we introduce the shorthand notation \begin{equation}
    I(r_,q_,\phi^{\pm},\phi^{\pm})=\int dy_1dy_2~r(y_1)q(y_2)~\phi^{\pm}(y_1)\phi^{\pm}(y_2),
\end{equation} where
\begin{align}
    r(y_1)\equiv\frac{-1}{y_1}G_k(x,y_1)A(k,y_1);\quad
    q(y_2)\equiv\frac{-1}{y_2}G_k(x,y_2)B(k,y_2).
\end{align} The correlator is then given by the four contributions
\begin{align*}
  \mathcal{C}_{k}^{\phi\phi}(N,N)
   &=|\alpha_k|^2I(r,r,\phi^{+},\phi^{-})+|\beta_k|^2I(r,r,\phi^{-},\phi^{+})+\alpha_k\beta_k^*I(r,r,\phi^+,\phi^{+})+\beta_k\alpha_k^*I(r,r,\phi^-,\phi^{-}),\\
   \mathcal{C}_{k}^{\pi\pi}(N,N)
   &=|\alpha_k|^2I(q,q,\phi_N^{+},\phi_N^{-})+|\beta_k|^2I(q,q,\phi_N^{-},\phi_N^{+})+\alpha_k\beta_k^*I(q,q,\phi_N^+,\phi_N^{+})+\beta_k\alpha_k^*I(q,q,\phi_N^-,\phi_N^{-}),\\
   \mathcal{C}_{k}^{\phi\pi}(N,N)
   &=|\alpha_k|^2I(r,q,\phi^{+},\phi_N^{-})+|\beta_k|^2I(r,q,\phi^{-},\phi_N^{+})+\alpha_k\beta_k^*I(r,q,\phi^+,\phi_N^{+})+\beta_k\alpha_k^*I(r,q,\phi^-,\phi_N^{-}),\\
     \mathcal{C}_{k}^{\pi\phi}(N,N)
   &=|\alpha_k|^2I(q,r,\phi_N^{+},\phi^{-})+|\beta_k|^2I(q,r,\phi_N^{-},\phi^{+})+\alpha_k\beta_k^*I(q,r,\phi_N^+,\phi^{+})+\beta_k\alpha_k^*I(q,r,\phi_N^-,\phi^{-}).
\end{align*} Then, collecting all the terms we get
\begin{align}\label{eq:usr_correlator}
    \mathcal{C}_k(N,N)&=
    |\alpha_k|^2\,\mathcal{I}_{+-} + |\beta_k|^2\,\mathcal{I}_{-+} + \alpha_k\beta_k^*\,\mathcal{I}_{++} + \beta_k\alpha_k^*\,\mathcal{I}_{--},
\end{align} with
\begin{align}
\mathcal{I}_{+-}&=I(r,r,\phi^+,\phi^-)+I(q,q,\phi_N^+,\phi_N^-)+I(r,q,\phi^+,\phi_N^-)+I(q,r,\phi_N^+,\phi^-),\\
\mathcal{I}_{++}&=I(r,r,\phi^+,\phi^+)+I(q,q,\phi_N^+,\phi_N^+)+I(r,q,\phi^+,\phi_N^+)+I(q,r,\phi_N^+,\phi^+),
\end{align} and note that $\mathcal{I}_{+-}=\bar{\mathcal{C}}_{k}(N,N)$ is the pre-transition correlator (for example
$I(r,r,\phi^{+},\phi^{-})=\bar{\mathcal{C}}_{k}^{\phi\phi}(N,N)$ is the $\phi\phi$ term of Eq.~(\ref{eq:dS_phiphi_term}) and
so on). Now, since $\mathcal{I}_{--}=(\mathcal{I}_{++})^{*}$, it remains only to evaluate $\mathcal{I}_{++}$ to obtain full correlator $\mathcal{C}_k(N,N).$
\paragraph{White noise case:} Evaluating it in the white noise case using Eq.~(\ref{eq:white_noise}) and the Green's function from Eq.~(\ref{eq:appendix_greens_fucntion_full}), we get
\begin{align*}
    \mathcal{I}_{++}&=\frac{H^2 e^{2 i x} (x+i)^2}{2 k^3}.
\end{align*} Then, using this in Eq.~(\ref{eq:usr_correlator}), the power spectrum can be written in a compact form
\begin{align}
    \mathcal{P}_{\delta\phi_{cg}}(k,N)&=\frac{k^3}{2\pi^2}\mathcal{C}_k(N,N)\\
    &=\frac{k^3}{2\pi^2}\Bigg[(|\alpha_k|^2+|\beta_k|^2)\Big(\frac{H^2}{2k^3}(1+x^2)\Big)+2\,\mathrm{Re}\left\{ \alpha_k\beta_k^{*} \frac{H^2}{2k^3}e^{2ix}(i+x)^2 \right\}\Bigg]\\
    &=\frac{H^2}{4\pi^2}\Big|\alpha_k(i+x)e^{+ix}+\beta_k(-i+x)e^{-ix}\Big|^2.
\end{align} For the case without the gradient term, using the Green's function in Eq.~(\ref{eq:dSGreen_noGrad}), we obtain
\begin{equation}
    \mathcal{I}_{++}  =\frac{H^2}{2k^3}e^{2i \sigma } \left(i+\sigma-\frac{i \sigma ^2}{3} +\frac{i x^3}{3 \sigma
    }\right)^2,
\end{equation} and the coarse-grain power spectrum is
\begin{equation}
 \mathcal{P}_{\delta\phi_{cg}}(k,N)  = \frac{H^2}{4\pi^2}\Big|\alpha_ke^{i \sigma }\left(i+\sigma-\frac{i \sigma ^2}{3} +\frac{i x^3}{3 \sigma
  }\right)+\beta_ke^{-i \sigma }\left(-i+\sigma+\frac{i \sigma ^2}{3}-\frac{i x^3}{3 \sigma }\right)\Big|^2.
\end{equation}
\paragraph{Color noise case:} To compute the coarse-grained correlator for the colored-noise case, we use the exponential and Gaussian window functions in Eqs.~(\ref{eq:color_filter_exponential}) and (\ref{eq:color_filter_gaussian}), together with the Green's function from Eq.~(\ref{eq:greensfunction_fullgrad}). The noise amplitudes are taken from the mode functions in Eq.~(\ref{eq:modefunction_pretransition}) and compute $\mathcal{I}_{++}$ to obtained :

\begin{align}
\mathcal{I}_{++}^{(1)} &=  \frac{H(N)^2}{32 k^3 \sigma^2 x_i^4} 
\, e^{-(x_i + 2 i \sigma x + x)/\sigma}  \Biggl[ \; 4 i \sigma (x+i) x_i^2 \, e^{x_i/(2\sigma) + 2 i x} 
- (x-i) (x_i+i)^2 \, e^{x/(2\sigma) + 2 i x_i} 
\nonumber\\
&\quad + (x+i) \, e^{\frac{1}{2} (\frac{1}{\sigma} + 4 i) x} 
\left(1 + (1 - 4 i \sigma) x_i^2 \right)
\Biggr]^2,
\\[1em]
\mathcal{I}_{++}^{(2)} &= 
- \frac{H^2}{8 x_i^2 k^3 \sigma^4} 
\, e^{-(x_i^2 + x^2)/\sigma^2 - 2 i (2 x_i + x)} 
\Biggl[
- (x-i) (x_i^2 + 2 i x_i \sigma^2 + 1) \, e^{x^2/(2\sigma^2) + 2 i x_i} 
\nonumber\\
&\quad + (x_i - i)^2 (x+i) \, e^{\frac{1}{2} x (\frac{x}{\sigma^2} + 4 i)} 
+ 2 x_i \sigma^2 (1 + i x) \, e^{\frac{1}{2} x_i (\frac{x_i}{\sigma^2} + 4 i)}
\Biggr]^2.
\end{align} for exponential (Gaussian) $n=1(2)$ smooth filters. As before, we assume that $x(N_i)\gg\sigma$ and simplify to get
\begin{align*}
 \lim_{x_i\to\infty}\mathcal{I}_{++}^{(1)}   &\simeq e^{-\frac{x}{\sigma}}\frac{ H^2}{2 k^3}e^{2 i x}
(i+x)^2,\qquad\lim_{x_i\to\infty}\mathcal{I}_{++}^{(2)}   \simeq e^{-(\frac{x}{\sigma})^2}\frac{ H^2}{2 k^3}e^{2 i x} (i+x)^2,
\end{align*} and using this in Eq.~(\ref{eq:post_transition_correlator}) as well as using Eq.~(\ref{eq:pre_color_spectrum}), the power spectrum for colored noise case can be written as
\begin{equation}
\mathcal{P}_{\delta\phi_{cg}}^{(n)}(k, N) 
\simeq e^{-(\frac{x}{\sigma})^n} \, \frac{H^2}{4 \pi^2} 
\Bigl| \, \alpha_k (i+x) e^{i x} + \beta_k (-i+x) e^{-i x} \, \Bigr|^2.
\end{equation}

\bibliographystyle{JHEP}
\bibliography{biblio.bib}

\end{document}